\documentclass[%
 reprint,
 amsmath,amssymb,
 aps,
]{revtex4-2}

\usepackage{graphicx}
\usepackage[utf8]{inputenc}
\usepackage{setspace}
\usepackage{enumitem}
\usepackage{latexsym,amsfonts}
\usepackage{etoolbox}
\usepackage{amsmath}
\usepackage{amssymb}
\usepackage[version=4]{mhchem}
\usepackage{caption}
\usepackage{subcaption}
\usepackage{xr}
\usepackage{hyperref}
\usepackage{graphicx}
\usepackage{dcolumn}
\usepackage{bm}
\usepackage[usenames,dvipsnames]{color}
\usepackage{stmaryrd} 

\usepackage{cancel}
\usepackage[nice]{nicefrac}

\usepackage[normalem]{ulem}
\usepackage{rotating}
\usepackage{filecontents}
\begin{document}
\setlength\parindent{0pt}
\preprint{AIP/123-QED}

\title{Accuracy of reaction coordinate based rate theories for modelling chemical reactions: insights from the thermal isomerization in retinal}%
\author{Simon Ghysbrecht}
\affiliation{ 
Freie Universität Berlin, Department of Biology, Chemistry and Pharmacy, Arnimallee 22, 14195 Berlin
}
\author{Luca Donati}
\affiliation{ 
Freie Universität Berlin, Department of Mathematics and Computer Science, Arnimallee 22, 14195 Berlin
}
\affiliation{ 
Zuse Institute Berlin, Takustra\ss e 7, 14195 Berlin
}
\author{Bettina G.~Keller}%
\email{bettina.keller@fu-berlin.de}
\affiliation{ 
Freie Universität Berlin, Department of Biology, Chemistry and Pharmacy, Arnimallee 22, 14195 Berlin
}
%

\date{\today}


\begin{abstract}
Modern potential energy surfaces have shifted attention to molecular simulations of chemical reactions.
While various methods can estimate rate constants for conformational transitions in molecular dynamics simulations, their applicability to studying chemical reactions remains uncertain due to the high and sharp energy barriers and complex reaction coordinates involved. 
This study focuses on the thermal cis-trans isomerization in retinal, employing molecular simulations and comparing rate constant estimates based on one-dimensional rate theories with those based on sampling transitions and grid-based models for low-dimensional collective variable spaces.
%
%
Even though each individual method to estimate the rate passes its quality tests, the rate constant estimates exhibit considerable disparities. 
Rate constant estimates based on one-dimensional reaction coordinates prove challenging to converge, even if the reaction coordinate is optimized. 
However, consistent estimates of the rate constant are achieved by sampling transitions and by multi-dimensional grid-based models.
\end{abstract}

\maketitle
\section{Introduction}

Elucidating chemical reaction mechanisms and rates is a central goal in computational chemistry.
Yet, calculating this type of dynamical properties remains significantly more challenging than obtaining structural or thermodynamic information.
Making precise predictions of reaction rates is particularly difficult.
The difficulties arise from two main sources: inaccuracies in the model of the potential energy surface (PES) \cite{vitalini2015dynamic}, and inaccuracies in the method to calculate the rate on this PES \cite{ghysbrecht2023thermal}.
Modelling a chemical reaction often requires a highly accurate PES based on explicitly evaluating the electronic structure at each nuclear configuration. 
Until recently, the computational cost of electronic structure methods has been so large that their use has been confined to single-point calculations \cite{harvey2019scope} or short simulations of small systems \cite{marx2000ab}.
Only few rate theories can work with so little information.
Among them Eyring transition state theory \cite{eyring1935activated} remains the most frequently used method. 
However, several extensions of Eyring transition state theory, such as variational transition state theory\cite{keck1960variational,truhlar1980variational} and Grote-Hynes theory \cite{grote1980stable}, have been introduced to account for recrossing and the influence of the solvent.
With recent advances in electronic structure methods \cite{schade2023breaking, gaus2011dftb3} 
and the advent of neural network potentials \cite{kocer2022neural, noe2020machine}, molecular-dynamics (MD) simulations of chemical reactions in complex environments become possible, allowing for the explicit treatment of solvent effects and entropic effects.
A wide variety of methods to estimate rates\cite{Peters2017, pietrucci2017strategies}, that have been developed in the context of MD simulations of soft-matter systems, can now be applied to chemical reactions in complex environments.
Soft-matter systems are characterized by rugged PES with multiple minima connected by energy barriers that are in the same range as the thermal energy.
Examples are peptide\cite{salvalaglio2014assessing} and protein dynamics\cite{lane2013milliseconds}, molecular binding\cite{ahmad2022enhanced} or crystal nucleation\cite{arjun2020molecular}.
The accuracy of simulation-based rate estimates in the context of chemical reactions, which usually feature a single high and sharp barrier, is still a matter of debate.
It is important to acknowledge that simulation-based rate estimates are founded on classical mechanics and therefore do not account for quantum tunneling or energy quantization.
While quantum tunneling is significant in proton transfer reactions, its rate diminishes exponentially with the square root of the reactant's mass and the barrier height. 
As a result, for reactions involving carbon or other medium-mass atoms, quantum tunneling is observable only when the reactant molecule is highly strained and consequently the reaction barrier is low \cite{borden2016reactions}. 
However, energy quantization of the vibrational degrees of freedom does have a noticeable effect in most reactions, in particular, if the reactant molecule is rigid. 
For the thermal isomerization of protonated Schiff bases, which are closely related to retinal, neglecting the energy quantization incurs an error in the reaction rate of about a factor of three at room temperature \cite{ghysbrecht2023thermal}.
It is worth noting that one can incorporate the effect of energy quantization into the potential energy and thereby achieve quantum-corrected classical dynamics \cite{zhang2023vibrational}.
Simulation-based rate estimates broadly fall into two distinct categories.
The first approach is based on counting transitions across the reaction barrier. 
Since for most chemical reactions, the mean first passage time exceeds the accessible simulation time by far, one employs dynamical reweighting techniques, in which the sampling is enhanced and the transition count is subsequently reweighted \cite{kieninger2020dynamical, bolhuis2002transition, zuckerman2017weighted}.
Infrequent metadynamics\cite{tiwary2013metadynamics} falls into this category.

The second approach is based on assuming an effective dynamics along a one-dimensional reaction coordinate, which requires the free-energy surface and diffusion constant of diffusion profile as a function of this reaction coordinate. 
The influence of the neglected degrees of freedom and the curvature of the reaction coordinate on the system's dynamics are captured by these two functions, which can be readily estimated from atomistic simulations of the full molecular system\cite{tribello2014plumed, daldrop2018butane, hummer2005position}. 
From the effective dynamics, one may then derive analytical expressions for the rate constants. 
Kramers' rate theory \cite{kramers1940brownian, Hanggi1990} falls into this second category.
The advantage of Kramers' theory is that, given a reaction coordinate, the individual steps of this approach are well-established and straight-forward. 
However, both the free-energy surface and the diffusion constant depend on the reaction coordinate and thus the accuracy of the rate estimate hinges on the quality of this coordinate.
Furthermore, Kramers' analytical expressions for the rate fall into three limiting cases (friction regimes), and it is essential to ensure the correct friction regime is applied.
Both the barrier height and the ``sharpness" of the barrier, represented by the barrier frequency, determine the friction regime. 
The high-friction regime is induced by high barriers (compared to thermal energy) and broad barriers (barrier frequency compared the friction due to the implicit degrees of freedom).
The low and intermediate friction regimes are induced by low and sharp barriers.
Chemical reactions with high and sharp barriers fall into a middle ground, where it is not a priori clear whether the high-friction regime applies. 
To investigate how these effects play out in a chemical reaction, we study the thermal cis-trans isomerization around the C$_{13}$=C$_{14}$ double bond of retinal coupled to a lysine in vacuum \cite{hayashi2002structural, malmerberg2011time}.
As PES, we use an empirical force field, whose computational efficiency permits a broad comparison of rate estimates.
For a cis-trans isomerization one may use an empirical force field, because the molecule's sigma bonds stay intact. 
Our goal is to explore whether classical MD in combination with Kramers' rate theory can model this reaction with quantitatively accurate reaction rates and mechanism (on a given PES). 
As comparison, we include rate estimates for overdamped Langevin dynamics along a one-dimensional reaction coordinate (Pontryagin's rate theory \cite{pontryagin1933statistical}), grid-based models \cite{Lie2013,donati2021markov} of an effective dynamics in a multidimensional collective variable space, and infrequent metadynamics\cite{tiwary2013metadynamics}.

%

%

\section{Theory}
\label{sec:Theory}

\subsection{Definitions}
The cis-trans isomerization of retinal is a unimolecular reaction 
\begin{eqnarray}
\label{eq:A_to_B}
	A \ce{->[k_{AB}]} B \, ,
\end{eqnarray}
where $A$ is the cis isomer, $B$ is the trans isomer, and $k_{AB}$ is the reaction rate constant.
The rate constant is related to the mean first-passage time (MFPT) $\tau_{AB}$ by
\begin{eqnarray}
    k_{AB} &=& \frac{1}{\tau_{AB}} \, .
\label{eq:MFPT}    
\end{eqnarray}
The configuration of the molecule is given by the positions of its $N$ atoms in Cartesian space: $\mathbf{x}\in\Gamma_x \subset \mathbb{R}^{3N}$, where $\Gamma_x$ is called configuration space.  
We model the dynamics within the Born-Oppenheimer approximation, where $V(\mathbf{x})$ represents the Born-Oppenheimer potential energy of the electronic ground state. 
Reactant state $A \subset \Gamma_x$ and the product state $B \subset \Gamma_x$ are regions around minima in $V(\mathbf{x})$, whereas the transition state ($TS$) corresponds to a saddle point in $V(\mathbf{x})$.

Collective variables are low-dimensional representations of the $3N$-dimensional atomic positions. 
A collective variable vector is a (possibly non-linear) function 
\begin{eqnarray}
    \mathbf{q}: \Gamma_x \rightarrow \mathbb{R}^m
\end{eqnarray}
which maps each position $\mathbf{x} \in \Gamma_x$ onto a low-dimensional vector $\mathbf{q} \in \mathbb{R}^m$, 
where $m \ll 3N$.
The free energy along $\mathbf{q}$ is defined as:
\begin{equation}
    F(\mathbf{q}) = - RT \ln \pi (\mathbf{q})
    \label{eq:free_energy}
\end{equation} 
where $\pi(\mathbf{q})$ is the configurational Boltzmann density marginalized to the collective variable space
\begin{equation}
    \pi(\mathbf{q}) = Z_{\mathrm{conf}}^{-1}\int_{\Gamma_x} \mathrm{d}\mathbf{x} \, \exp\left(-\frac{V(\mathbf{x})}{RT}\right)\delta\left[\mathbf{q}(\mathbf{x}) - \mathbf{q}\right] \, .
    \label{eq:eq_distribution}
\end{equation}
Here, $\delta\left[\mathbf{q}(\mathbf{x}) - \mathbf{q}\right]$ is the Dirac delta function and $Z_{\mathrm{conf}}$ is the configurational part of the classical partition function
$Z_{\mathrm{conf}}  = \int_{\Gamma_x} \mathrm{d}\mathbf{x} \, \exp\left(\nicefrac{-V(\mathbf{x})}{RT}\right)$.

A reaction coordinate is a one-dimensional collective variable that scales monotonously between reactant state $A$ and product state $B$:
\begin{eqnarray}
q: \Gamma_x \rightarrow [0,1] \, .
\label{eq:reactionCoordinate}
\end{eqnarray}
$q$ is zero for the reactant state $A$ and one for product state $B$. 
In this manner, $q$ represents the progress of the reaction. 
Other intervals are also possible, but can be rescaled to $\left[0,1\right]$. 
The free energy $F(q)$ along the reaction coordinate is defined analogous to eqs.~\ref{eq:free_energy} and \ref{eq:eq_distribution}.

In eqs.~\ref{eq:free_energy} and ~\ref{eq:eq_distribution}, $R$ is the ideal gas constant and $T$ is the temperature. 
We calculate and report potential and free energies in units of J/mol, correspondingly the thermal energy is also reported as a molar quantity: $RT$. 
If units of energy are used for potential and free energies, $R$ should be replaced by the Boltzmann constant $k_B = R/N_A$ in eqs.~\ref{eq:free_energy} and ~\ref{eq:eq_distribution} and all of the following equations. 
$N_A$ is the Avogadro constant.
Equations of motion for the effective dynamics for $q$ and $\mathbf{q}$ (underdamped Langevin dynamics, overdamped Langevin dynamics with and without position dependent diffusion), as well as the associated Fokker-Planck operators are reported in section \ref{supp-sec:SI_theory} of the SI.
The equations of motion for the effective dynamics require an effective mass (molar) $\mu_q$, which can be estimated from the equipartition theorem\cite{daldrop2018butane}
\begin{eqnarray}
\langle E_{\mathrm{kin}}\rangle = \frac{1}{2} \mu_q \langle v^2\rangle  = \frac{1}{2}RT
\label{eq:effective_mass}
\end{eqnarray}
where $\langle v^2\rangle$ is the average squared velocity along $q$.

The rate theories introduced in the following, with the exception of the grid-based models, all assume separation of time scales. 
That is, on average, the system should fully sample the local equilibrium distribution within $A$, before it escapes over the transition state $TS$. 
This is only the case if the free energy barrier $F^\ddagger_{AB}$ of the reaction is much larger than the thermal energy: $F^\ddagger_{AB}  \gg RT$.

\begin{figure*}
\includegraphics[scale=1]{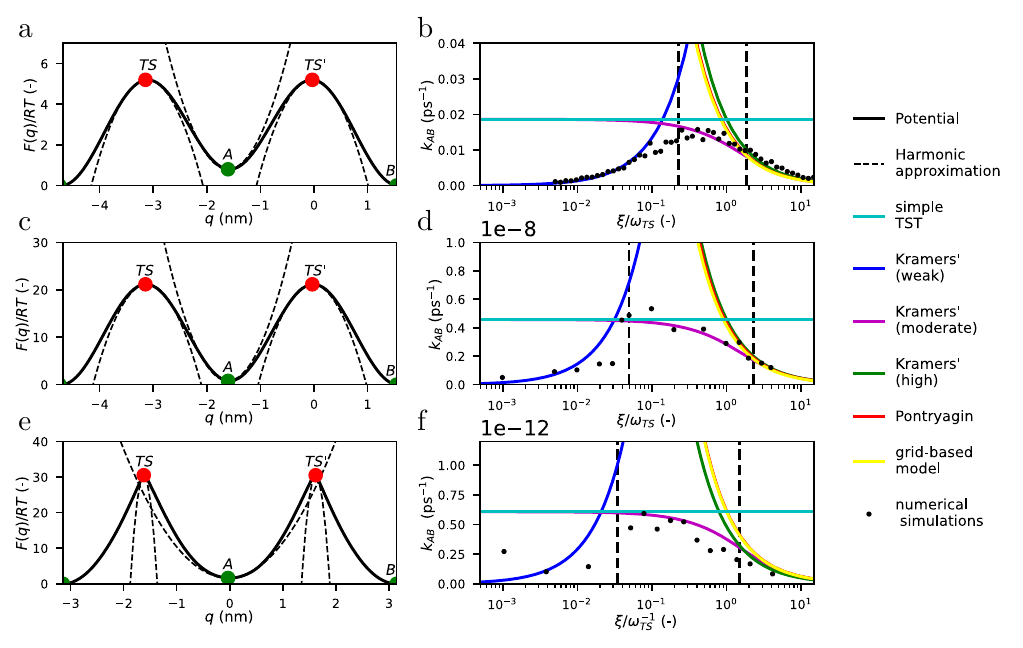}
\caption{One-dimensional model systems and corresponding rate constants from $A$ to $B$ as a function of $\xi/\omega_{TS}$: 
\textbf{a, b:} low barriers, \textbf{c, d:} high and broad barriers, \textbf{e, f:} high and sharp barriers.
Rates have been calculated by 
simple TST (eq.~\ref{eq:simpleTST}), 
Kramers' weak friction (eq.~\ref{eq:KramersWeakLimit}), 
Kramers' moderate friction (eq.~\ref{eq:KramersModerate}), 
Kramers' high friction (eq.\ref{eq:KramersHigh}), 
Pontryagin (eq.\ref{eq:Pontryagin}), 
grid-based model (eq.~\ref{eq:Qij_SqRA_1}), 
and from numerical simulations.
Threshold between weak and moderate friction is $\xi/\omega_{TS}=RT/F^\ddagger_{AB}$. 
Threshold between moderate and high friction is at the value of $\xi/\omega_{TS}$ where eq.~\ref{eq:KramersModerate}) and eq.~\ref{eq:KramersHigh} deviate less than five percent.
}
\label{fig:1Dmodel_rates}
\end{figure*}
\subsection{Simple transition state theory}
In simple TST \cite{Hanggi1990, Peters2017} (or equivalently: harmonic TST or Vineyard TST), one uses a one-dimensional reaction coordinate $q$ and the free energy $F(q)$ along this reaction coordinate. 
$A$ then corresponds to the region around a minimum on the one-dimensional free energy surface, whereas $TS$ is a point $q_{TS}$ along the reaction coordinate that separates reactant state $A$ ($q<q_{TS}$) and product state $B$ ($q>q_{TS}$). 
Usually $TS$ is positioned at the maximum of the free energy barrier.
The rate is derived by considering the probability flux across $TS$ (see SI section \ref{supp-sec:SI_theory})
\begin{equation}
    k_{AB} = \kappa \cdot \frac{\omega_A}{2\pi}\exp\left(-\frac{F^\ddagger_{AB}}{RT} \right) \, .
    \label{eq:simpleTST}
\end{equation}
The free energy barrier is
\begin{eqnarray}
    F^\ddagger_{AB} &=& F(q_{TS}) - F(q_A) \cr
    F^\ddagger_{BA} &=& F(q_{TS}) - F(q_B) 
\end{eqnarray}
where $F^\ddagger_{AB}$ is measured from the free energy minimum of $A$ to $TS$,
and, analogously, $F^\ddagger_{BA}$ is measured from the free energy minimum of $B$ to $TS$.
$\omega_A$ in eq.~\ref{eq:simpleTST} is the angular frequency of the harmonic approximation of the reactant state minimum. 
$\kappa \in [0,1]$ is the transmission factor, which accounts for the fraction of molecules that proceed from $TS$ to the product state $B$.
Molecules, that revert to $A$ after they have already passed $TS$, recross the transition state region.
At this point, $\kappa$ is an ad-hoc correction to the rate constant.
In this contribution, we will set $\kappa=1$ when applying eq.~\ref{eq:simpleTST}, meaning that all molecules that reach $TS$ complete the reaction, and recrossing can be neglected.
%

\subsection{Kramers' rate theory}
In Kramers' rate theory\cite{kramers1940brownian, Hanggi1990}, one uses a  one-dimensional reaction coordinate $q$.
One models the effective dynamics along $q$ by underdamped Langevin dynamics, where the free energy $F(q)$ takes the role of the potential energy governing the drift and the neglected degrees act as a thermal bath. 
The interaction with this thermal bath is modelled by a friction and a random force, where the friction force can be scaled by a friction coefficient or collision rate $\xi$ (with units time$^{-1}$).
Thus, two thermal parameters enter Kramers' model: $\xi$ and $T$.
One models $F(q)$ as a double well function, where the minima correspond to reactant ($A$) and product ($B$) state, and the barrier corresponds to the transition state ($TS$). 
Around each of the three states, $F(q)$ is approximated by a harmonic function 
\begin{equation}
F(q) = 
\begin{cases}
F(q_A) + \frac{1}{2} \mu_{q}\, \omega_A^2 \left(q - q_A\right)^2   
~&\text{ if }~ q\approx q_A \, \cr
F(q_{TS}) - \frac{1}{2} \mu_{q}\, \omega_{TS}^2 \left(q - q_{TS}\right)^2   
~&\text{ if }~ q\approx q_{TS}  \cr
F(q_B) + \frac{1}{2} \mu_{q}\, \omega_B^2 \left(q - q_B\right)^2   
~&\text{ if }~ q\approx q_B \, ,
\end{cases}
\, ,
 \label{eq:BarrierApprox}
\end{equation}
where $q_A$, $q_B$ and $q_{TS}$ are positions of the extrema,  
$\omega_A$, $\omega_{B}$ and $\omega_{TS}$  are the angular frequencies of the harmonic approximation around the extrema.
Fig.~\ref{fig:1Dmodel_rates}.a, c, e show the harmonic approximation for double wells on a circular coordinate.

In total, five parameters originating from the free energy surface govern Kramers' model: 
$\omega_A$, $\omega_B$, $\omega_{TS}$, $F^\ddagger_{AB}$ and $F^\ddagger_{BA}$.
To obtain the rate constant, the thermal parameters are compared to the free energy parameters. 
Three limiting cases are classified according to  the thermal energy $RT/F^\ddagger_{AB}$ and the friction $\xi/\omega_{TS}$
(See Fig.~15 in ref.~\citenum{Hanggi1990}).
%

%
%
%
The weak friction limit  (or sometimes: diffusion limited regime) is defined by $\xi /\omega_{TS} < RT / F^\ddagger_{AB}$.
In this regime, the deterministic forces (due to the free energy) dominate the diffusive forces (friction and the thermal noise terms). 
Thus, the underdamped Langevin dynamics is quasi-Hamiltonian.
The rare interactions with the heat bath cause the total energy of the system to slowly oscillate and the rate constant is derived by considering the time evolution for the energy probability density \cite{Hanggi1990}.
One obtains
\begin{equation}
    k_{AB} = 
   \frac{I(F^\ddagger_{BA})}{I(F^\ddagger_{AB}) + I(F^\ddagger_{BA})}\
    \cdot \xi \frac{I(F^\ddagger_{AB})}{R T}
    \cdot \frac{\omega_A}{2 \pi} \exp\left(-\frac{F^\ddagger_{AB}}{RT} \right)\, .
    \label{eq:KramersWeakLimit}
\end{equation}
where     
\begin{eqnarray}
I(F^\ddagger_{AB})  
&=& 
\oint_{H(q,p)=F^\ddagger_{AB}} p \, \mathrm{d}q \cr
&=&
2 \int_{q^{-}_{AB}}^{q^{+}_{AB}}
\sqrt{2\mu_q \left( F^\ddagger_{AB} - F(q) \right)}\, \mathrm{d}q \cr
&=&
\frac{2\pi F^\ddagger_{AB}}{\omega_A}
\label{eq:I_FAB}
\end{eqnarray}
is an integral over closed orbits of the phase space corresponding respectively to the total energy $F^\ddagger_{AB}$. 
$I(F^\ddagger_{BA})$ is defined analogously.
The limits of the integrals are obtained by setting $p=0$ in the Hamiltonian function: $q^{\pm}_{AB} = q_A\pm\sqrt{ \nicefrac{2 F^\ddagger_{AB}}{\mu_q \omega_A^2}}$ (and equivalent $q^{\pm}_{BA} = q_B\pm \sqrt{ \nicefrac{2 F^\ddagger_{BA}}{\mu_q \omega_B^2}}$).
The resulting formula is the reduced action of the harmonic oscillator at an energy $F^\ddagger_{AB}$ (and equivalent for $F^\ddagger_{BA}$).
A sharp peak at the transition state corresponds to a large value of $\omega_{TS}$, and thus might induce the weak friction limit.
%

%
%
%
The moderate-to-high friction limit is defined by $\xi /\omega_{TS} > RT / F^\ddagger_{AB}$. 
The diffusive forces are stronger than the deterministic forces, but not by orders of magnitude.
In this regime, one assumes a steady state probability flux from state $A$ across the a transition state region \cite{Hanggi1990}. 
This assumption replaces the requirement for thermal equilibrium between reactant and transition state in transition state theory.
This yields 
\begin{align}
k_{AB} 
=
\frac{\xi}{\omega_{TS}} \left(\sqrt{\frac{1}{4} + \frac{\omega_{TS}^2}{\xi^2}} - \frac{1}{2} \right)
\cdot \frac{\omega_A}{2 \pi} \exp\left(-\frac{F^\ddagger_{AB}}{RT}   \right) \, .
\label{eq:KramersModerate}
\end{align}
%
    
%
%
%
The high friction limit is defined by  $\xi /\omega_{TS} \gg RT / F^\ddagger_{AB}$.
The diffusive forces dominate the deterministic forces.
At high values of $\xi$, the prefactor in eq.~\ref{eq:KramersModerate} can be approximated as $\omega_{TS}/\xi$ (see SI section \ref{supp-sec:SI_theory}), yielding
\begin{eqnarray}
    k_{AB} 
    &=& \frac{\omega_{TS}}{\xi}\cdot
    \frac{\omega_A}{2\pi}\exp\left(-\frac{F^\ddagger_{AB}}{RT} \right) \, .\label{eq:KramersHigh}
\end{eqnarray}
This regime is also called the spatial diffusion limited regime, or the diffusive regime.

%
%
%
The rate constants for the three friction regimes (eqs.~\ref{eq:KramersWeakLimit}, \ref{eq:KramersModerate}, \ref{eq:KramersHigh}) have the same functional form as in simple TST (eq.~\ref{eq:simpleTST}), but in addition they provide explicit expressions for the transmission factor $\kappa$. 
Kramers' rate theory provides a model for recrossing in terms of the shape of the free energy surface, the temperature and the strength of the heat bath.
%

\subsection{Pontryagin's rate theory} 
The following rate model is often quite generically introduced as a means to calculate the mean first-passage time (MFPT) $\tau_{AB}$ or escape rate $k_{AB}$ for diffusion over a barrier.
It is derived  from the Fokker-Planck equation for overdamped Langevin dynamics (Smoluchowski equation).
Here, we will refer to it as the Pontryagin's rate theory \cite{pontryagin1933statistical}.
In Pontryagin' rate theory \cite{pontryagin1933statistical, Hanggi1990}, one uses a  one-dimensional reaction coordinate $q$ and models the effective dynamics along $q$ by overdamped Langevin dynamics, which is the high friction limit of underdamped Langevin dynamics. 
In this rate theory, the friction coefficient $\xi(q)$ may vary along the reaction coordinate $q$.
This generalization is important because the fluctuations of the neglected degrees of freedom may vary along $q$ \cite{hummer2005position}, and additionally the curvature of $q$ may give rise to a position dependent friction.
Conventionally, Pontryagin's rate constant is not formulated in terms of $\xi(q)$ but in terms of the closely related position dependent diffusion profile
\begin{equation}
D(q) = \frac{RT}{\mu_q\xi(q)} \, .
\label{eq:Einstein}
\end{equation}
where $\mu_q$ is a molar mass.
The rate constant is then given by the following nested integral
\begin{align}
k_{AB}  = \left\lbrace 
\int_{q_A}^{q_B}\mathrm{d} q'\left[ \frac{1}{D(q')} e^{\beta F(q')}
\int_{-\infty}^{q'} \mathrm{d}q'' \, e^{-\beta F(q'')}
\right]
\right\rbrace^{-1} \, .
\label{eq:Pontryagin}
\end{align} 
with $\beta = 1/RT$.
A closed-form version is not available, but computing the nested integral numerically is straightforward. 
This expression for the rate constant does not make any assumptions on the shape of the reactant state and transition barrier and includes the full position dependent diffusion profile.
Inserting the harmonic approximation and assuming constant diffusion in eq.~\ref{eq:Pontryagin} yields Kramers' rate equations in the high friction limit (eq.~\ref{eq:KramersHigh}).
%

\subsection{Grid-based models}
In grid-based models\cite{Lie2013,donati2021markov}, one uses a multidimensional collective variable $\mathbf{q} \in \mathbb{R}^m$ and models the effective dynamics in this collective variable space by overdamped Langevin dynamics with position dependent diffusion.
The collective variable space is discretized into $n$ disjoint cells. 
The cells are divided into three sets $\mathcal{A}$, $\mathcal{B}$, and $\mathcal{I}$, where $\mathcal{A}$ represents the reactant state $A$,  $\mathcal{B}$ represents the product state $B$, and $\mathcal{I}$ the intermediate region.
Independent of the assignment to the three sets, the transition rate $Q_{ij}$ from cell $i$ to cell $j$ is
\begin{align}
        Q_{ij} = \left\{
        \begin{array}{ll}
            Q_{ij}
            & \text{if } i\neq j \text{ and cells adjacent}\\
            0
            & \text{if } i\neq j \text{ and cells not adjacent}\\
            -\sum_{l=1,l\neq i}^{n} Q_{il}
            & \text{if } i = j
        \end{array}
    \right.
    \label{eq:rateMatrix}
\end{align}
Eq.~\ref{eq:rateMatrix} defines a $n\times n$ row-normalized rate matrix $\mathbf{Q}$ with elements $Q_{ij}$.
$\mathbf{Q}$ is a discretization of the Fokker-Planck operator for overdamped Langevin dynamics, where we assumed that  the free energy is constant within each grid cell. 
$Q_{ij}$ between adjacent cells can be calculated as\cite{Bicout1998, Lie2013,donati2021markov}
\begin{eqnarray}
    Q_{ij} &=& D_{ij}\frac{\mathcal{S}_{ij}}{\mathcal{V}_ih_{ij}}\cdot \frac{\sqrt{\pi(\mathbf{q}_j) \pi(\mathbf{q}_i)}}{\pi(\mathbf{q}_i)} \, .
\label{eq:Qij_SqRA_1}    
\end{eqnarray}
where 
$\mathbf{q}_i$ and $\mathbf{q}_j$ are the centers of the adjacent grid cells, 
$\pi(\mathbf{q})$ is given by eq.~\ref{eq:eq_distribution},  
$h_{ij}=\left\Vert \mathbf{q}_j-\mathbf{q}_i\right\Vert$ is the Euclidean distance between the centers of the cells, 
$\mathcal{S}_{ij}$ is the area of the intersecting surface between cells $i$ and $j$, 
$\mathcal{V}_i$ is the volume of the Voronoi cell $i$, 
and $D_{ij}$ is the diffusion on the boundary between cells $i$ and $j$, which we approximate as $D_{ij} = \frac{1}{2}\left(D(\mathbf{q}_i)+D(\mathbf{q}_j)\right)$.
Because of the square root in eq.~\ref{eq:Qij_SqRA_1}, the approach is called Square Root Approximation of the Fokker-Planck equation (FP-SqRA) \cite{Lie2013,donati2021markov}.
In eq.~\ref{eq:Qij_SqRA_1}, the probability density at the cell boundary between adjacent cells is approximated by the geometric mean of the Boltzmann weights of the cells\cite{Lie2013,donati2021markov}.
Using a harmonic mean instead leads to the Harmonic Averaging Approximation of the Fokker-Planck equation (FP-HAA):
\begin{eqnarray}
    Q_{ij} &=&   D_{ij}
            \frac{\mathcal{S}_{ij}}{\mathcal{V}_ih_{ij}}\cdot
            \frac{1}{\pi(\mathbf{q}_j)}
            \frac{2\pi(\mathbf{q}_j)\pi(\mathbf{q}_i)}{\pi(\mathbf{q}_i)+\pi(\mathbf{q}_j)} \, 
\label{eq:Qij_HAA_1}            
\end{eqnarray}
and has improved convergence properties \cite{heida2021consistency}.
Mean first-passage times $\tau_{i\rightarrow B}$ from any cell $i$ to the to the product state $B$ can be computed by solving\cite{berezhkovskii2019committors}
\begin{equation}
    \mathbf{Q}\,\boldsymbol{\tau}_B = - \boldsymbol{1} \, 
    \label{eq:solve_Q}
\end{equation} 
for $\boldsymbol{\tau}_B=\left[\tau_{1\rightarrow B},\dots,\tau_{n\rightarrow B}\right]^T$. 
This vector contains MFPTs for all cells $i$ to the product state $B$.
To enforce this, eq.~\ref{eq:solve_Q} must be solved while imposing the boundary condition that $\tau_{k\rightarrow B} = 0$ for all  $k \in \mathcal{B}$.
The MFPT from $A$ to $B$ is then obtained by averaging over the state-wise MFPTs
\begin{equation}
    \tau_{AB} = \sum_{i\in \mathcal{A}} \pi_{A,i} \tau_{i\rightarrow B} 
\label{eq:Q_MFPT}    
\end{equation}
where
$
\pi_{A,i} = \pi_i/\sum_{i\in \mathcal{A}} \pi_i
$
and $\pi_i=\int_{\mathbf{q} \in \,\mathrm{cell}\, i}\mathrm{d}\mathbf{q} \, \pi(\mathbf{q})$.
The rate constant is the inverse of this MFPT (eq.~\ref{eq:MFPT}).
%

\subsection{Rates from sampling transitions}
\label{sec:InMetaD}
The system is simulated on $V(\mathbf{x})$, and the reaction rate $k_{AB}$ is obtained as a statistical estimate of the observed transitions between $A$ and $B$. 
It is sufficient to define $A \subset \Gamma_x$ and $B \subset \Gamma_x$ as regions in the configurational space, a transition state does not need to be defined. 
The first-passage times from $A$ to $B$ are recorded in a series of $n$ simulations, whose initial states are located in $A$ and which are terminated once they reach $B$. 
This yields a series if first-passage times $(\tau_{AB,1}, \tau_{AB,2} \dots \tau_{AB,n})$. 
The MFPT $\tau_{AB}$ can be calculated as the arithmetic mean of these first-passage times, or - with better statistical accuracy - by fitting a the cumulative distribution function of a Poisson process\cite{salvalaglio2014assessing}
\begin{equation}
\label{eq:TCDF}
    P(\tau_{AB,i}) = 1 - \exp\left(-\frac{\tau_{AB,i}}{ \tau_{AB}}\right).
\end{equation} 
to the cumulative distribution histogram of these fist passage times. 
In eq.~\ref{eq:TCDF}, $\tau_{AB}$ is the MFPT and acts as a fitting parameter, which is inserted into eq.~\ref{eq:MFPT} to obtain the reaction rate.
For reactions with high energy barriers the transition times are orders of magnitude longer than the accessible simulation times.  
Therefore in infrequent metadynamics simulations \cite{tiwary2013metadynamics, barducci2008well}, a time dependent bias function $U(\mathbf{x},t)$ is introduced that increases in strength as the simulation proceeds and pushes the system over the barrier into state $B$.
Each accelerated first-passage time is then reweighted to the corresponding physical first-passage time by a discretized time-integral over the length of the trajectory \cite{grubmuller1995predicting,voter1997hyperdynamics,tiwary2013metadynamics}
\begin{eqnarray}
    \tau_{AB,i} = \Delta t \sum_{k=1}^{T_i}  \exp\left(\frac{U(\mathbf{x}_{i,k}, k\Delta t)}{R T} \right)\, .
\label{eq:InMetaD}   
\end{eqnarray}
where $\Delta t$ is the time step of the trajectory, $T_i$ is the total number of time steps in the $i$th trajectory, $\mathbf{x}_{i,k}$ is the $k$th configuration in this trajectory, and $t=k\Delta t$ is the corresponding time.
This reweighting assumes that no bias has been deposited on the transition state, which is approximately ensured by the slow deposition of the infrequent metadynamics protocol.

\section{Results}

\subsection{Friction regimes}
\label{sec:1D_model}

To study the effect of the curvature of the free energy surface on the friction regime independently from the choice of the reaction coordinate, we devised one-dimensional model systems with  a circular reaction coordinate $q \in [-\pi, +\pi]$. 
As in the actual retinal molecule, the free energy functions $F(q)$ for these models exhibits two energy barriers and two minima.
%
The models differ in the height and the ``sharpness" of the barriers, 
where the first model has a low and broad free-energy barriers, 
the second model has high and broad free-energy barriers.
The third model is the actual free-energy function along the C$_{13}$=C$_{14}$ torsion angle of retinal and exhibits sharp and high free-energy barriers.
Figs.~\ref{fig:1Dmodel_rates}.a, c, e show the free energy functions along with the harmonic approximations for the minima and the barriers.
Tab.~\ref{tab:1D_parameters} reports the corresponding parameters.
We set $T=300\,\mathrm{K}$, and thus the thermal energy is $RT = 2.49$ kJ/mol.

With increasing barrier height the rate constant due to simple TST drops by orders of magnitude from $k_{AB} \sim 10^{-2}\, \mathrm{ps}$ to $k_{AB} \sim 10^{-9}\, \mathrm{ps}$ and $k_{AB}\sim 10^{-13}\, \mathrm{ps}$ (cyan lines in Fig.~\ref{fig:1Dmodel_rates}.b, d, f).
However, comparison to the numerical simulations (black dots in Fig.~\ref{fig:1Dmodel_rates}.b, d, f) shows that simple TST is a crude approximation and severely overestimates the rate constants in the low- and high-friction regimes.
The numerical simulations reproduce Kramers' turnover\cite{Hanggi1990,tiwary2016kramers}, i.e.~the bell curve characterized by low rates in the weak friction regime, high rates in the moderate friction region, and low rates again in the high friction region (see SI Tab.~\ref{supp-tab:rates1D} for representative numerical values).
Kramers' rate theory models this turnover by devising a seperate rate equation for each of the three friction regimes 
(eqs.~\ref{eq:KramersWeakLimit}, ~\ref{eq:KramersModerate}, and~\ref{eq:KramersHigh}).
The theory requires that $F^\ddagger_{AB} \gg RT$, which is well fulfilled for the second ($F^\ddagger_{AB}=20.4 \, RT$) and the third model ($F^\ddagger_{AB}=29.0 \, RT$)  and to a lesser extent for the first model ($F^\ddagger_{AB}=4.4 \, RT$). 
The friction regime is determined by the relative sizes of the ratios $RT/ F^\ddagger_{AB}$ and $\xi/\omega_{TS}$. 
The ratio $RT/ F^\ddagger_{AB}$ compares the thermal energy to the free energy barrier.
Within the assumptions of Kramers' theory, $RT/ F^\ddagger_{AB} \ll 1$.
The ratio $\xi/\omega_{TS}$ compares the time it takes to cross the transition state region, $1/\omega_{TS}$, to the average time between two interactions with the thermal bath.
$\xi/\omega_{TS} > 1$  means that, on average, several interactions with the thermal bath occur while the system crosses the transition state region, implying a high friction regime.
$\xi/\omega_{TS} < 1$ means that, on average, no interaction with the thermal bath occurs while the system crosses the transition state region, implying a weak friction regime.
For $\xi/\omega_{TS} \approx 1$, transition time and interaction with the thermal bath occur on the same timescale.

All other parameters being equal, an increase in the curvature of the free-energy barrier leads to an increase in $\omega_{TS}$ and thus might shift the effective dynamics into the weak or intermediate friction regime.
In our model systems, the $\omega_{TS}$ increases across the models from 
$\omega_{TS} \approx 5 \,\mathrm{ps}^{-1}$ to $\omega_{TS} \approx 10 \,\mathrm{ps}^{-1}$ and finally reaching $\omega_{TS} = 48.38\,\mathrm{ps}^{-1}$, and $\omega_{TS'} = 46.21\,\mathrm{ps}^{-1}$ for the model representing the actual retinal.
Simultaneously, the free-energy barrier increases across the models. 
The resulting boundaries between the friction regimes are shown as vertical dashed lines in Fig.~\ref{fig:1Dmodel_rates}.b, d, f. 
Kramers' rate constants $k_{AB}$ as a function of $\xi/\omega_{TS}$ are represented as blue, purple and green lines for the three friction limits (Fig.~\ref{fig:1Dmodel_rates}.b, d, f).
Each of the three rate equations agrees well with the numerical results when applied within the appropriate friction regime. 
Outside of their respective friction regime, the three rate equations yield very inaccurate results. 
In particular, the rate equation for the high-friction regime vastly overestimates the rates in the weak and the weak-to-high friction regime.

Additionally, we report the results from Pontryagin's rate theory (red line, eq.~\ref{eq:Pontryagin}) and the grid-based model (yellow line, eq.~\ref{eq:Qij_SqRA_1}), which both assume overdamped Langevin dynamics. 
For a position-independent friction coefficient $\xi$, these models closely align with the high friction regime of Kramers' rate theory, and equally overestimate the rate constant in the weak and weak-to-moderate friction regime.
These results underlines the importance of determining the system's friction regime and choosing the appropriate rate model.
For the free-energy function of retinal (third model system), the moderate friction regime ranges from  $\xi = 1.45 \, \mathrm{ps}^{-1}$ to $\xi = 72.57 \, \mathrm{ps}^{-1}$.
The friction coefficient $\xi$ of the effective dynamics along $q$ is not a parameter that can be chosen freely, but it is determined by the influence of the neglected degrees of freedom and is calculated from the diffusion constant $D(q)$ (eq.~\ref{eq:Einstein}) and the effective mass $\mu_p$ (eq.~\ref{eq:effective_mass}).
This is explored in the following section.

%
\begin{table}
    \centering
    \begin{tabular}{c  c  c  c  c  c}
     & 
     & \textbf{Small}   
     & \textbf{High} 
     & \multicolumn{2}{c}{\textbf{Inter-}}
     \\ 
     &
     & \textbf{barrier}   
     & \textbf{barrier}
     & \multicolumn{2}{c}{\textbf{polated}}
     \\
     &&&& $TS$ & $TS'$
     \\ 
    \hline
    \hline
    $R T$ & [$\mathrm{kJ\,mol^{-1}}$]                 & $2.49$    & $2.49$    & \multicolumn{2}{c}{$2.49$}  \\
    $F^\ddagger_{AB}$  & [$\mathrm{kJ\,mol^{-1}}$]      & $10.98$   & $50.84$   & $72.11$& $72.10$ \\
    $F^\ddagger_{BA}$ & [$\mathrm{kJ\,mol^{-1}}$]       & $12.96$   & $52.74$   & $76.15$& $76.14$\\
    $\omega_A$          & [$\mathrm{ps^{-1}}$]          & $4.78$    & $10.13$   & \multicolumn{2}{c}{$6.83$}\\
    $\omega_{B}$ & [$\mathrm{ps^{-1}}$]                 & $4.98$    & $10.21$   & \multicolumn{2}{c}{$7.60$}\\
    $\omega_{TS}$ &  [$\mathrm{ps^{-1}}$]               & $4.89$    & $10.17$   & $48.38$       &$46.21$\\
    \hline
    \multicolumn{6}{c}{energy ratio}\\
    \hline
    $R T/F^\ddagger_{AB}$ & [-]                       & $0.23$    & $0.05$    & $0.03$ & $0.03$\\
    \hline
    \multicolumn{6}{c}{threshold between weak and moderate friction}\\
    \hline
    $\xi/\omega_{TS}$   & [-]                       & $0.23$    & $0.05$    & $0.03$ & $0.03$\\
    $\xi$               & [$\mathrm{ps^{-1}}$]      & $1.12$    & $0.51$    & $1.45$ & $1.39$\\
    \hline
    \multicolumn{6}{c}{threshold between moderate and high friction}\\
    \hline
    $\xi/\omega_{TS}$   & [-]                       & $1.90$    & $2.30$    & $1.50$ &$1.57$\\
    $\xi$               & [$\mathrm{ps^{-1}}$]      & $9.29$    & $23.39$    & $72.57$ &$72.55$\\
    \hline
    \end{tabular}
    \caption{Parameters for one-dimensional rate theories calculated for the one-dimensional systems.}
    \label{tab:1D_parameters}
\end{table}

\subsection{Atomistic model of retinal}
\label{sec:retinal_atomistic_model}
\begin{figure}[ht]
\includegraphics[scale=1]{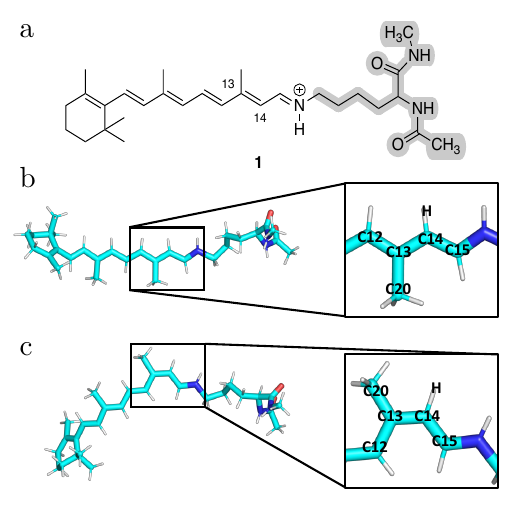}\\
\caption{\textbf{a:} Retinal covalently linked to a capped lysine residue. Position restraints have been applied to heavy atoms highlighted in gray.
\textbf{b:} trans-configuration. \textbf{c:} cis-configuration.
}
\label{fig:retinal_structures}
\centering
\end{figure}
Our model of retinal (Fig.~\ref{fig:retinal_structures}.a) consists of the retinal molecule, which is covalently bound to a capped lysine residue via a protonated Schiff base \cite{schulten2014quantum}.
This corresponds to the chemical structure of retinal in a protein environment. 
Since the lysine residue cannot move freely in a protein environment, we placed position restraints on all heavy atoms of lysine (backbone and side chain) as well as on the atoms in the caps. 
All atoms in retinal including the cyclohexene ring were allowed to move freely.
A potential energy function of this molecule we use an empirical atomistic force field. 
Our goal is to calculate the reaction rate constants of the thermal cis-trans isomerization around the  
C$_{13}$=C$_{14}$ double bond (Fig.~\ref{fig:retinal_structures}.b-c), 
where the cis-configuration is the reactant state $A$ and the trans-configuration is the product state $B$.
%

\subsection{C$_{13}$=C$_{14}$-torsion angle as reaction coordinate}
\label{sec:RC_torsion_angle}
%
%
\begin{figure}
\includegraphics[scale=1]{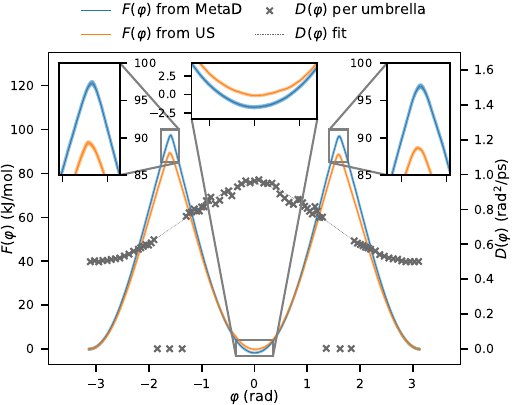} \\
\caption{
Free energy surfaces $F(\varphi)$ and diffusion profiles $D(\varphi)$ estimated from 
umbrella sampling (US) and metadynamics (MetaD) by biasing the C$_{13}$=C$_{14}$ torsion angle $\varphi$.
Statistical standard errors given by the thickness of the curves.
}
\label{fig:FES_phi}
\centering
\end{figure}

As initial reaction coordinate for the one-dimensional rate models, we choose the torsion angle $\varphi$ constituted by the chain of carbon atoms C$_{12}$-C$_{13}$=C$_{14}$-C$_{15}$ (Fig.~\ref{fig:retinal_structures}).
Fig.~\ref{fig:FES_phi} shows two free energy functions along this reaction coordinate, which were numerically calculated by well-tempered metadynamics simulations\cite{barducci2008well} (blue line) and by umbrella sampling simulations\cite{torrie1977nonphysical} combined with weighted histogram analysis\cite{kumar1992weighted} (orange line).
The statistical uncertainty of the free energy profiles are shown as shaded areas in Fig.~\ref{fig:FES_phi}, but they are only about as large as the linewidth.
The figure also shows the position dependent diffusion $D(\varphi)$ obtained from umbrella sampling simulations following Ref.~\onlinecite{hummer2005position}.
Because of the sharp barriers, the diffusion profile could not be estimated in the transition regions, and we relied on the interpolation (dotted line in Fig.~\ref{fig:FES_phi}) in these regions.
Both $F(\varphi)$ have minima at the cis-configuration ($\varphi=0\,\mathrm{rad}$) and at the trans-configuration ($\varphi=\pm\pi\,\mathrm{rad}$), where we set the trans state to $F(\varphi) = F(\pi) = 0$. 
Cis- and trans-configuration have the same free energy in $F(\varphi)$ from umbrella sampling, whereas, in $F(\varphi)$ from metadynamics, the cis-configuration is about $1.6\,\mathrm{kJ/mol}$ lower than the trans-configuration.
The minima are separated by two free energy barriers $TS$ and $TS'$ corresponding to rotating clockwise and counterclockwise around $\varphi$, respectively.
Both methods, umbrella sampling and metadynamics, predict that the barriers $TS$ and $TS'$ are equal in height. 
Umbrella sampling yields a barrier height of 
$F_{\mathrm{c} \rightarrow \mathrm{t}}^{\ddagger} = 89\, \mathrm{kJ/mol}$, whereas metadynamics yields barriers that are about $10\,\mathrm{kJ/mol}$ higher ($F_{\mathrm{c} \rightarrow \mathrm{t}}^{\ddagger} = 99\, \mathrm{kJ/mol}$).

Even though we monitored the convergence of the two free energy methods carefully, the difference in the predicted free energy barrier is sizeable. 
At at room temperature, the difference corresponds to about four times the thermal energy of $RT =  2.49\, \mathrm{kJ/mol}$, and in absolute terms it is well above the limit for chemical accuracy of $1\,\mathrm{kcal/mol} = 4.2\,\mathrm{kJ/mol}$.
Because the free energy difference enters exponentially in each of the rate models, this difference strongly affects the predicted rate. 
We return to this discussion in section \ref{sec:USvsMetaD}, but for now will discuss rates based on the umbrella sampling $F(\varphi)$.
The parameters for the one-dimensional rate theories for $F(\varphi)$ from umbrella sampling and from metadyamics are reported in SI Tabs.~\ref{supp-tab:dihedral_constants_us} and \ref{supp-tab:dihedral_constants_metad}.

Next, we determine the friction regime by comparing the energy ratio is $RT/ F_{\mathrm{c} \rightarrow \mathrm{t}}^{\ddagger} = 0.028$ to the friction retio $\xi_{TS} / \omega_{TS}$.
The friction coefficient of the effective dynamics along $\varphi$ is $\xi_{TS} = 131 \, \mathrm{ps}^{-1}$ for transitions via $TS$ (determined via eqs.~\ref{eq:Einstein} and ~\ref{eq:effective_mass}).
The curvature of $TS$ is $\omega_{TS} = 244 \, \mathrm{ps}^{-1}$, yielding the friction ratio
$\xi_{TS} / \omega_{TS} = 0.54$. 
The corresponding ratio for the other barrier is $\xi_{TS'} / \omega_{TS'} = 0.50$.
Both friction ratios are much higher than the energy ratio, and therefore the effective dynamics along $\varphi$ fall into the moderate-to-high or even high-friction regime.
Tab.~\ref{tab:rates} shows the rate constants derived from one-dimensional rate theories for the moderate and high friction regime.
Methods that assume high friction (Kramers' (high friction), Pontryagin, grid-based) all yield a rate constant of $k_{\mathrm{cis}\rightarrow\mathrm{trans}} \approx 0.009 - 0.015\, \mathrm{s}^{-1}$ for the cis-trans transition.
The two methods that are based on overdamped Langevin dynamics (Pontryagin and grid-based models) yield indistinguishable rate constants  ($k_{\mathrm{cis}\rightarrow\mathrm{trans}} \approx 0.009 \, \mathrm{s}^{-1}$), which is slightly lower than the high friction limit of Kramers' rate theory ($k_{\mathrm{cis}\rightarrow\mathrm{trans}} \approx 0.015\, \mathrm{s}^{-1}$). 
The high friction Kramers' rate constant ($k_{\mathrm{cis}\rightarrow\mathrm{trans}} \approx 0.015 \, \mathrm{s}^{-1}$) is higher than the one form the moderate friction regime ($k_{\mathrm{cis}\rightarrow\mathrm{trans}} \approx 0.006 \, \mathrm{s}^{-1}$). 
Since the two methods would coincide in the high friction region, this indicates, that the effective dynamics along $\varphi$ fall into the moderate friction regime and are best described Kramers' rate theory for moderate friction. 
%
%

%
Simple TST is a reasonable approximation in the moderate friction regime and yields a rate constant of $k_{\mathrm{cis}\rightarrow\mathrm{trans}} \approx 0.008 \, \mathrm{s}^{-1}$, only slightly overestimating Kramers' rate constant for moderate friction.
The rate constant of the reverse reaction, $k_{\mathrm{trans}\rightarrow\mathrm{cis}}$, are reported in Tab.~\ref{tab:rates} and show the same effects.

\subsection{Comparison to infrequent metadynamics}

Even though the results from one-dimensional rate theories (using $\varphi$ as reaction coordinate) seem consistent, they deviate drastically from rate constants estimated from molecular simulations (Tab.~\ref{tab:rates}).
We used infrequent metadynamics and biased along $\varphi$ to simulate the cis-trans isomerization.
The resulting rate constant, $k_{\mathrm{cis} \rightarrow\mathrm{trans}} = 2.23\cdot 10^{-5} \, \mathrm{s}^{-1}$, is more than two orders of magnitude smaller than the most appropriate one-dimensional rate theory $k_{\mathrm{cis}\rightarrow\mathrm{trans}} = 5.83\cdot 10^{-3} \, \mathrm{s}^{-1}$ (Kramers' with moderate friction).
By moving from a one-dimensional system (Fig.~\ref{fig:1Dmodel_rates}) to a high-dimensional system (Fig.~\ref{fig:retinal_structures}) we have lost the agreement between one-dimensional rate theories and numerical simulations.
The deviation between Kramers' rate theory and numerical simulation for retinal is in stark contrast to the good agreement between the two approaches for the one-dimensional systems. 
It is, however, in line with the results from Ref.~\citenum{ghysbrecht2023thermal}, where we found a similar disparity between simulated rate constants and the moderate friction limit of Kramers' rate theory. 

Two known error sources of infrequent metadynamics are ($i$) slow processes that occur orthogonal to the biased coordinate, e.g.~due to sub-minima in the reactant state \cite{dickson2018erroneous,khan2020fluxional}, and ($ii$) perturbation of the transitions state region because bias is deposited there\cite{tiwary2013metadynamics,salvalaglio2014assessing}. 
Retinal is a very rigid molecule and does not exhibit sub-minima within the cis- or within the trans-configuration, making the first error source unlikely.
Our rate constants are insensitive to the variations in the bias deposition rate, which confirms that the transition state region is unperturbed (Fig.~\ref{supp-fig:inmetad}). 
Thus, for this specific system the limitations of infrequent dynamics metadynamics do not explain the deviation results of one-dimensional rate theories.
A second error source might be a wrong choice of the friction regime. However, the analysis in section \ref{sec:RC_torsion_angle} shows that, for this system, the difference between the high-friction and the intermediate friction results are small and do not explain the discrepancy with the simulation results.Specifically,
high friction:  $k_{\mathrm{cis}\rightarrow\mathrm{trans}} = 1.45\cdot 10^{-2} \, \mathrm{s}^{-1}$,
moderate friction:  $k_{\mathrm{cis}\rightarrow\mathrm{trans}} = 5.83\cdot 10^{-3} \, \mathrm{s}^{-1}$,
simulation:  $k_{\mathrm{cis}\rightarrow\mathrm{trans}} = 2.23\cdot 10^{-5} \, \mathrm{s}^{-1}$.
We conclude that the disparity between Kramers rate constant for moderate friction and the simulated results is based in the high-dimensionality of the system.
One-dimensional rate theories are sensitive to the choice of reaction coordinate\cite{peters2016reaction}.
To explain the gap between the simulated rate constant and Kramers' rate constant, we will next optimize the reaction coordinate. 

%
\begin{table*}
    \centering
    \begin{tabular}{|l | c | c | c | c | c | c | c ||}

     &  & & \multicolumn{2}{c|}{\textbf{$F(\mathbf{q})$ via US}} & \multicolumn{2}{c||}{\textbf{$F(\mathbf{q})$ via  MetaD}}\\
     \textbf{method}      & \textbf{ eq. } & \textbf{CV}  & $k_{\mathrm{trans}\rightarrow\mathrm{cis}}$ [s$^{-1}$] & $k_{\mathrm{cis}\rightarrow\mathrm{trans}}$ [s$^{-1}$]& $k_{\mathrm{trans}\rightarrow\mathrm{cis}}$ [s$^{-1}$] & $k_{\mathrm{cis}\rightarrow\mathrm{trans}}$ [s$^{-1}$]\\ 
    \hline
    \multicolumn{7}{c}{\textbf{C$_{13}$=C$_{14}$-torsion angle as reaction coordinate}} \\
    \hline  
    Simple TST                &\ref{eq:simpleTST}
                                & $\varphi$
                                & $  5.17 \times 10^{-3} $
                                & $  7.55 \times 10^{-3} $
                                & $  2.05 \times 10^{-4} $
                                & $  1.60 \times 10^{-4} $\\
    Kramers (moderate friction) &\ref{eq:KramersModerate}
                                & $\varphi$
                                & $  3.99 \times 10^{-3} $
                                & $  5.83 \times 10^{-3} $
                                & $  1.67 \times 10^{-4} $
                                & $  1.30 \times 10^{-4} $\\
    Kramers (high friction) &\ref{eq:KramersHigh}
                                & $\varphi$
                                & $  9.90 \times 10^{-3} $
                                & $  1.45 \times 10^{-2} $
                                & $  4.96 \times 10^{-4} $
                                & $  3.86 \times 10^{-4} $\\
    Pontryagin                  &\ref{eq:Pontryagin}
                                & $\varphi$
                                & $  1.08 \times 10^{-2} $
                                & $  8.93 \times 10^{-3} $
                                & $  4.68 \times 10^{-4} $
                                & $  2.17 \times 10^{-4} $\\
    Grid-based                  &\ref{eq:Qij_HAA_1}
                                & $\varphi$
                                & $  1.07 \times 10^{-2} $
                                & $  8.95 \times 10^{-3} $ 
                                & $  4.67 \times 10^{-4} $
                                & $  2.17 \times 10^{-4} $\\
    \hline           
    \multicolumn{7}{c}{\textbf{Optimized reaction coordinate $\sigma_s$}} \\    
    \hline
    Simple TST                &\ref{eq:simpleTST}
                                & path
                                & $  2.41 \times 10^{-4} $
                                & $  2.57 \times 10^{-4} $
                                & $  9.56 \times 10^{-6} $
                                & $  1.19 \times 10^{-5} $\\
    Kramers (moderate friction) &\ref{eq:KramersModerate}
                                & path
                                & $  4.45 \times 10^{-5} $
                                & $  4.81 \times 10^{-5} $
                                & $  1.85 \times 10^{-6} $
                                & $  2.26 \times 10^{-6} $\\
                                
    Kramers (high friction)     &\ref{eq:KramersHigh}
                                & path
                                & $  4.61 \times 10^{-5} $
                                & $  4.98 \times 10^{-5} $
                                & $  1.92 \times 10^{-6} $
                                & $  2.34 \times 10^{-6} $\\
                                
    Pontryagin                  &\ref{eq:Pontryagin}
                                & path
                                & $  5.80 \times 10^{-5} $
                                & $  3.66 \times 10^{-5} $
                                & $  2.51 \times 10^{-6} $
                                & $  1.73 \times 10^{-6} $\\
                                
    Grid-based                  &\ref{eq:Qij_HAA_1}
                                & path
                                & $  5.78 \times 10^{-5}$
                                & $  3.66 \times 10^{-5}$
                                & $  2.53 \times 10^{-6}$
                                & $  1.75 \times 10^{-6}$\\
    \hline
    \multicolumn{7}{c}{\textbf{Grid-based model for  multidimensional collective variables}} \\        
    \hline    
    Grid-based (diffusion \textit{grid1})   
                                &\ref{eq:Qij_HAA_1}
                                & $\varphi$, $\chi_1$, $\chi_2$
                                & $  (5.66 \times 10^{-6}) $
                                & $  (7.05 \times 10^{-6}) $
                                & $  8.13 \times 10^{-6} $
                                & $  1.23 \times 10^{-5} $
                                \\
    Grid-based (averaged \textit{grid1})  
                                &\ref{eq:Qij_HAA_1}
                                & $\varphi$, $\chi_1$, $\chi_2$
                                & $  (1.14 \times 10^{-5}) $
                                & $  (1.44 \times 10^{-5}) $
                                & $  1.58 \times 10^{-5} $
                                & $  2.40 \times 10^{-5} $
                                \\
    Grid-based (diffusion \textit{grid2}) 
                                &\ref{eq:Qij_HAA_1}
                                & $\varphi$, $\chi_1$, $\chi_2$
                                & $  (7.14 \times 10^{-6}) $                                
                                & $  (9.05 \times 10^{-6}) $
                                & $  1.03 \times 10^{-5} $
                                & $  1.56 \times 10^{-5} $
                                \\
    Grid-based (averaged \textit{grid2})  
                                &\ref{eq:Qij_HAA_1}
                                & $\varphi$, $\chi_1$, $\chi_2$
                                & $  (1.02 \times 10^{-5}) $
                                & $  (1.35 \times 10^{-5}) $
                                & $  1.40 \times 10^{-5} $
                                & $  2.13 \times 10^{-5} $
                                \\
    \hline
    \multicolumn{7}{c}{\textbf{Sampling}} \\       
         \textbf{method}      & \textbf{ eq. } & \textbf{CV}  & $k_{\mathrm{trans}\rightarrow\mathrm{cis}}$ [s$^{-1}$] & $k_{\mathrm{cis}\rightarrow\mathrm{trans}}$ [s$^{-1}$]\\ 
    \cline{1-5\emph{}}
    InMetaD                     &\ref{eq:InMetaD}
                                & $\varphi$
                                & $   2.18  \times 10^{-5}$
                                & $   2.23  \times 10^{-5}$ \\
    InMetaD                     &\ref{eq:InMetaD}
                                & $\varphi$, $\chi_1$, $\chi_2$
                                & $   2.22  \times 10^{-5}$
                                & $   2.60  \times 10^{-5}$ \\
    \cline{1-5\emph{}}
    \cline{1-5\emph{}}
    \end{tabular}
    \caption{Rate constants determined through different methodologies for the thermal cis-trans isomerization over the C$_{13}$=C$_{14}$ double bond in retinal. $(...)$: results sensitive to the grid.  
    }
    \label{tab:rates}
\end{table*}
%

\subsection{Optimized reaction coordinate}
\label{sec:optimized_rc}
\begin{figure*}
\includegraphics[scale=1]{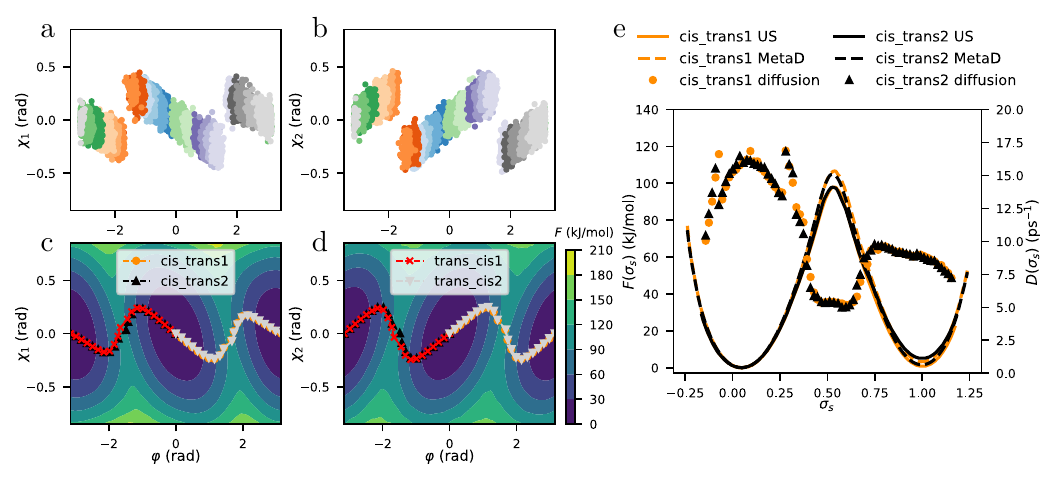}
\caption{
\textbf{a:} Scatter plots of the umbrella sampling simulations (one color per umbrella) for dihedral $\varphi$ vs. improper dihedral $\chi_1$.
\textbf{b:} The same for dihedral $\phi$ vs. improper dihedral $\chi_2$.
\textbf{c:} 3-dimensional free energy surface $F(\varphi, \chi_1, \chi_2)$ from metadynamics projected into the $(\varphi, \chi_1)$-space. Lines show optimized reaction coordinates.
\textbf{d:} The same but projected into $(\varphi, \chi_2)$-space
\textbf{e:} Free energy profiles from metadynamics and umbrella sampling as well as diffusion profiles for optimized reaction coordinate for the cis-to-trans isomerization.
 }
\label{fig:2D_correlations_FES_paths}
\centering
\end{figure*}
The reaction coordinate $q=\phi$ cleanly separates the reactant and the product state and thus fulfills an important criterion for a good reaction coordinate. 
However, closer inspections shows that other degrees of freedom besides the torsion angle participate in the cis-trans isomerization. 
The bonding environments around C$_{13}$ and C$_{14}$ are planar when retinal is in the cis- or trans-configuration, but 
both C$_{13}$ and C$_{14}$ bend out of plane in the vicinity of the transition state \cite{bondar2011ground,ghysbrecht2023thermal}.
The out-of-plane motion around C$_{13}$ is captured by the improper dihedral $\chi_1$ defined by $\left\{\mathrm{C}_{13},\mathrm{C}_{14}, \mathrm{C}_{12}, \mathrm{C}_{20}\right\}$. 
Likewise, the out-of-plane motion around C$_{14}$ is captured by the improper dihedral $\chi_2$ defined by $\left\{\mathrm{C}_{14},\mathrm{C}_{15}, \mathrm{C}_{13}, \mathrm{H}\right\}$. 
The correlation between $\varphi$ and the two improper dihedrals has been demonstrated at the levels of DFT/B3LYP and DFTB\cite{bondar2011ground,ghysbrecht2023thermal}, and is also captured by our umbrella sampling simulations using an empirical force field.
Fig.~\ref{fig:2D_correlations_FES_paths}.a shows the configurations of a series of umbrella sampling simulations projected into the two-dimensional space spanned by $\varphi$ and $\chi_1$. 
These distributions seem to ``jump" at the transition states ($\varphi = \pm \frac{\pi}{2}$). 
The projection into the space spanned by $\varphi$ and $\chi_2$ shows a similar behaviour (Fig.~\ref{fig:2D_correlations_FES_paths}.b). 
Note that the amplitude of the ``jump" is not very large, only $\pm 0.4\, \mathrm{rad}$, compared to the range of $\varphi$ itself.
(In SI Fig.~\ref{supp-fig:unscaled_2D} the zoom on $\chi_1$ and $\chi_2$ has been removed to gives a more realistic impression of the amplitude.) 
We optimized nonlinear reaction paths $\mathbf{s}\left(\sigma_s\right)$ in the space spanned by $\varphi$, $\chi_1$ and $\chi_2$ using the path finding algorithm from Ref.~\citenum{leines2012path}.
The paths are parametrized by a path progression parameter $\sigma_s$ which can be used as a reaction coordinate in rate theories: $q = \sigma_s$.
In total, we optimized four reaction paths: 
two reaction paths for the transition from cis to trans, each rotating in a different direction, and similarly two reaction paths for the transition from trans to cis (Fig.~\ref{fig:2D_correlations_FES_paths}.c,d).
The progress of the optimization is shown in SI Fig.~\ref{supp-fig:2D_path_convergence}. 
The optimized reaction coordinates are correlated to $\chi_1$ and $\chi_2$ but do not exhibit any sudden jumps in the two-dimensional distributions (SI Fig.~\ref{supp-fig:sss_correlation}).
To employ one-dimensional rate theories on these optimized reaction coordinates, we calculated free energy functions $F(\sigma_s)$, using umbrella sampling and metadynamics, as well as diffusion profiles (see Fig.~\ref{fig:2D_correlations_FES_paths}.e for paths from cis to trans and Fig.~\ref{supp-fig:path_profiles_trans_cis} in the SI for paths from trans to cis).
We will discuss the rate constant derived from the umbrella sampling for the reaction cis $\rightarrow$ trans in detail. 
The rate constants for the reverse reaction have similar values and show the same trends (see Tab.~\ref{tab:rates}).
For the optimized reaction coordinate umbrella sampling yields a barrier height of 
$F_{\mathrm{cis} \rightarrow \mathrm{trans}}^{\ddagger} = 98\, \mathrm{kJ/mol}$, which is $9\, \mathrm{kJ/mol}$ higher than the free energy barrier for $\varphi$.
Due to this higher free energy barrier, all one-dimensional rate theories yield lower rates for $\sigma_s$ than for $\varphi$ and therefore are in much better agreement with the numerical results. 
Nonetheless, a discussion of the friction regime is worthwhile.
Despite the increase in the free energy barrier, the energy ratio is only slightly lower than for $\varphi$:
$RT / F_{\mathrm{cis} \rightarrow \mathrm{trans}}^{\ddagger} = 0.025$.
By contrast the friction ratio for $\sigma_s$ is about ten times higher than for $ \varphi$: namely $\xi_{TS} / \omega_{TS} = 5.09$ (for path \textit{cis\_trans1}).
This is caused by an increased friction coefficient of the effective dynamics and a broader free energy barrier ($\xi_{TS} = 785 \, \mathrm{ps}^{-1}$  and $\omega_{TS} = 154 \, \mathrm{ps}^{-1}$ for path \textit{cis\_trans1}).  
Consequently, the effective dynamics along $q=\sigma_s$ fall into the high friction regime.
This is also reflected by the values for Kramers' rate constants for moderate friction regime and for the high friction regime. 
For $q=\sigma_s$, these two equation yield almost the same value ($k_{\mathrm{cis}\rightarrow\mathrm{trans}} = 4.81\cdot 10^{-5} \, \mathrm{s}^{-1}$ and
$k_{\mathrm{cis}\rightarrow\mathrm{trans}} = 4.98\cdot 10^{-5} \, \mathrm{s}^{-1}$,
see Tab.~\ref{tab:rates}), which is only the case in the high friction regime. 
Another consequence of the higher friction regime is that simple TST considerably overestimates the rate constant 
($k_{\mathrm{cis}\rightarrow\mathrm{trans}} = 2.57\cdot 10^{-4} \, \mathrm{s}^{-1}$).
Pontryagin's rate theory and the grid-based model yield the same rate constant ($k_{\mathrm{cis}\rightarrow\mathrm{trans}} = 3.66\cdot 10^{-5} \, \mathrm{s}^{-1}$), which is lower than the result from Kramers' rate theory.
Since the effective dynamics fall into the high friction regime, this deviation is not likely caused by the assumption of overdamped Langevin dynamics in these theories. 
The more likely cause is that Kramers' rate theory assumes a position independent diffusion constant, whereas both Pontryagin and grid-based models account for variations in the diffusion constant along the reaction coordinate. 
In summary, optimizing the reaction coordinate had two effects on the one-dimensional rate models: the free-energy barrier increased, and the friction ratio $\xi_{TS} / \omega_{TS}$ increased.
Both effects lower the estimate of the rate constant and thus improve the agreement with the simulation results.

\subsection{Umbrella sampling vs metadynamics}
\label{sec:USvsMetaD}
For both reaction coordinates, $q=\varphi$ and $q= \sigma_s$, we find that the free energy barriers from 
metadynamics are consistently 7 to 10 kJ/mol higher than the free energy barriers from umbrella sampling (SI Tab.~\ref{supp-tab:path_constants_us} and \ref{supp-tab:path_constants_metad}). 
Consequently, the rate constants based on metadynamics are about an order of magnitude lower than those based on umbrella sampling. 
The sampling for both methods is generous, such that the statistical uncertainty is negligible 
(Fig.~\ref{fig:FES_phi} and \ref{fig:2D_correlations_FES_paths}.e). 
The free energy functions do not change noticeably when we vary the parameters of the method (force constant and positioning of the umbrella potentials, width of the Gaussian bias potentials in metadynamics, SI Fig.~\ref{supp-fig:us_metad_together}).
However, in metadynamics, the error estimated by block analysis\cite{bussi2019analyzing} as well as the free energy difference between the cis- and the trans-configuration converged only slowly for both reaction coordinates (SI Fig.~\ref{supp-fig:metad_convergence}).
This could indicate that $\varphi$ and also the optimized $\sigma_s$ is correlated to further degrees of freedom. 
Candidates are the torsion around the neighboring single bonds, i.e. C$_{12}$-C$_{13}$ and  C$_{13}$-C$_{14}$.
Projecting the configurations into the space spanned by these torsion angles and $\varphi$, we find similar ``jumps" as in Fig.~\ref{fig:2D_correlations_FES_paths}, albeit less pronounced (SI Fig.~\ref{supp-fig:phi_proper_correlation}). 
The optimized reaction coordinate $\sigma_s$ is still correlated to the torsion around these single bonds but does not exhibit any sudden jumps in the two-dimensional distributions
(SI Fig.~\ref{supp-fig:sss_correlation}), even though these torsion angles were not explicitly part of the optimization process. 
Thus, it seems unlikely that the discrepancy between umbrella sampling and metadynamics can be attributed to the correlation to these single bonds.
%

\subsection{Multidimensional collective variables}
An alternative to one-dimensional rate theories are grid-based models in multidimensional collective variable spaces.
We calculated the three-dimensional free energy function $F(\varphi, \chi_1, \chi_2)$ using metadynamics with three-dimensional Gaussian bias functions, as well as using umbrella sampling with three-dimensional harmonic constraints.
The position dependent diffusion profile for the diffusion in each of the three directions were calculated using umbrella sampling with three-dimensional harmonic restraining potentials on a coarse grid (\textit{grid1}) and a fine grid (\textit{grid2}). 
See SI Figs. \ref{supp-fig:3D_diffusion1} and \ref{supp-fig:3D_diffusion2}. 
The projection of $F(\varphi, \chi_1, \chi_2)$ into the two-dimensional spaces $(\varphi, \chi_1)$ and $(\varphi, \chi_2)$ are shown in Fig.~\ref{fig:2D_correlations_FES_paths}.c and d, and explain the sudden ``jumps" in the two-dimensional distributions in Fig.~\ref{fig:2D_correlations_FES_paths}.a and b. 
The free energy minima of the cis- and the trans-configuration are slanted in the two-dimensional space. 
Specifically, the configurations overlap for values of $\varphi$ near the barrier, and thus $\varphi$ does not cleanly discriminate between cis- and the trans-configuration.
In Fig.~\ref{fig:2D_correlations_FES_paths}.c, when going form negative values of $\varphi$ to positive values across the cis-minimum, $\chi_1$ steadily decreases from +0.3 rad to -0.3 rad.
At the transition state, the value of $\chi_1$ is restored to $\chi_1 = +0.3\, \mathrm{rad}$ within a short interval of $\varphi$, giving rise to ``jumps" in the two-dimensional distribution. 
The correlation of $\varphi$ to $\chi_2$ shows a similar behaviour (Fig.~\ref{fig:2D_correlations_FES_paths}.D).
The optimized path follows this sudden change in $\chi_1$ and $\chi_2$ by zigzagging through the three-dimensional space. 
To obtain our grid-based rate model, we discretized the three-dimensional space $(\varphi, \chi_1, \chi_2)$, and calculated the rate matrix $\mathbf{Q}$ from the free energy surface and the diffusion profiles  using eq.~\ref{eq:Qij_HAA_1}, which then yielded the reaction rate constants via eq.~\ref{eq:solve_Q} and \ref{eq:Q_MFPT}.
Convergence of the rates with respect to different discretizations of $(\varphi, \chi_1, \chi_2)$ is better for metadynamics than for umbrella sampling. (SI Fig.~\ref{supp-fig:k_discretization}). 
The rate constants of the three-dimensional models are in good agreement with the rate constants from the simulations (Tab.~\ref{tab:rates}).
Most importantly, in the three-dimensional models, the results from metadynamics and from umbrella sampling are in excellent agreement. 
The rate estimates are somewhat sensitive to the model of the diffusion profile. 
In particular, using a uniform diffusion profile along each of the three collective variables (averaged \textit{grid1} and \textit{grid2}) yields rate constants that are slightly closer to the simulated results than when estimating a fully position dependent diffusion profile (diffusion \textit{grid1} and \textit{grid2}). 
This might be caused by numerical effects when using the fully position dependent diffusion profile.

We additionally repeated the infrequent metadynamics simulations using three-dimensional Gaussian bias functions  in the space spanned by $\varphi$, $\chi_1$ and $\chi_2$.
The resulting rate constants are very close to those obtained from infrequent metadynamics with one-dimensional biasing (Tab.~\ref{tab:rates}).
%

\section{Computational methods}
\subsection{One-dimensional model systems}
The model free energy functions on the circular reaction coordinate $q\in\left[-\pi,\pi\right]$ in Fig.~\ref{fig:1Dmodel_rates}.a and \ref{fig:1Dmodel_rates}.b are defined by
\begin{eqnarray}
F(q) = a \cos 2q -  b\sin q  \, ,
\label{eq:pot1D}
\end{eqnarray}
in units $\mathrm{kJ\,mol^{-1}}$. 
We set $a = 2.4 R T$ and $b = -1\, \mathrm{kJ\, mol^{-1}}$ for Fig.~\ref{fig:1Dmodel_rates}.a, and 
$a = 10.4 \, \beta^{-1} \,  \mathrm{kJ \, mol}^{-1}$ and $b = -1\, \mathrm{kJ\, mol^{-1}}$ 
for Fig.~\ref{fig:1Dmodel_rates}.b.
The free energy function in Fig.~\ref{fig:1Dmodel_rates}.c was prescribed by a spline interpolation of a metadynamics profile measured along $\varphi$ of the retinal system studied in this work.
Numerical simulation was carried out by implementing the ISP integrator\cite{Izaguirre2010} for underdamped Langevin dynamics (SI eq.~\ref{supp-eq:ISP}) for a particle with effective mass $m = 1 \, \mathrm{amu\cdot nm^2 \cdot rad^{-2}}$ and using a time step of $\Delta t = 0.001\, \mathrm{ps}$.
The temperature of the system was $T = 300\, \mathrm{K}$, and the gas constant $R =8.314463 \, \mathrm{J} \, \mathrm{mol}^{-1} \, \mathrm{K}^{-1}$ for all simulations.
We varied the value of the friction coefficient $\xi$ in ranges that matched the free energy barrier of the model potentials: 
Fig.~\ref{fig:1Dmodel_rates}.a: $\xi = 0.002 \, \mathrm{ps}^{-1}$ to $\xi = 72 \, \mathrm{ps}^{-1}$; 
Fig.~\ref{fig:1Dmodel_rates}.b: $\xi = 0.005 \, \mathrm{ps}^{-1}$ to $\xi = 150 \, \mathrm{ps}^{-1}$;
Fig.~\ref{fig:1Dmodel_rates}.c: $\xi = 0.02 \, \mathrm{ps}^{-1}$ to $\xi = 713 \, \mathrm{ps}^{-1}$.
For the model systems in Fig.~\ref{fig:1Dmodel_rates}.b and Fig.~\ref{fig:1Dmodel_rates}.c, we used infrequent metadynamics \cite{tiwary2013metadynamics} to enhance the sampling. 
For Fig.~\ref{fig:1Dmodel_rates}.b,  Gaussian bias functions of height 
$0.1\,\mathrm{kJ \, mol}^{-1}$ and width $0.6\,\mathrm{rad}$ were deposited every 300 time steps (weak friction regime); 
$0.05\,\mathrm{kJ \, mol}^{-1}$ and width $0.4\,\mathrm{rad}$ every 150 time steps (moderate and high friction regime).
For Fig.~\ref{fig:1Dmodel_rates}.c, Gaussian bias functions of height $0.8\,
\mathrm{kJ \, mol}^{-1}$ and width $0.1\,\mathrm{rad}$ were deposited every 100 time steps (weak friction regime); and of height $0.5\, \mathrm{kJ \, mol}^{-1}$ and width $0.1\,\mathrm{rad}$ every 100 time steps (moderate and high friction regime).
Well-tempering has been enforced using a biasing factor of 100.
Forces were calculated by adding the gradient of the free energy profile to the gradient of the biasing potential $U(q,t)$.
The transition rates in Fig.~\ref{fig:1Dmodel_rates}.a and Fig.~\ref{fig:1Dmodel_rates}.b were estimated by realizing 100 simulations starting at the left minimum $q_A = -1.6\, \mathrm{rad}$ and measuring the first-passage time to reach the barrier at $q_{TS}= 0\, \mathrm{rad}$ or $q_{TS'}= -\pi\, \mathrm{rad}$.
The reciprocal of the mean and standard deviation of the first-passage times gives the escape rate with its uncertainty.
The transition rates in Fig.~\ref{fig:1Dmodel_rates}.b were estimated by the same procedure, but simulations started at $q_A = 0\, \mathrm{rad}$ and stopped at $q_{TS} = 1.6 \, \mathrm{rad}$ or $q_{TS'}= -1.6\, \mathrm{rad}$.
Transition rates for simple TST formula and Kramers' rate theory in the moderate and high friction limit were calculated by applying of eqs.~\ref{eq:simpleTST}, \ref{eq:KramersModerate}, \ref{eq:KramersHigh}, where $\omega_A$, $\omega_B$ or $\omega_{TS}$ were calculated from the second derivative of the free energy profile.
The integrals in Kramers' rate theory in the weak friction regime (eq.~\ref{eq:KramersWeakLimit}) and  Pontryagin's rate theory (eq.~\ref{eq:Pontryagin}) were evaluated by discretizing the interval $[-\pi, +\pi]$ in 100 subsets of equal length and employing the trapezoidal rule.
The same discretization was used for grid-based model (eq.~\ref{eq:Qij_SqRA_1}).
%

\subsection{Atomistic model of retinal}
Retinal parameters for atomistic force field calculations were taken from DFT studies on the protonated Schiff base\cite{hayashi2002structural}, adapted to GROMACS format\cite{malmerberg2011time}, while the connecting amino acid was modelled using the AMBER99SB*-ILDN forcefield\cite{lindorff2010improved}. 
The starting structure was obtained by cutting out the lysine amino acid and  retinal cofactor from a recent crystal structure\cite{volkov2017structural}.
All simulations are carried out at $300\,\mathrm{K}$ in vacuum using stochastic dynamics with GROMACS\cite{abraham2015gromacs} version 2019.4 built in Langevin integrator with a $2\,\mathrm{fs}$ step size and an inverse friction coefficient of $2\,\mathrm{ps}$.
For the path collective variables occasionally lower time steps were used. 
Position restraints of $10000\,\mathrm{kJ}\,\mathrm{mol}^{-1}\mathrm{nm}^{-2}$ were put on all heavy atoms of the peptide chain as well as on the lysine chain carbon atoms (Fig.~\ref{fig:retinal_structures}), while the LINCS constraint algorithm\cite{hess1997lincs} was applied to all hydrogen bonds. 
Before all simulations, energy minimization and NVT equilibration were performed.
Metadynamics \cite{huber1994local, grubmuller1995predicting, voter1997hyperdynamics,laio2002escaping}
and umbrella sampling \cite{torrie1977nonphysical} simulations were carried out by plugging PLUMED\cite{tribello2014plumed} with GROMACS.
Diffusion profiles were calculated by following Ref~\onlinecite{hummer2005position}.
The reaction coordinate was optimized using the PLUMED implementation of the adaptive path CV method in Ref.~\citenum{leines2012path} in combination with metadynamics.
Effective masses of the reactant states were calculated by measuring the average squared velocity along the reaction coordinate and applying eq.~\ref{eq:effective_mass}.
Frequencies $\omega_A$, $\omega_B$, $\omega_{TS}$ and $\omega_{TS'}$ were calculated from spring constants obtained by harmonically fitting the corresponding wells or barriers.
Free energy barriers are measured from the FES directly.
One-dimensional rate methods (simple TST, Kramers', Pontryagin) can then be applied straightforwardly.
For grid-based methods, 500 cells were used for one-dimensional discretizations, while a discretization of (31,23,23) was used in the 3D CV space ($\varphi$,$\chi_1$,$\chi_2$), with $\chi_1$ and $\chi_2$ being discretized in the region between -1 and 1 radians for metadynamics and -0.7 and 0.7 radians for umbrella sampling.
Rates from direct numerical simulation were obtained from infrequent metadynamics runs, where acceleration factors were calculated directly by PLUMED.
See SI section \ref{supp-sec:SI_comp_det} for a complete description of the computational details.

%

\section{Discussion and Conclusion}

\begin{figure}
\includegraphics[scale=1]{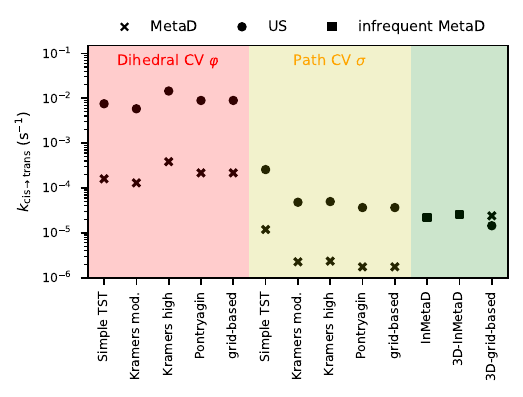}
\caption{
Rate constants determined through different methodologies for the thermal cis-trans isomerization over the C$_{13}$=C$_{14}$ double bond in retinal.
}
\label{fig:RateOverview}
\centering
\end{figure}
%
%
Fig.~\ref{fig:RateOverview} summarizes the results of this study.
For the thermal cis-trans isomerization in retinal, different methods to calculate the rate constant yield drastically different results. 
Specifically, reaction-coordinate based rate estimates using the torsion angle $q=\varphi$ as an intuitive reaction coordinate differ by about two orders of magnitude from the infrequent metadynamics results, which is based on counting transitions.
Furthermore within the reaction-coordinate based estimates, the results are very sensitive to the method of calculating the free-energy profile: 
the results with an umbrella-sampling FES differ systematically form those with a metadynamic FES.
These deviations are not primarily caused by a poor choice of the friction regime in the Kramers' rate estimates.
For $q=\varphi$ the effective dynamics falls into the intermediate friction regime, but using rate equations for overdamped friction instead changes the rate constant only by about a factor of 2. 
Thus, for particular choice for calculating $F(q)$, all one-dimensional rate theories yield similar results. 
The same is true for the optimized reaction coordinate, whose effective dynamics falls into the high friction regime. 
However, it remains crucial to confirm the friction regime and use the appropriate formula, because Kramers' rate for high-friction scenarios may significantly overestimate the rate constant when applied in the wrong friction regime (Fig.~\ref{fig:1Dmodel_rates}).
Optimizing the reaction coordinate lowered the rate estimates of the one-dimensional rate theories by two orders of magnitude compared to $q=\varphi$.
These lower rate constants are likely more accurate, since the free energy function of a poor reaction coordinate underestimates the true reaction barrier.
It is surprising at first that, for the cis-trans isomerization, an improved reaction coordinate has such a massive effect on the accuracy of the rate constant.
Cis and trans configuration are defined by the  torsion angle $\varphi$ and  therefore $q=\varphi$ cleanly separates reactant and product state \cite{bolhuis2002transition}, which is a crucial criterion for an optimal reaction coordinate.
However, the optimization of $q$ revealed that the intuitive reaction coordinate fails another important criterion. 
In the transition state, the optimized reaction coordinate forms a large angle to the intuitive reaction coordinate $q=\varphi$ (Fig.~\ref{fig:2D_correlations_FES_paths}.c and d).
Consequently, the probability flux across the barrier is nearly orthogonal to $q=\varphi$, rather than parallel as expected for an optimal reaction coordinate.
\cite{bolhuis2002transition}.
This curvature of the optimal reaction coordinate arises, because at the transition state the C$_{13}$ and C$_{14}$ slightly bend out of plane and thus the reaction coordinate takes a short detour into otherwise rigid degrees of freedoms, namely the improper dihedral angles $\chi_1$ and $\chi_2$.
This detour is possible because, at the transition state, the electronic structure changes.
In this case, the $p$-orbitals of C$_{13}$ and C$_{14}$ do no longer overlap. 
This effect is captured by DFT-calculations \cite{ghysbrecht2023thermal} and reproduced by the empirical force field used in this study.
Since a change in the electronic structure a the transition state is a hallmark of chemical reactions, we suspect that such short detours into orthogonal degrees of freedom (with respect to an intuitive reaction coordinate) will be the rule rather than the exception when modelling chemical reactions.
However, finding such a curved optimal reaction coordinate is not trivial, even if an initial reaction coordinate and candidates for further correlated degrees of freedom are known, as in the case of retinal \cite{bondar2011ground,ghysbrecht2023thermal}. 
Besides the path-based method \cite{leines2012path} we used in our study, a wide range of other methods to identify optimal reaction coordinates have been proposed \cite{vanden2010transition, roux2021string, palacio2021free}, including recent approaches based on neural networks \cite{shmilovich2023girsanov, jung2023machine}.
An alternative to optimizing the reaction coordinate is to improve the estimate of the rate constant for a sub-optimal reaction coordinate $q$ by including non-Markovian effects into the effective dynamics along $q$.
The corresponding equations are based on the generalized Langevin equation (GLE). 
Here, non-Markovian behaviour arises from the memory kernel, which is a time-integral over the time-dependent friction coefficient \cite{mori1965transport, zwanzig1961memory, zwanzig2001nonequilibrium}. 
Memory kernels are notoriously hard to predict, but recently multiple methods have emerged to model them\cite{dominic2023memory,vroylandt2022likelihood,ayaz2021non}.
In addition, Grote-Hynes theory provides an equation for the  memory kernel\cite{grote1980stable}.
The resulting rate equation has the same functional form as Kramers rate equation for the moderate friction regime (eq.~\ref{eq:KramersModerate}), where the Markovian friction $\xi$ is replaced by the Laplace transform of the time-dependent friction coefficient \cite{Peters2017}.
In general, the closer the reaction coordinate follows the probability flux of the reaction, the smaller are the non-Markovian effects \cite{zwanzig1961memory}. 
Although non-Markovian rate theories provide accurate rate estimates even for imperfect reaction coordinates, using these suboptimal reaction coordinates risks obscuring important mechanistic details needed for understanding the reaction.
For example in retinal, the out-of-plane bending of C$_{13}$ and C$_{14}$ near the transition state is not captured by the initial reaction coordinate $q=\varphi$.
For our system, despite using an optimized reaction coordinate, metadynamics and umbrella sampling produced different free-energy barriers, leading to significantly different rate estimates, as shown in Fig.~\ref{fig:RateOverview}.
This is likely caused by the strong curvature of the optimized reaction coordinate and might indicate that $q=\sigma_s$ is not yet fully optimal. 
Grid-based models in a multidimensional collective variable space offer an alternative to optimizing the one-dimensional reaction coordinate or including memory effects. 
Using the torsion angle $\phi$ and two improper torsion angles at C$_{13}$ and C$_{14}$ as collective variables, we obtained rate estimates that are in very good agreement with the simulation results (Fig.~\ref{fig:RateOverview}).
Moreover, for these multidimensional models, the free-energy functions derived from metadynamics and umbrella sampling agree, leading to similar rate constants.
Multidimensional models have additional advantages: they can be applied to multistate dynamics, do not assume timescale separation, and they yield information on all of the slow processes in the system\cite{Lie2013,donati2021markov}.
The trade-off is the need to estimate a multidimensional free energy surface.

Furthermore, methods that model the reaction rate by envisioning a flux over a dividing surface \cite{wigner1938transition} rather than a maximum in an energy landscape can be considered. 
In variational transition state theory (VTST), different approaches are used to optimize the dividing surface and minimize the TST reaction rate\cite{keck1960variational,truhlar1980variational,bao2017variational}.
The reactive flux method \cite{chandler1978statistical}, links the flux across a dividing surface to a correlation function which can be estimated from molecular simulations. 
Modern methods based on this framework include transition path sampling\cite{dellago1999calculation,bolhuis2002transition}, transition interface sampling\cite{van2003novel} and forward flux sampling\cite{allen2006forward}.
Our results show that rate constants for chemical reactions can be determined with high accuracy (within the classical approximation) from molecular simulations.
The caveat is that the methods need to be carefully chosen for the system at hand. 
Of the various parameters that influence the rate constant, the curvature of the reaction coordinate at the transition state emerges as the most critical one. 
%

%
%
\section{Acknowledgements}
This research has been funded by Deutsche Forschungsgemeinschaft (DFG) 
through grant SFB 1114 "Scaling Cascades in Complex Systems" - project number 235221301,
as well as by the Cluster of Excellence MATH+, project AA1-15 ``Math-powered drug-design''.
S.G.~acknowledges funding by the Einstein Center of Catalysis/BIG-NSE.

%
%
\section{References}

\bibliographystyle{unsrt}
\bibliography{literature}

\begin{thebibliography}{10}

\bibitem{vitalini2015dynamic}
Francesca Vitalini, Antonia~SJS Mey, Frank No{\'e}, and Bettina~G Keller.
\newblock Dynamic properties of force fields.
\newblock {\em The Journal of Chemical Physics}, 142(8), 2015.

\bibitem{ghysbrecht2023thermal}
Simon Ghysbrecht and Bettina~G Keller.
\newblock Thermal isomerization rates in retinal analogues using ab-initio molecular dynamics.
\newblock {\em Journal of Computational Chemistry}, 45(16):1390--1403, 2024.

\bibitem{harvey2019scope}
Jeremy~N Harvey, Fahmi Himo, Feliu Maseras, and Lionel Perrin.
\newblock Scope and challenge of computational methods for studying mechanism and reactivity in homogeneous catalysis.
\newblock {\em Acs Catalysis}, 9(8):6803--6813, 2019.

\bibitem{marx2000ab}
Dominik Marx and Jurg Hutter.
\newblock Ab initio molecular dynamics: Theory and implementation.
\newblock {\em Modern methods and algorithms of quantum chemistry}, 1(301-449):141, 2000.

\bibitem{eyring1935activated}
Henry Eyring.
\newblock The activated complex in chemical reactions.
\newblock {\em The Journal of Chemical Physics}, 3(2):107--115, 1935.

\bibitem{keck1960variational}
James~C Keck.
\newblock Variational theory of chemical reaction rates applied to three-body recombinations.
\newblock {\em The Journal of Chemical Physics}, 32(4):1035--1050, 1960.

\bibitem{truhlar1980variational}
Donald~G Truhlar and Bruce~C Garrett.
\newblock Variational transition-state theory.
\newblock {\em Acc. Chem. Res.}, 13(12):440--448, 1980.

\bibitem{grote1980stable}
Richard~F Grote and James~T Hynes.
\newblock The stable states picture of chemical reactions. ii. rate constants for condensed and gas phase reaction models.
\newblock {\em The Journal of Chemical Physics}, 73(6):2715--2732, 1980.

\bibitem{schade2023breaking}
Robert Schade, Tobias Kenter, Hossam Elgabarty, Michael Lass, Thomas~D K{\"u}hne, and Christian Plessl.
\newblock Breaking the exascale barrier for the electronic structure problem in ab-initio molecular dynamics.
\newblock {\em The International Journal of High Performance Computing Applications}, 37(5):530--538, 2023.

\bibitem{gaus2011dftb3}
Michael Gaus, Qiang Cui, and Marcus Elstner.
\newblock {DFTB3}: Extension of the self-consistent-charge density-functional tight-binding method {(SCC-DFTB)}.
\newblock {\em Journal of Chemical Theory and Computation}, 7(4):931--948, 2011.

\bibitem{kocer2022neural}
Emir Kocer, Tsz~Wai Ko, and J{\"o}rg Behler.
\newblock Neural network potentials: A concise overview of methods.
\newblock {\em Annual review of physical chemistry}, 73:163--186, 2022.

\bibitem{noe2020machine}
Frank No{\'e}, Alexandre Tkatchenko, Klaus-Robert M{\"u}ller, and Cecilia Clementi.
\newblock Machine learning for molecular simulation.
\newblock {\em Annual review of physical chemistry}, 71:361--390, 2020.

\bibitem{Peters2017}
Baron Peters.
\newblock {\em Reaction Rate Theory and Rare Events}.
\newblock Elsevier, 1st edition, 2017.

\bibitem{pietrucci2017strategies}
Fabio Pietrucci.
\newblock Strategies for the exploration of free energy landscapes: Unity in diversity and challenges ahead.
\newblock {\em Reviews in Physics}, 2:32--45, 2017.

\bibitem{salvalaglio2014assessing}
Matteo Salvalaglio, Pratyush Tiwary, and Michele Parrinello.
\newblock Assessing the reliability of the dynamics reconstructed from metadynamics.
\newblock {\em Journal of Chemical Theory and Computation}, 10(4):1420--1425, 2014.

\bibitem{lane2013milliseconds}
Thomas~J Lane, Diwakar Shukla, Kyle~A Beauchamp, and Vijay~S Pande.
\newblock To milliseconds and beyond: challenges in the simulation of protein folding.
\newblock {\em Current opinion in structural biology}, 23(1):58--65, 2013.

\bibitem{ahmad2022enhanced}
Katya Ahmad, Andrea Rizzi, Riccardo Capelli, Davide Mandelli, Wenping Lyu, and Paolo Carloni.
\newblock Enhanced-sampling simulations for the estimation of ligand binding kinetics: current status and perspective.
\newblock {\em Frontiers in molecular biosciences}, 9, 2022.

\bibitem{arjun2020molecular}
A~Arjun and PG~Bolhuis.
\newblock Molecular understanding of homogeneous nucleation of {CO2} hydrates using transition path sampling.
\newblock {\em The Journal of Physical Chemistry B}, 125(1):338--349, 2020.

\bibitem{borden2016reactions}
Weston~Thatcher Borden.
\newblock Reactions that involve tunneling by carbon and the role that calculations have played in their study.
\newblock {\em Wiley Interdisciplinary Reviews: Computational Molecular Science}, 6(1):20--46, 2016.

\bibitem{zhang2023vibrational}
Yuzhe Zhang, Yiwen Wang, Xi~Xu, Zehua Chen, and Yang Yang.
\newblock Vibrational spectra of highly anharmonic water clusters: Molecular dynamics and harmonic analysis revisited with constrained nuclear-electronic orbital methods.
\newblock {\em Journal of Chemical Theory and Computation}, 19(24):9358--9368, 2023.

\bibitem{kieninger2020dynamical}
Stefanie Kieninger, Luca Donati, and Bettina~G Keller.
\newblock Dynamical reweighting methods for {M}arkov models.
\newblock {\em Current opinion in structural biology}, 61:124--131, 2020.

\bibitem{bolhuis2002transition}
Peter~G Bolhuis, David Chandler, Christoph Dellago, and Phillip~L Geissler.
\newblock Transition path sampling: Throwing ropes over rough mountain passes, in the dark.
\newblock {\em Annual Review of Physical Chemistry}, 53(1):291--318, 2002.

\bibitem{zuckerman2017weighted}
Daniel~M Zuckerman and Lillian~T Chong.
\newblock Weighted ensemble simulation: review of methodology, applications, and software.
\newblock {\em Annual review of biophysics}, 46:43--57, 2017.

\bibitem{tiwary2013metadynamics}
Pratyush Tiwary and Michele Parrinello.
\newblock From metadynamics to dynamics.
\newblock {\em Physical Review Letters}, 111(23):230602, 2013.

\bibitem{tribello2014plumed}
Gareth~A Tribello, Massimiliano Bonomi, Davide Branduardi, Carlo Camilloni, and Giovanni Bussi.
\newblock {PLUMED} 2: New feathers for an old bird.
\newblock {\em Computer Physics Communications}, 185(2):604--613, 2014.

\bibitem{daldrop2018butane}
Jan~O Daldrop, Julian Kappler, Florian~N Br{\"u}nig, and Roland~R Netz.
\newblock Butane dihedral angle dynamics in water is dominated by internal friction.
\newblock {\em Proceedings of the National Academy of Sciences}, 115(20):5169--5174, 2018.

\bibitem{hummer2005position}
Gerhard Hummer.
\newblock Position-dependent diffusion coefficients and free energies from {B}ayesian analysis of equilibrium and replica molecular dynamics simulations.
\newblock {\em New Journal of Physics}, 7(1):34, 2005.

\bibitem{kramers1940brownian}
Hendrik~Anthony Kramers.
\newblock Brownian motion in a field of force and the diffusion model of chemical reactions.
\newblock {\em Physica}, 7(4):284--304, 1940.

\bibitem{Hanggi1990}
Peter H\"anggi, Peter Talkner, and Michal Borkovec.
\newblock Reaction-rate theory: fifty years after kramers.
\newblock {\em Rev. Mod. Phys.}, 62:251–341, 1990.

\bibitem{hayashi2002structural}
Shigehiko Hayashi, Emad Tajkhorshid, and Klaus Schulten.
\newblock Structural changes during the formation of early intermediates in the bacteriorhodopsin photocycle.
\newblock {\em Biophysical Journal}, 83(3):1281--1297, 2002.

\bibitem{malmerberg2011time}
Erik Malmerberg, Ziad Omran, Jochen~S Hub, Xuewen Li, Gergely Katona, Sebastian Westenhoff, Linda~C Johansson, Magnus Andersson, Marco Cammarata, Michael Wulff, et~al.
\newblock Time-resolved {WAXS} reveals accelerated conformational changes in iodoretinal-substituted proteorhodopsin.
\newblock {\em Biophysical Journal}, 101(6):1345--1353, 2011.

\bibitem{pontryagin1933statistical}
Lev Pontryagin, Aleksandr Andronov, and Aleksandr Vitt.
\newblock On the statistical investigation of dynamic systems.
\newblock {\em Zh. Eksp. Teor. Fiz}, 3:165, 1933.

\bibitem{Lie2013}
Han~Cheng Lie, Konstantin Fackeldey, and Marcus Weber.
\newblock A square root approximation of transition rates for a {M}arkov state model.
\newblock {\em SIAM. J. Matrix Anal. Appl.}, 34:738–756, 2013.

\bibitem{donati2021markov}
Luca Donati, Marcus Weber, and Bettina~G Keller.
\newblock Markov models from the square root approximation of the {F}okker--{P}lanck equation: Calculating the grid-dependent flux.
\newblock {\em Journal of Physics: Condensed Matter}, 33(11):115902, 2021.

\bibitem{Bicout1998}
D.~J. Bicout and A.~Szabo.
\newblock Electron transfer reaction dynamics in non-{D}ebye solvents.
\newblock {\em Journal of Chemical Physics}, 109:10.1063/1.476800, 1998.

\bibitem{heida2021consistency}
Martin Heida, Markus Kantner, and Artur Stephan.
\newblock Consistency and convergence for a family of finite volume discretizations of the {F}okker--{P}lanck operator.
\newblock {\em ESAIM: Mathematical Modelling and Numerical Analysis}, 55(6):3017--3042, 2021.

\bibitem{berezhkovskii2019committors}
Alexander~M Berezhkovskii and Attila Szabo.
\newblock Committors, first-passage times, fluxes, {M}arkov states, milestones, and all that.
\newblock {\em The Journal of Chemical Physics}, 150(5):054106, 2019.

\bibitem{barducci2008well}
Alessandro Barducci, Giovanni Bussi, and Michele Parrinello.
\newblock Well-tempered metadynamics: a smoothly converging and tunable free-energy method.
\newblock {\em Physical Review Letters}, 100(2):020603, 2008.

\bibitem{grubmuller1995predicting}
Helmut Grubm{\"u}ller.
\newblock Predicting slow structural transitions in macromolecular systems: Conformational flooding.
\newblock {\em Physical Review E}, 52(3):2893, 1995.

\bibitem{voter1997hyperdynamics}
Arthur~F Voter.
\newblock Hyperdynamics: Accelerated molecular dynamics of infrequent events.
\newblock {\em Physical Review Letters}, 78(20):3908, 1997.

\bibitem{tiwary2016kramers}
Pratyush Tiwary and BJ~Berne.
\newblock Kramers turnover: From energy diffusion to spatial diffusion using metadynamics.
\newblock {\em The Journal of Chemical Physics}, 144(13):134103, 2016.

\bibitem{schulten2014quantum}
Klaus Schulten and Shigehiko Hayashi.
\newblock Quantum biology of retinal.
\newblock In {\em Quantum Effects in Biology}, pages 237--263. Cambridge University Press, 2014.

\bibitem{torrie1977nonphysical}
Glenn~M Torrie and John~P Valleau.
\newblock Nonphysical sampling distributions in {M}onte {C}arlo free-energy estimation: Umbrella sampling.
\newblock {\em Journal of Computational Physics}, 23(2):187--199, 1977.

\bibitem{kumar1992weighted}
Shankar Kumar, John~M Rosenberg, Djamal Bouzida, Robert~H Swendsen, and Peter~A Kollman.
\newblock The weighted histogram analysis method for free-energy calculations on biomolecules. {I.} {T}he method.
\newblock {\em Journal of Computational Chemistry}, 13(8):1011--1021, 1992.

\bibitem{dickson2018erroneous}
Bradley~M Dickson.
\newblock Erroneous rates and false statistical confirmations from infrequent metadynamics and other equivalent violations of the hyperdynamics paradigm.
\newblock {\em Journal of Chemical Theory and Computation}, 15(1):78--83, 2018.

\bibitem{khan2020fluxional}
Salman~A Khan, Bradley~M Dickson, and Baron Peters.
\newblock How fluxional reactants limit the accuracy/efficiency of infrequent metadynamics.
\newblock {\em The Journal of Chemical Physics}, 153(5), 2020.

\bibitem{peters2016reaction}
Baron Peters.
\newblock Reaction coordinates and mechanistic hypothesis tests.
\newblock {\em Annual Review of Physical Chemistry}, 67:669--690, 2016.

\bibitem{bondar2011ground}
Ana-Nicoleta Bondar, Michaela Knapp-Mohammady, S{\'a}ndor Suhai, Stefan Fischer, and Jeremy~C Smith.
\newblock Ground-state properties of the retinal molecule: from quantum mechanical to classical mechanical computations of retinal proteins.
\newblock {\em Theoretical Chemistry Accounts}, 130:1169--1183, 2011.

\bibitem{leines2012path}
Grisell~D{\'\i}az Leines and Bernd Ensing.
\newblock Path finding on high-dimensional free energy landscapes.
\newblock {\em Physical Review Letters}, 109(2):020601, 2012.

\bibitem{bussi2019analyzing}
Giovanni Bussi and Gareth~A Tribello.
\newblock Analyzing and biasing simulations with {PLUMED}.
\newblock In {\em Biomolecular Simulations}, pages 529--578. Springer, 2019.

\bibitem{Izaguirre2010}
J.~A~. Izaguirre, C.~R. Schweet, and V.S. Pande.
\newblock Multiscale dynamics of macromolecules using normal mode {L}angevin.
\newblock {\em Pacific Symposium on Biocomputing}, {15}:240--251, 2010.

\bibitem{lindorff2010improved}
Kresten Lindorff-Larsen, Stefano Piana, Kim Palmo, Paul Maragakis, John~L Klepeis, Ron~O Dror, and David~E Shaw.
\newblock Improved side-chain torsion potentials for the {Amber ff99SB} protein force field.
\newblock {\em Proteins: Structure, Function, and Bioinformatics}, 78(8):1950--1958, 2010.

\bibitem{volkov2017structural}
Oleksandr Volkov, Kirill Kovalev, Vitaly Polovinkin, Valentin Borshchevskiy, Christian Bamann, Roman Astashkin, Egor Marin, Alexander Popov, Taras Balandin, Dieter Willbold, et~al.
\newblock Structural insights into ion conduction by channelrhodopsin 2.
\newblock {\em Science}, 358(6366), 2017.

\bibitem{abraham2015gromacs}
Mark~James Abraham, Teemu Murtola, Roland Schulz, Szil{\'a}rd P{\'a}ll, Jeremy~C Smith, Berk Hess, and Erik Lindahl.
\newblock {GROMACS}: High performance molecular simulations through multi-level parallelism from laptops to supercomputers.
\newblock {\em SoftwareX}, 1:19--25, 2015.

\bibitem{hess1997lincs}
Berk Hess, Henk Bekker, Herman~JC Berendsen, and Johannes~GEM Fraaije.
\newblock {LINCS}: A linear constraint solver for molecular simulations.
\newblock {\em Journal of Computational Chemistry}, 18(12):1463--1472, 1997.

\bibitem{huber1994local}
Thomas Huber, Andrew~E Torda, and Wilfred~F Van~Gunsteren.
\newblock Local elevation: a method for improving the searching properties of molecular dynamics simulation.
\newblock {\em Journal of computer-aided molecular design}, 8:695--708, 1994.

\bibitem{laio2002escaping}
Alessandro Laio and Michele Parrinello.
\newblock Escaping free-energy minima.
\newblock {\em Proceedings of the National Academy of Sciences}, 99(20):12562--12566, 2002.

\bibitem{vanden2010transition}
Eric Vanden-Eijnden et~al.
\newblock Transition-path theory and path-finding algorithms for the study of rare events.
\newblock {\em Annual review of physical chemistry}, 61:391--420, 2010.

\bibitem{roux2021string}
Beno{\^\i}t Roux.
\newblock String method with swarms-of-trajectories, mean drifts, lag time, and committor.
\newblock {\em The Journal of Physical Chemistry A}, 125(34):7558--7571, 2021.

\bibitem{palacio2021free}
Karen Palacio-Rodriguez and Fabio Pietrucci.
\newblock Free energy landscapes, diffusion coefficients and kinetic rates from transition paths.
\newblock {\em arXiv preprint arXiv:2106.05415}, 2021.

\bibitem{shmilovich2023girsanov}
Kirill Shmilovich and Andrew~L Ferguson.
\newblock Girsanov reweighting enhanced sampling technique (grest): On-the-fly data-driven discovery of and enhanced sampling in slow collective variables.
\newblock {\em The Journal of Physical Chemistry A}, 127(15):3497--3517, 2023.

\bibitem{jung2023machine}
Hendrik Jung, Roberto Covino, A~Arjun, Christian Leitold, Christoph Dellago, Peter~G Bolhuis, and Gerhard Hummer.
\newblock Machine-guided path sampling to discover mechanisms of molecular self-organization.
\newblock {\em Nature Computational Science}, pages 1--12, 2023.

\bibitem{mori1965transport}
Hazime Mori.
\newblock Transport, collective motion, and brownian motion.
\newblock {\em Progress of theoretical physics}, 33(3):423--455, 1965.

\bibitem{zwanzig1961memory}
Robert Zwanzig.
\newblock Memory effects in irreversible thermodynamics.
\newblock {\em Physical Review}, 124(4):983, 1961.

\bibitem{zwanzig2001nonequilibrium}
Robert Zwanzig.
\newblock {\em Nonequilibrium statistical mechanics}.
\newblock Oxford university press, 2001.

\bibitem{dominic2023memory}
Anthony~J Dominic~III, Siqin Cao, Andr{\'e}s Montoya-Castillo, and Xuhui Huang.
\newblock Memory unlocks the future of biomolecular dynamics: Transformative tools to uncover physical insights accurately and efficiently.
\newblock {\em Journal of the American Chemical Society}, 145(18):9916--9927, 2023.

\bibitem{vroylandt2022likelihood}
Hadrien Vroylandt, Ludovic Gouden{\`e}ge, Pierre Monmarch{\'e}, Fabio Pietrucci, and Benjamin Rotenberg.
\newblock Likelihood-based non-{M}arkovian models from molecular dynamics.
\newblock {\em PNAS}, 119(13):e2117586119, 2022.

\bibitem{ayaz2021non}
Cihan Ayaz, Lucas Tepper, Florian~N Br{\"u}nig, Julian Kappler, Jan~O Daldrop, and Roland~R Netz.
\newblock Non-markovian modeling of protein folding.
\newblock {\em Proceedings of the National Academy of Sciences}, 118(31):e2023856118, 2021.

\bibitem{wigner1938transition}
Eugene Wigner.
\newblock The transition state method.
\newblock {\em Transactions of the Faraday Society}, 34:29--41, 1938.

\bibitem{bao2017variational}
Junwei~Lucas Bao and Donald~G Truhlar.
\newblock Variational transition state theory: theoretical framework and recent developments.
\newblock {\em Chemical Society Reviews}, 46(24):7548--7596, 2017.

\bibitem{chandler1978statistical}
David Chandler.
\newblock Statistical mechanics of isomerization dynamics in liquids and the transition state approximation.
\newblock {\em The Journal of Chemical Physics}, 68(6):2959--2970, 1978.

\bibitem{dellago1999calculation}
Christoph Dellago, Peter~G Bolhuis, and David Chandler.
\newblock On the calculation of reaction rate constants in the transition path ensemble.
\newblock {\em The Journal of Chemical Physics}, 110(14):6617--6625, 1999.

\bibitem{van2003novel}
Titus~S Van~Erp, Daniele Moroni, and Peter~G Bolhuis.
\newblock A novel path sampling method for the calculation of rate constants.
\newblock {\em The Journal of chemical physics}, 118(17):7762--7774, 2003.

\bibitem{allen2006forward}
Rosalind~J Allen, Daan Frenkel, and Pieter~Rein Ten~Wolde.
\newblock Forward flux sampling-type schemes for simulating rare events: Efficiency analysis.
\newblock {\em The Journal of chemical physics}, 124(19), 2006.

\end{thebibliography}


\begin{thebibliography}{10}

\bibitem{risken1996fokker}
Hannes Risken and Hannes Risken.
\newblock {\em Fokker-Planck equation}.
\newblock Springer, 1996.

\bibitem{Izaguirre2010}
J.~A~. Izaguirre, C.~R. Schweet, and V.S. Pande.
\newblock Multiscale dynamics of macromolecules using normal mode {L}angevin.
\newblock {\em Pacific Symposium on Biocomputing}, {15}:240--251, 2010.

\bibitem{Hanggi1990}
Peter H\"anggi, Peter Talkner, and Michal Borkovec.
\newblock Reaction-rate theory: fifty years after kramers.
\newblock {\em Rev. Mod. Phys.}, 62:251–341, 1990.

\bibitem{Peters2017}
Baron Peters.
\newblock {\em Reaction Rate Theory and Rare Events}.
\newblock Elsevier, 1st edition, 2017.

\bibitem{pelzer1932speed}
H~Pelzer and E~Wigner.
\newblock The speed constansts of the exchange reactions.
\newblock {\em Z. Phys. Chem. B}, 15:445--552, 1932.

\bibitem{vineyard1957frequency}
George~H Vineyard.
\newblock Frequency factors and isotope effects in solid state rate processes.
\newblock {\em Journal of Physics and Chemistry of Solids}, 3(1-2):121--127,
  1957.

\bibitem{tajkhorshid1999role}
Emadeddin Tajkhorshid, B{\'e}la Paizs, and Sandor Suhai.
\newblock Role of isomerization barriers in the p{K}a control of the retinal
  {S}chiff base: a density functional study.
\newblock {\em The Journal of Physical Chemistry B}, 103(21):4518--4527, 1999.

\bibitem{hayashi2002structural}
Shigehiko Hayashi, Emad Tajkhorshid, and Klaus Schulten.
\newblock Structural changes during the formation of early intermediates in the
  bacteriorhodopsin photocycle.
\newblock {\em Biophysical Journal}, 83(3):1281--1297, 2002.

\bibitem{tajkhorshid2000molecular}
Emadeddin Tajkhorshid, J{\'e}r{\^o}me Baudry, Klaus Schulten, and Sandor Suhai.
\newblock Molecular dynamics study of the nature and origin of retinal's
  twisted structure in bacteriorhodopsin.
\newblock {\em Biophysical Journal}, 78(2):683--693, 2000.

\bibitem{malmerberg2011time}
Erik Malmerberg, Ziad Omran, Jochen~S Hub, Xuewen Li, Gergely Katona, Sebastian
  Westenhoff, Linda~C Johansson, Magnus Andersson, Marco Cammarata, Michael
  Wulff, et~al.
\newblock Time-resolved {WAXS} reveals accelerated conformational changes in
  iodoretinal-substituted proteorhodopsin.
\newblock {\em Biophysical Journal}, 101(6):1345--1353, 2011.

\bibitem{lindorff2010improved}
Kresten Lindorff-Larsen, Stefano Piana, Kim Palmo, Paul Maragakis, John~L
  Klepeis, Ron~O Dror, and David~E Shaw.
\newblock Improved side-chain torsion potentials for the {Amber ff99SB} protein
  force field.
\newblock {\em Proteins: Structure, Function, and Bioinformatics},
  78(8):1950--1958, 2010.

\bibitem{volkov2017structural}
Oleksandr Volkov, Kirill Kovalev, Vitaly Polovinkin, Valentin Borshchevskiy,
  Christian Bamann, Roman Astashkin, Egor Marin, Alexander Popov, Taras
  Balandin, Dieter Willbold, et~al.
\newblock Structural insights into ion conduction by channelrhodopsin 2.
\newblock {\em Science}, 358(6366), 2017.

\bibitem{van2005gromacs}
David Van Der~Spoel, Erik Lindahl, Berk Hess, Gerrit Groenhof, Alan~E Mark, and
  Herman~JC Berendsen.
\newblock {GROMACS}: fast, flexible, and free.
\newblock {\em Journal of Computational Chemistry}, 26(16):1701--1718, 2005.

\bibitem{abraham2015gromacs}
Mark~James Abraham, Teemu Murtola, Roland Schulz, Szil{\'a}rd P{\'a}ll,
  Jeremy~C Smith, Berk Hess, and Erik Lindahl.
\newblock {GROMACS}: High performance molecular simulations through multi-level
  parallelism from laptops to supercomputers.
\newblock {\em SoftwareX}, 1:19--25, 2015.

\bibitem{wiki:Dihedral_angle}
Wikipedia.
\newblock {Dihedral angle} --- {W}ikipedia{,} {The Free Encyclopedia}.
\newblock
  \url{http://en.wikipedia.org/w/index.php?title=Dihedral\%20angle&oldid=1155191517},
  2023.
\newblock [Online; accessed 08-June-2023].

\bibitem{github:plumed2}
The~{PLUMED} consortium.
\newblock Plumed2.
\newblock \url{https://github.com/plumed/plumed2}, 2019.

\bibitem{wiki:Atan2}
Wikipedia.
\newblock {Atan2} --- {W}ikipedia{,} {The Free Encyclopedia}.
\newblock
  \url{http://en.wikipedia.org/w/index.php?title=Atan2&oldid=1156958741}, 2023.
\newblock [Online; accessed 08-June-2023].

\bibitem{bonomi2009plumed}
Massimiliano Bonomi, Davide Branduardi, Giovanni Bussi, Carlo Camilloni, Davide
  Provasi, Paolo Raiteri, Davide Donadio, Fabrizio Marinelli, Fabio Pietrucci,
  Ricardo~A Broglia, et~al.
\newblock {PLUMED}: A portable plugin for free-energy calculations with
  molecular dynamics.
\newblock {\em Computer Physics Communications}, 180(10):1961--1972, 2009.

\bibitem{tribello2014plumed}
Gareth~A Tribello, Massimiliano Bonomi, Davide Branduardi, Carlo Camilloni, and
  Giovanni Bussi.
\newblock {PLUMED} 2: New feathers for an old bird.
\newblock {\em Computer Physics Communications}, 185(2):604--613, 2014.

\bibitem{plumed2019promoting}
{The PLUMED consortium}.
\newblock Promoting transparency and reproducibility in enhanced molecular
  simulations.
\newblock {\em Nature Methods}, 16(8):670--673, 2019.

\bibitem{barducci2008well}
Alessandro Barducci, Giovanni Bussi, and Michele Parrinello.
\newblock Well-tempered metadynamics: a smoothly converging and tunable
  free-energy method.
\newblock {\em Physical Review Letters}, 100(2):020603, 2008.

\bibitem{branduardi2012metadynamics}
Davide Branduardi, Giovanni Bussi, and Michele Parrinello.
\newblock Metadynamics with adaptive {G}aussians.
\newblock {\em Journal of Chemical Theory and Computation}, 8(7):2247--2254,
  2012.

\bibitem{bonomi2009reconstructing}
Massimiliano Bonomi, Alessandro Barducci, and Michele Parrinello.
\newblock Reconstructing the equilibrium {B}oltzmann distribution from
  well-tempered metadynamics.
\newblock {\em Journal of Computational Chemistry}, 30(11):1615--1621, 2009.

\bibitem{tiwary2015time}
Pratyush Tiwary and Michele Parrinello.
\newblock A time-independent free energy estimator for metadynamics.
\newblock {\em The Journal of Physical Chemistry B}, 119(3):736--742, 2015.

\bibitem{bussi2020using}
Giovanni Bussi and Alessandro Laio.
\newblock Using metadynamics to explore complex free-energy landscapes.
\newblock {\em Nature Reviews Physics}, 2(4):200--212, 2020.

\bibitem{henin2022enhanced}
J{\'e}r{\^o}me H{\'e}nin, Tony Leli{\`e}vre, Michael~R Shirts, Omar Valsson,
  and Lucie Delemotte.
\newblock Enhanced sampling methods for molecular dynamics simulations.
\newblock {\em arXiv preprint arXiv:2202.04164}, 2022.

\bibitem{bussi2019analyzing}
Giovanni Bussi and Gareth~A Tribello.
\newblock Analyzing and biasing simulations with {PLUMED}.
\newblock In {\em Biomolecular Simulations}, pages 529--578. Springer, 2019.

\bibitem{tan2012theory}
Zhiqiang Tan, Emilio Gallicchio, Mauro Lapelosa, and Ronald~M Levy.
\newblock Theory of binless multi-state free energy estimation with
  applications to protein-ligand binding.
\newblock {\em The Journal of Chemical Physics}, 136(14):04B608, 2012.

\bibitem{hummer2005position}
Gerhard Hummer.
\newblock Position-dependent diffusion coefficients and free energies from
  {B}ayesian analysis of equilibrium and replica molecular dynamics
  simulations.
\newblock {\em New Journal of Physics}, 7(1):34, 2005.

\bibitem{efron1982jackknife}
Bradley Efron.
\newblock {\em The jackknife, the bootstrap and other resampling plans}.
\newblock SIAM, 1982.

\bibitem{gatz1995standard}
Donald~F Gatz and Luther Smith.
\newblock The standard error of a weighted mean concentration—-{I}.
  {B}ootstrapping vs other methods.
\newblock {\em Atmospheric Environment}, 29(11):1185--1193, 1995.

\bibitem{hub2010g_wham}
Jochen~S Hub, Bert~L De~Groot, and David Van Der~Spoel.
\newblock g\_wham a free weighted histogram analysis implementation including
  robust error and autocorrelation estimates.
\newblock {\em Journal of Chemical Theory and Computation}, 6(12):3713--3720,
  2010.

\bibitem{daldrop2018butane}
Jan~O Daldrop, Julian Kappler, Florian~N Br{\"u}nig, and Roland~R Netz.
\newblock Butane dihedral angle dynamics in water is dominated by internal
  friction.
\newblock {\em Proceedings of the National Academy of Sciences},
  115(20):5169--5174, 2018.

\bibitem{mpmath}
Fredrik Johansson.
\newblock mpmath.
\newblock \url{https://github.com/fredrik-johansson/mpmath}.

\bibitem{FLINT}
Fredrik Johansson.
\newblock python-flint.
\newblock \url{https://github.com/fredrik-johansson/python-flint}.

\bibitem{salvalaglio2014assessing}
Matteo Salvalaglio, Pratyush Tiwary, and Michele Parrinello.
\newblock Assessing the reliability of the dynamics reconstructed from
  metadynamics.
\newblock {\em Journal of Chemical Theory and Computation}, 10(4):1420--1425,
  2014.

\bibitem{leines2012path}
Grisell~D{\'\i}az Leines and Bernd Ensing.
\newblock Path finding on high-dimensional free energy landscapes.
\newblock {\em Physical Review Letters}, 109(2):020601, 2012.

\bibitem{perez2019adaptive}
Alberto P{\'e}rez~de Alba~Ort{\'\i}z, Jocelyne Vreede, and Bernd Ensing.
\newblock The adaptive path collective variable: a versatile biasing approach
  to compute the average transition path and free energy of molecular
  transitions.
\newblock In {\em Biomolecular Simulations}, pages 255--290. Springer, 2019.

\bibitem{ortiz2021simultaneous}
Alberto P{\'e}rez de~Alba Ort{\'\i}z and Bernd Ensing.
\newblock Simultaneous sampling of multiple transition channels using adaptive
  paths of collective variables.
\newblock {\em arXiv preprint arXiv:2112.04061}, 2021.

\bibitem{byrd1995limited}
Richard~H Byrd, Peihuang Lu, Jorge Nocedal, and Ciyou Zhu.
\newblock A limited memory algorithm for bound constrained optimization.
\newblock {\em SIAM Journal on scientific computing}, 16(5):1190--1208, 1995.

\bibitem{ghysbrecht2023thermal}
Simon Ghysbrecht and Bettina~G Keller.
\newblock Thermal isomerization rates in retinal analogues using ab-initio
  molecular dynamics.
\newblock {\em Journal of Computational Chemistry}, 45(16):1390--1403, 2024.

\bibitem{ochterski1999vibrational}
Joseph~W Ochterski.
\newblock Vibrational analysis in {G}aussian.
\newblock {\em Gaussian Inc}, 1999.

\end{thebibliography}

\makeatletter\@input{xx.tex}\makeatother
\end{document}


\preprint{AIP/123-QED}

\title{Supplementary Information: Accuracy of reaction coordinate based rate theories for modelling chemical reactions: insights from the thermal isomerization in retinal}%
\author{Simon Ghysbrecht}
\author{Luca Donati}
\author{Bettina G.~Keller}%

\date{\today}

\maketitle

\section{Extensions to Theory}
\label{sec:SI_theory}
\subsection{Effective dynamics}
\label{sec:SI_effective_dynamics}
\subsubsection{One-dimensional reaction coordinates}
The effective dynamics of the system along a one-dimensional reaction coordinate $q$ can be modelled by underdamped Langevin dynamics \cite{risken1996fokker}
\begin{equation}
\begin{cases}
\dot q_t  & =  \frac{1}{\mu_q}p_t \\
\dot p_t  & =  - \frac{d }{d q} F(q) - \xi p_t + \sqrt{2 R T \xi \mu_q} \, \eta_t \, ,
\end{cases}
\label{eq:langevin}
\end{equation}
where $q_t = q(x_t)$ denotes the position of the system along the reaction coordinate at time $t$, 
$p_t$ is the conjugate momentum, 
$\mu_q$ is an effective mass, $\xi$ is the effective friction coefficient or collision frequency (units: [$\mathrm{time^{-1}}$]), and $F(q)$ is the free energy profile defined in eq.~\ref{main-eq:free_energy}.
As in the main part of the manuscript, the free energies has units of J/mol, and correspondingly the thermal energy is also formulated as a molar quantity $RT$, where $T$ us the temperature and $R$ is the ideal gas constant. 
The last term in eq.~\ref{eq:langevin} is the random force, where $\eta_t$ is a Gaussian white noise with $\langle \eta_t \rangle = 0$ and $\langle \eta_0, \, \eta_t \rangle = \delta_t$.
Eq.~\ref{eq:langevin} samples the Boltzmann distribution
\begin{eqnarray}
    \pi(q,p) &=&  Z_{\mathrm{conf}}^{-1}\exp\left(-\frac{F(q)}{RT}\right)  
                  \cdot \sqrt{\frac{1}{2\pi \mu_q RT }} \exp\left( -\frac{1}{2 \mu_q RT}p^2\right) \, . 
\end{eqnarray}
The first factor is the configurational Boltzmann distribution 
\begin{eqnarray}
    \pi(q) &=& Z_{\mathrm{conf}}^{-1}\exp\left(-\frac{F(q)}{RT}\right) 
\label{eq:config_Boltzmann_density}    
\end{eqnarray}
where $Z_\mathrm{conf}$ the configurational partition function, which normalizes the configurational Boltzmann distribution.

Let us now consider the Langevin equation defined in eq.~\ref{eq:langevin} and assume to have a trajectory realised with a very fine time discretization in $\Delta t$ timesteps.
If we counted the number of collisions between the molecular system and the solvent molecules, whose action is represented by the friction term and the noise term, we would observe few collisions in the time unit $\Delta t$.
Imagine now to enlarge the time unit $\Delta t$ by a unitless factor $g>1$, we would observe more collisions and the time-averaged acceleration over the timestep $g\cdot\Delta t$ would be zero.
In other words, by increasing the number of collisions in the unit time, the velocity reaches a steady-state.
Then, by coarse-graining the time, the term $\dot{p}_t$ on the left-hand side of the Langevin equation can be deleted.
Instead of enlarging the time unit, to increase the number of observed collisions in the unit time, we can act on the parameter $\xi$, i.e. the friction coefficient.
Increasing $\xi\rightarrow g\cdot \xi$ is in fact equivalent to increasing the number of collisions in the unit time $\Delta t$. 
This allows us, in a completely equivalent manner, to delete the term on the left-hand side of eq.~\ref{eq:langevin} and write the so-called Langevin equation for the high friction regime:
\begin{eqnarray}
\dot{q}_t & =&  - \frac{1}{\mu_q\xi}\frac{d }{d q} F(q_t)  + \sqrt{\frac{2R T}{ \xi \mu_q}} \, \eta_t \, .
\label{eq:overdamped_Langevin1}
\end{eqnarray}
When modeling rare events and transitions across large free energy barriers, the constant friction $\xi$ is often replaced by a position dependent friction coefficient $\xi(q)$.
Eq.~\ref{eq:overdamped_Langevin1} can then be written as 
\begin{eqnarray}
\dot{q}_t = -\frac{D(q_t)}{RT}  
\frac{d }{d q} F(q_t)
+
\frac{d }{d q} D(q_t) 
+
\sqrt{2 D(q_t)} \, \eta_t 
\label{eq:overdamped_Langevin2}
\end{eqnarray}
where we introduced the position dependent diffusion profile $D(q)$ which is defined via the Einstein relation
\begin{equation}
\label{eq:einstein_relation}
D(q) = \frac{RT}{\mu_q\xi(q)} \, .
\end{equation}
Eq.~\ref{eq:overdamped_Langevin2} can be derived by applying Ito's formula to a higher-dimensional Langevin equation with constant diffusion.
Both eq.~\ref{eq:overdamped_Langevin1} and eq.~\ref{eq:overdamped_Langevin2} sample the same configurational equilibrium density
(eq.~\ref{eq:config_Boltzmann_density}).

Numerical simulations of the eq.~\ref{eq:langevin} can be realized using the ISP algorithm\cite{Izaguirre2010}
\begin{align}
\begin{cases}
\begin{aligned}
v_{k+1}  =
  \exp &\left( - \xi \, \Delta t \right) \, v_k  \\
  &- \bigg[ 1 - \exp \left( - \xi \, \Delta t\right) \bigg] \, \frac{\nabla F(q_k)}{\xi m}  
  \\
  &+ \, \sqrt{\frac{R T}{m} \, \bigg[ 1 - \exp \left( - 2 \xi \, \Delta t\right) \bigg]} \, \eta_k 
\end{aligned}\cr
q_{k+1}  =  q_k + v_{k+1} \Delta t
\end{cases}
\label{eq:ISP}
\end{align}
where $q_k$ and $v_k$ denote respectively the position and the velocity of the particle at time $t_k$, $\Delta t = t_{k+1} - t_k= 0.001 \, \mathrm{ps}^{-1}$ is the integrator time step, and $\eta_k$ are independent and uncorrelated random numbers drawn from a standard Gaussian distribution.

\subsubsection{Multidimensional collective variables}
The effective dynamics in this $m$-dimensional collective variable space can be modelled as overdamped Langevin dynamics with position-dependent diffusion.
\begin{align}\dot{\mathbf{q}}_t = -\frac{1}{RT} D(\mathbf{q}_t) \nabla F(\mathbf{q}_t)
+
\nabla \cdot  D(\mathbf{q}_t)
+
\sqrt{2 D(\mathbf{q}_t)}  \boldsymbol{\eta}_t \, ,
\label{eq:overdamped_Langevin3}
\end{align}
where $\nabla = (\partial / \partial q_1, \dots \partial / \partial q_m)^{\top}$ is the gradient with respect to $\mathbf{q}$, 
$\boldsymbol{\eta}_t$ is a $m$-dimensional  Gaussian white noise with $\langle \boldsymbol{\eta}_t \rangle = \boldsymbol{0}$ and $\langle \boldsymbol{\eta}_0, \, \boldsymbol{\eta}_t \rangle = \delta_t$.
$D(\mathbf{q})$ is a $m\times m$ diagonal matrix whose $i$th element represents the diffusion profile along the $i$th collective variable.

\subsubsection{White noise vs. Wiener process}
\label{sec:SI_eta}
Eq.~\ref{eq:langevin} contains $\eta_t$ as a symbol for a Gaussian white noise. 
The use of a white noise process is problematic, because it does not have a clear physical interpretation. 
Formally, one can define $\eta$ as the time derivative of a Wiener process $W_t$, i.e.~$\eta_t = \dot{W_t}$.
Unfortunately, the Wiener process is not differentiable and the derivative is only defined in a finite difference sense $ \dot{W_t} \approx (W_{t+h} - W_t)/h$, for small time increments $h$.
A mathematically more rigorous way to formulate eq.~\ref{eq:langevin} is to use increments of the Wiener process rather than time derivatives:
\begin{equation}
\begin{cases}
dq_t  & =  \frac{1}{\mu_p} p_t dt \\
d p_t  & =  - \frac{d }{d q} F(q) dt - \xi p_t dt + \sqrt{2 R T \xi \mu_p} \, dW_t \, .
\end{cases}
\end{equation}
The same discussion applies to eqs.~\ref{eq:overdamped_Langevin1}, \ref{eq:overdamped_Langevin2} and \ref{eq:overdamped_Langevin3}.

\subsubsection{Fokker-Planck equations}
\label{sec:SI_FP}
Associated to each of the stochastic equations of motion (eqs.~\ref{eq:langevin}, \ref{eq:overdamped_Langevin1}, \ref{eq:overdamped_Langevin2} and \ref{eq:overdamped_Langevin3})  there exists a Fokker-Planck equation. 
The Fokker-Planck equation is a deterministic partial differential equation which describes how the probability density $\rho(p,q,t)$, for eq.~\ref{eq:langevin}, or $\rho(q,t)$, for eqs.~\ref{eq:overdamped_Langevin1} and \ref{eq:overdamped_Langevin2}, or $\rho(\mathbf{q},t)$ for eq.~\ref{eq:overdamped_Langevin3} evolves with time:
\begin{eqnarray}
    \frac{\partial}{\partial t} \rho(t) &=& \mathcal{Q} \rho(t) \, .
\label{eq:FP}
\end{eqnarray}
$\mathcal{Q}$ is Fokker-Planck operator. 
The Fokker-Planck equation for underdamped Langevin dynamics (eq.~\ref{eq:langevin}) is called Klein-Kramers equation and $\mathcal{Q} $ is given as
\begin{align}
    \mathcal{Q} =  - \frac{p}{m}\frac{\partial}{\partial q} 
+ \xi m\frac{\partial}{\partial p} \left(\frac{p}{m} + R T \frac{\partial}{\partial p} \right) 
+ \frac{\partial F(q)}{\partial q}  \frac{\partial}{\partial p}
\label{eq:KleinKramers}
\end{align}

The Fokker-Planck equation for overdamped Langevin dynamics (eq.~\ref{eq:overdamped_Langevin1}) is called Smoluchowski equation. 
$\mathcal{Q}$ given as
\begin{align}
\mathcal{Q} = D \frac{\partial^2}{\partial q^2}  + \xi^{-1}m^{-1}\frac{\partial}{\partial q}\frac{\partial F(q)}{\partial q}
\label{eq:Smoluchowski} 
\end{align}

The Fokker-Planck operator for overdamped Langevin dynamics with position-dependent diffusion (eq.~\ref{eq:overdamped_Langevin1}) is
\begin{align}
\mathcal{Q}
&=
\frac{\partial^2}{\partial q^2} D(q)
+
\frac{\partial }{\partial q}\left( \beta D(q)  \frac{\partial F(q)}{\partial q} - \frac{\partial D(q)}{\partial q} \right)
 \label{eq:Smoluchowski2_2}
\\
&=
\frac{\partial}{\partial q} D(q)e^{-\beta F(q)} \frac{\partial}{\partial q} e^{\beta F(q)}
\, .
\label{eq:Smoluchowski2}
\end{align}

The Fokker-Planck equation for overdamped Langevin dynamics with position-dependent diffusion in a multidimensional space (eq.~\ref{eq:overdamped_Langevin3}) is given as
 \begin{align}
\mathcal{Q}
&=
\nabla \cdot D(\mathbf{q})e^{-\beta F(\mathbf{q})} \nabla e^{\beta F(\mathbf{q})}
\label{eq:Smoluchowski3}
\end{align}
(if $D(\mathbf{q})$ is a diagonal matrix).
We used the following convention to denote differential operators: derivatives written as operators ($\frac{\partial}{\partial q}$,$\frac{\partial}{\partial p}$,$\frac{\partial^2}{\partial q^2}$ and $\nabla$) should be applied to anything that follows behind it, while derivatives written as functions ($\frac{\partial F(q)}{\partial q}$ and $\frac{\partial D(q)}{\partial q}$) should be considered stand-alone functions, i.e.~the derivative only applies to the function ($F(q)$ or $D(q)$ respectively) directly and not what comes after it.

\subsection{Simple transition state theory}
\label{sec:SI_TST}
In simple TST \cite{Hanggi1990, Peters2017} one defines the transition state $TS$ as a point $q_{TS}$ along the reaction coordinate that separates reactant state $A$ ($q<q_{TS}$) and product state $B$ ($q>q_{TS}$). 
In the full dimensional configurational space $\Gamma_x$, this point corresponds to an isosurface on which the value of the reaction coordinate is constant. 
Using the Dirac delta function, the surface is defined by $\delta\left(q(\mathbf{x})-q_{TS}\right)$ and separates the reactant configurations from the product configurations.
The TST rate constant is derived from the one-directional flux across the dividing surface assuming the reactant and transition state are in equilibrium\cite{pelzer1932speed,Peters2017}:
\begin{equation}
\label{eq:generalizedTST}
    k_{AB} = \kappa \cdot \frac{1}{2}\left<\left|\dot{q}\right|\right>_{TS} \cdot l_q^{-1} \exp\left(-\frac{F(q_{TS})-F_A}{RT} \right) \, .
\end{equation} 
The variable 
\begin{eqnarray}
    F_A = -RT \ln\left(l_q^{-1}\int_A \mathrm{d}q \exp\left(\frac{-F(q)}{RT}\right)\right)
\label{eq:FA}    
\end{eqnarray}
denotes the free energy of the entire reactant state, not just its minimum.
The factor $l_{q}^{-1}\exp\left(-\frac{F(q_{TS})-F_A}{RT}\right)$ in eq.~\ref{eq:generalizedTST} is the relative probability density of finding the system at the transition state, where $l_q$ is the unit of length along coordinate $q$.
The factor $\left<\left|\dot{q}\right|\right>_{TS}$ is the averaged absolute velocity along $q$ at the transition state $TS$. 
The factor $1/2$ accounts for the fact that only half of all systems in an ensemble move in the forward direction. 
$\kappa$ is again the transmission factor to correct for the fact that in reality not all systems that cross the dividing surface proceed to state $B$, but instead revert to $A$ (recrossing).
Since transition state theory assumes the transition state to be in thermal equilibrium with the reactant state, the absolute velocity $\left|\dot{q}\right|$ can be averaged using the Maxwell-Boltzmann distribution, giving $\left<\left|\dot{q}\right|\right>_{TS} = \sqrt{\frac{2RT}{\pi\mu_q}}$, where $\mu_q$ is the effective mass.
Furthermore, the reactant state $A$ can be approximated by a harmonic potential around the reactant state minimum $q_A$,
\begin{equation}
\label{eq:F_A_harmonic}
    F(q) = F(q_A) + \frac{1}{2}\mu_q \omega_A^2 (q-q_A)^2 ~\text{ if }~ q\approx q_A \, ,
\end{equation}
where $\omega_A$ is the angular frequency associated to harmonic approximation, $\mu_q$ the reduced mass, and $F(q_A)$ is the free energy at the minimum of the reactant state.
Carrying out the integral in eq.~\ref{eq:FA} for eq.~\ref{eq:F_A_harmonic} and inserting the and the result for $\left<\left|\dot{q}\right|\right>_{TS}$ into eq.~\ref{eq:generalizedTST} yields
\begin{equation}
    k_{AB} = \kappa \cdot \frac{\omega_A}{2\pi}\exp\left(-\frac{F^\ddagger_{AB}}{RT} \right) \, .
    \label{eq:simpleTST}
\end{equation}
In Ref.~\onlinecite{Peters2017}, eq.~\ref{eq:generalizedTST} is called the generalized TST approach, and eq.~\ref{eq:simpleTST} is called one-dimensional Vineyard TST\cite{vineyard1957frequency}.
In this work, we follow Ref.~\citenum{Hanggi1990} where the result in eq.~\ref{eq:simpleTST} is called simple transition state theory.

\subsection{Kramers' rate theory: from moderate to high friction}
\label{sec:SI_Kramers_moderate_to_high_friction}

In the main part of the article, eq.~\ref{main-eq:KramersHigh} is derived from eq.~\ref{main-eq:KramersModerate} as follows:
\begin{align}
    \frac{\xi}{\omega_{TS}} \left(\sqrt{\frac{1}{4} + \frac{\omega_{TS}^2}{\xi^2}} - \frac{1}{2} \right)
    &= 
    \frac{\xi}{\omega_{TS}} \left(\sqrt{\frac{1}{4}} \sqrt{ 1+ 4\frac{\omega_{TS}^2}{\xi^2}} - \frac{1}{2} \right) \cr
    &\approx \frac{\xi}{\omega_{TS}} \left(\frac{1}{2} \left( 1+ \frac{1}{2} 4\frac{\omega_{TS}^2}{\xi^2} \right) - \frac{1}{2} \right) \cr
    &= \frac{\xi}{\omega_{TS}} \left( \frac{1}{2}+ \frac{\omega_{TS}^2}{\xi^2} - \frac{1}{2} 
    \right) \cr
    &= \frac{\omega_{TS}}{\xi} \, ,
\end{align}
where in the second line, we approximated the square-root by a power series
\begin{align}
    \sqrt{1 + a} &= 1 + \frac{1}{2}a - \frac{1}{8}a^2 + \frac{1}{16}a^3 - \frac{5}{128}a^4 + \dots \,\quad |a| \le 1
\end{align}
with $a = 4 \omega_{TS}^2/\xi^2$, and truncated after the second term.

\clearpage
\newpage


\section{Computational details}
\label{sec:SI_comp_det}
%
%

%
%
\subsection{Classical MD with atomistic force field}
\label{sec:methods_MD}

%
%
\subsubsection{Dynamics}
Retinal parameters for atomistic force field calculations were taken from DFT studies on the protonated Schiff base\cite{tajkhorshid1999role,hayashi2002structural,tajkhorshid2000molecular}, adapted to GROMACS format\cite{malmerberg2011time}, while the connecting amino acid was modelled using the AMBER99SB*-ILDN forcefield\cite{lindorff2010improved}. 
The starting structure was obtained by cutting out the lysine amino acid and retinal cofactor from a recent crystal structure\cite{volkov2017structural}, while the ends of the lysine were capped with methyl groups as shown in Fig.~\ref{main-fig:retinal_structures}.
All simulations are carried out at $300\,\mathrm{K}$ in vacuum and are done using stochastic dynamics with GROMACS\cite{van2005gromacs,abraham2015gromacs} version 2019.4 built in Langevin integrator with a $2\,\mathrm{fs}$ timestep and an inverse friction coefficient of $2\,\mathrm{ps}$, except when using path collective variables, where, when explicitly mentioned, lower time steps were used. 
Strong position restraints of $10000\,\mathrm{kJ}\,\mathrm{mol}^{-1}\mathrm{nm}^{-2}$ were put on all heavy atoms of the peptide chain as well as on the lysine chain carbon atoms (Fig.~\ref{main-fig:retinal_structures}), while the LINCS constraint algorithm was applied to all hydrogen bonds. 
Before all simulations, energy minimization and NVT equilibration were performed.
%

%
%
\subsubsection{Free energy and diffusion constant calculation along $\varphi$}
%
As initial reaction coordinate for the one-dimensional rate models, we choose the dihedral angle $\varphi$ constituted by the retinal chain atoms C$_{12}$-C$_{13}$=C$_{14}$-C$_{15}$.
For four atoms with indices $i$, $j$, $k$ and $l$, the vectors connecting the atoms are $\mathbf{r}_{ji}=\mathbf{r}_j-\mathbf{r}_i$, $\mathbf{r}_{kj}=\mathbf{r}_k-\mathbf{r}_j$ and $\mathbf{r}_{lk}=\mathbf{r}_l-\mathbf{r}_k$.
The general dihedral angle $\phi$ is then defined \cite{wiki:Dihedral_angle,github:plumed2} by the angle between two planes, one
constituted by vectors $\mathbf{r}_{ji}$ and $\mathbf{r}_{kj}$ and the other constituted by vectors $\mathbf{r}_{kj}$ and $\mathbf{r}_{lk}$:
\begin{subequations}
\begin{eqnarray}
    \cos\varphi &=& \frac{\left(\mathbf{r}_{ji}\times\mathbf{r}_{kj}\right)\cdot\left(\mathbf{r}_{kj}\times\mathbf{r}_{lk}\right)}{\left|\mathbf{r}_{ji}\times\mathbf{r}_{kj}\right|\left|\mathbf{r}_{kj}\times\mathbf{r}_{lk}\right|} \\
    \sin\varphi &=& \frac{\left[\left(\mathbf{r}_{ji}\times\mathbf{r}_{kj}\right)\times\left(\mathbf{r}_{kj}\times\mathbf{r}_{lk}\right)\right]\cdot\mathbf{r}_{kj}}{\left|\mathbf{r}_{kj}\right|\left|\mathbf{r}_{ji}\times\mathbf{r}_{kj}\right|\left|\mathbf{r}_{kj}\times\mathbf{r}_{lk}\right|} \, .
\end{eqnarray}
\label{eq:dihedral_cos_sin}
\end{subequations}
The torsion angle can be obtained using the atan2 function\cite{wiki:Atan2}:
\begin{align}
\label{eq:dihedral_atan2}
    \varphi = \mathrm{atan2}
    \bigl(-&\left[\left(\mathbf{r}_{kj}\times\mathbf{r}_{ji}\right)\times\left(\mathbf{r}_{lk}\times\mathbf{r}_{kj}\right)\right]\cdot\mathbf{r}_{kj}  \, , \notag\\
    &\left|\mathbf{r}_{kj}\right|\left(\mathbf{r}_{ji}\times\mathbf{r}_{kj}\right)\cdot\left(\mathbf{r}_{kj}\times\mathbf{r}_{lk}\right)\bigr) \, .
\end{align}
This implies a certain convention with regards to the sign and phase of $\varphi$.
In general, $\varphi$ is zero for the case where the the dihedral corresponds to a cis/syn state, and $\pm\pi$ when the dihedral corresponds to a anti/trans state.
Increasing values of $\varphi$ correspond to a clockwise rotation of the plane constituted by vectors $\mathbf{r}_{kj}$ and $\mathbf{r}_{lk}$ with regards to the plane constituted by vectors $\mathbf{r}_{ji}$ and $\mathbf{r}_{kj}$ when looking along the $\mathbf{r}_{kj}$ vector, i.e.~similar to conventions in stereochemistry\cite{wiki:Dihedral_angle}.
For the case of retinal in Fig.~\ref{main-fig:retinal_structures}, the dihedral angle $\varphi$ is defined by matching indices $i$, $j$, $k$ and $l$ with atoms C$_{12}$, C$_{13}$, C$_{14}$ and C$_{15}$ respectively.
Metadynamics (MetaD) and umbrella sampling (US) were carried out by plugging PLUMED\cite{bonomi2009plumed,tribello2014plumed,plumed2019promoting} with the GROMACS software package\cite{van2005gromacs,abraham2015gromacs}.
Before production runs, the model system was energy minimized and NVT equilibrated over $400\,\mathrm{ps}$.
Subsequently, $2\,\mathrm{\mu s}$ of well-tempered metadynamics\cite{barducci2008well} were run biasing $\varphi$ at a pace of $1\,\mathrm{ps}$ using Gaussians with a height of $1.2\,\mathrm{kJ/mol}$, a standard deviation of $0.05\,\mathrm{radians}$ while the bias factor was 10. 
Unbiasing weights for the trajectory were calculated using the bias potential obtained at the end as described in Ref.~\citenum{branduardi2012metadynamics}. 
Free energy surfaces can then be calculated after building a weighted histogram from the trajectory starting at a simulation time where the bias can be considered converged. 
On account of the large simulation time, there is no significant change in the free energy profile depending on whether we build the histogram on the full trajectory or only after a certain time at which we consider the bias converged. 
There was also no considerable difference when calculating FES after reweighting with a time-dependent bias as in Refs.~\citenum{bonomi2009reconstructing} and \citenum{tiwary2015time}, and FES from reweighted trajectories were always close to the free energy estimated from the upside-down bias potential $F(\varphi) = -\frac{\gamma}{\gamma-1}V_b(\varphi)$, where $V_b(\varphi)$ is the biasing potential at the end of the well-tempered metadynamics simulation\cite{barducci2008well}, and $\gamma$ is the bias factor\cite{bussi2020using,henin2022enhanced}.
%

\begin{figure}[h]
\centering
\includegraphics[scale=1]{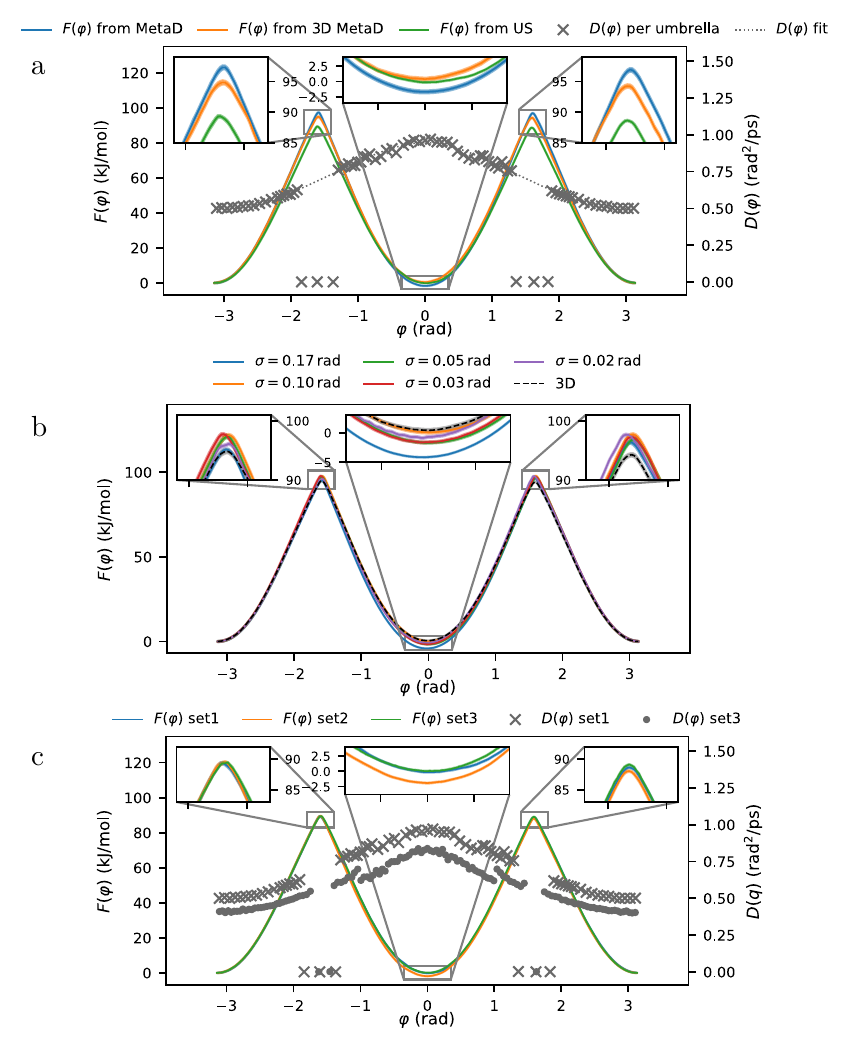}
\caption{\textbf{a:} Free energy surfaces $F(\varphi)$ and diffusion profiles $D(\varphi)$ estimated from MetaD and US biasing C$_{13}$=C$_{14}$ torsion angle $\varphi$ including standard errors.\\
\textbf{b:} Free energy profiles for metadynamics simulations biasing the C$_{13}$=C$_{14}$ dihedral angle $\varphi$ for different Gaussian standard deviations (in radians), as well as profile reweighted from 3D metadynamics.
One-dimensional metadynamics simulations (colored) were run for $2\,\mathrm{\mu s}$ with a deposition pace of $1\,\mathrm{ps}$ using Gaussians with a height of $1.2\,\mathrm{kJ/mol}$ and a biasing factor of 10.
3D metadynamics simulation (black, dashed) was run for $1\,\mathrm{\mu s}$ with a deposition pace of $1\,\mathrm{ps}$ using Gaussians with a height of $1.2\,\mathrm{kJ/mol}$ and a width of $0.07\,\mathrm{rad}$ in each dimension and a biasing factor of 12. \\
\textbf{c:} Free energy profiles for US simulations biasing the C$_{13}$=C$_{14}$ dihedral angle $\varphi$. The statistical uncertainty of the free energy profiles are shown as shaded areas, but they are so small, that they are hardly discernible.}
\label{fig:us_metad_together}
\centering
\end{figure}

%
Monitoring the evolution of the metadynamics simulations can be done by following the free energy difference $\Delta F$ between the trans and cis state as estimated from the upside-down bias potential as a function of simulation time as in Fig.~\ref{fig:metad_convergence}.a.
See SI section \ref{sec:US_vs_metad} below for more details.
It is apparent from the oscillating free energy differences that the biasing potentials are still undergoing changes with time.
Consequently, dynamics along the dihedral angle do not reach a point of being completely diffusive, which is a first indication of hidden motion not being included in the collective variable used here, i.e.~the dihedral angle $\varphi$.
To test the sensitivity of metadynamics to the width of the deposited Gaussians, additional sets of simulations were performed using the same simulation and metadynamics parameters as before but changing the standard deviation of the deposited Gaussians. 
The resulting free energy profiles can be seen in Fig.~\ref{fig:us_metad_together}.b.
The free energy surfaces appear to have a small dependency on the width of the Gaussians used, which can in part be explained by the biasing potentials still evolving due to hidden motion as explained above.
That being said, Gaussians of standard deviation $0.17\,\mathrm{radians}$ seem too wide for accurately reweighting the shapes of barrier peaks and reactant wells.
Error estimates for free energy profiles obtained from metadynamics reweighting can be determined using the block analysis technique\cite{bussi2019analyzing} on the reweighted trajectory. 
To check convergence of the free energy profile, one commonly plots the average error as a function of block size. 
Because data from an MD trajectory are generally correlated, the average error will be underestimated for small block sizes in which case the error analysis of the free energy profile will not represent an accurate evaluation of the quality of the free energy surface. 
When sufficiently large blocks are used, the average error will converge to a plateau value suggesting the data has decorrelated and indicating the error analysis can now be trusted. 
In cases where the average error does not converge even for very large block sizes, correlated effects should be considered too strong and the trajectory too short to truthfully capture them, and thus the accuracy of the computed free energy surface and its error analysis can be questioned. 
Block analysis was carried out using the example code on the PLUMED website\cite{bussi2019analyzing}. 
Average errors of the energy profile as a function of block size are shown for different metadynamics simulations in Fig.~\ref{fig:metad_convergence}.a. 
The average errors appear to be converging for large block sizes.
Notice that FES could still depend on the parameters chosen for the metadynamics simulations, and errors are only estimated within a certain parameter set.
%

%
Umbrella sampling was carried out by running 83 trajectories of $12\,\mathrm{ns}$ for a total of $996\,\mathrm{ns}$ of simulation time.
Each trajectory was restrained with a harmonic potential of spring constant $400\,\mathrm{kJ}.\mathrm{mol}^{-1}.\mathrm{rad}^{-2}$ at different values of $\varphi$:
\begin{itemize}
    \item 63 umbrellas were positioned at regular 0.1 radian intervals between -3.1 and 3.1 radians
    \item 10 umbrellas were positioned at regular 0.1 radian intervals between -1.95 and -1.05 radians
    \item 10 umbrellas were positioned at regular 0.1 radian intervals between 1.05 and 1.95 radians.
\end{itemize}
For each trajectory, a two step equilibration procedure was carried out before each production runs.
First, a $20\,\mathrm{ps}$ NVT equilibration was carried out at a lower spring constant of $100\,\mathrm{kJ}.\mathrm{mol}^{-1}.\mathrm{rad}^{-2}$ starting from an energy minimized structure.
Second, another $20\,\mathrm{ps}$ NVT equilibration was carried out at the same spring constant of the production runs, i.e.~at $400\,\mathrm{kJ}.\mathrm{mol}^{-1}.\mathrm{rad}^{-2}$.
In this way, the production runs start from configurations which can be considered equilibrated within their respective umbrella sampling restraints. 
From the umbrella sampling trajectories, binless WHAM\cite{tan2012theory,bussi2019analyzing} was used to reconstruct the free energy profile. 
For each trajectory, a value for the diffusion coefficient was calculated using Hummer's formulation of position dependent diffusion coefficients\cite{hummer2005position}:
\begin{equation}
    D(\varphi = \left< \varphi \right>) = \frac{\mathrm{var}(\varphi)}{\tau_\varphi}
    \label{eq:diffusion_coefficient}
\end{equation}
where $\tau_\varphi = \int_{0}^{\infty} \left< \delta \varphi(t)\delta \varphi (0) \right> \mathrm{d}t/\mathrm{var}(\varphi)$ with $\delta \varphi(t) = \varphi(t) - \left<\varphi\right>$. 
The diffusion coefficient as a continuous function of $\varphi$ was obtained using cubic spline interpolation on all resulting diffusion data points excluding data points near the transition state where Hummer's formula cannot be applied directly and the diffusion coefficient is underestimated.
Accordingly, all data points with values under $0.4\,\mathrm{rad^{2}/ps}$ were ignored for interpolation. 
The corresponding profiles can be found in Fig.~\ref{main-fig:FES_phi} as well as in Fig.~\ref{fig:us_metad_together}.c under the label \textit{set1}. 
Additional sets have been run and are also shown:
\begin{itemize}
    \item \textit{set2} has the same parameter setup as \textit{set1}.
    \item For \textit{set3},  125 trajectories of $12\,\mathrm{ns}$ were run with harmonic spring constant \\ $750\,\mathrm{kJ}\,\mathrm{mol}^{-1}\mathrm{rad}^{-2}$ positioned in $0.05\,\mathrm{rad}$ intervals between $-3.1$ and $3.1\,\mathrm{rad}$. 
\end{itemize}
Computation of the reweighted histograms was done applying kernel density estimation (KDE) with Gaussian kernels of bandwidth $0.01\,\mathrm{radians}$ for all metadynamics runs as well as for umbrella sampling sets \textit{set1} and \textit{set2}.
For umbrella sampling set \textit{set3}, it turned out to be challenging to find a good choice of bandwidth for KDE, and therefore conventional discrete histograms were utilized instead. 
Error estimates for the free energy profiles obtained from umbrella sampling can be computed using the bootstrapping method\cite{efron1982jackknife}. 
For each umbrella, the trajectory was split in 20 blocks of equal length. 
A `new' trajectory of the same length as the original is then constructed by taking combinations of these 20 blocks with the possibility of repetition. 
After doing this for all umbrellas, the free energy surface is recalculated using WHAM. 
This procedure is repeated 200 times, producing 200 free energy surfaces which allows calculation of standard deviations which can be shown to be good estimates of standard errors on the free energy surface\cite{gatz1995standard}. 
Notice the standard errors might be underestimated because of correlations between blocks within each trajectory\cite{hub2010g_wham}.
Free energy and diffusion profiles including error estimates for all umbrella sampling sets can be found in Fig.~\ref{fig:us_metad_together}.c.

%
%
\subsubsection{Rate calculations along dihedral reaction coordinate}
\label{section:rates}
Rates along the $\varphi$ reaction coordinate were calculated using the free energy profiles in Fig.~\ref{main-fig:FES_phi}, both for metadynamics ($\sigma=0.05\,\mathrm{rad}$ and umbrella sampling (\textit{set1}), see Table \ref{main-tab:rates}.
Diffusion coefficients were taken from the diffusion profile from umbrella sampling \textit{set1}.
Free energy barriers $F^\ddagger$ were measured directly from the free energy profile by subtracting the minimum free energy value at the reactant side of the isomerization under consideration from the maximum value at the corresponding peak. 
Notice we denote the peak at negative $\varphi$ as $TS$ and the peak at positive $\varphi$ as $TS'$, similar as in Fig.~\ref{main-fig:1Dmodel_rates}.e.
In this fashion, four energy barriers per free energy surface $F^\ddagger_{\mathrm{t}\rightarrow\mathrm{c},TS}$, $F^\ddagger_{\mathrm{t}\rightarrow\mathrm{c},TS'}$, $F^\ddagger_{\mathrm{c}\rightarrow\mathrm{t},TS}$ and $F^\ddagger_{\mathrm{c}\rightarrow\mathrm{t},TS'}$ are obtained. 
Masses in reduced dimensions for reactant states $\mu_\mathrm{trans}$ and $\mu_\mathrm{cis}$ were calculated by running unbiased $12\,\mathrm{ns}$ runs in the corresponding states, calculating the average kinetic energy in the reduced dimension (i.e.~the dihedral angle) and comparing to temperature using the equipartition theorem
\begin{equation}
\label{eq:reduced_mass}
    \mu_A = \frac{k_B T}{\left< v_\varphi^2 \right>_{A}}
\end{equation}
similar as in Ref.~\citenum{daldrop2018butane}.
In principle, applying the equipartition theorem here is an approximation, since it cannot be used for collective variables obtained from nonlinear transformations of Cartesian coordinates.
Since the free energy surface is nearly harmonic at the reactants states, however, we expect it to be a good approximation. 
The reactant state dihedral velocities $\omega_A$ (where $A$ denotes cis or trans) can then be calculating using 
\begin{equation}
\label{eq:omega_A}
    \omega_A = \sqrt{\frac{\kappa_A}{\mu_A}}
\end{equation}
where spring constant $\kappa_A$ is obtained by fitting the free energy surface to a harmonic potential $\frac{1}{2}\kappa_A(\varphi-\varphi_A)^2$ where $\varphi_A$ corresponds to the free energy minimum at the corresponding reactant state $A$. 
Fits for the trans and cis free energy wells show close agreement with harmonic potentials at the bottom, which validates the harmonic assumptions of the reactant and product states in the formulations for simple TST and Kramers' equations (eqs.~\ref{main-eq:simpleTST}, \ref{main-eq:KramersWeakLimit}, \ref{main-eq:KramersModerate} and \ref{main-eq:KramersHigh}).
Alternatively, one can calculate a period $T_A$ from the unbiased trajectories by choosing two cutoff values for $\varphi$ above and below its value for minimal free energy (e.g. above and below approximately zero radians for the cis state) and by counting transitions of the trajectory dihedral angle between these cutoffs as a function of time. 
Angular velocities calculated from this period $\omega_A = 2\pi/T_A$ gave similar results to the ones obtained from the harmonic fit in combination with the equipartition theorem above. 
%
%
Given the free energy barrier heights and the reactant state angular frequency, simple TST rates can be calculated directly for each barrier using eq.~\ref{main-eq:simpleTST}. 
Notice that calculating reaction constants for full processes requires taking into account transitions over both peaks:
\begin{subequations}
\begin{align}
     k_{\mathrm{trans}\rightarrow\mathrm{cis}} &= k_{\mathrm{t}\rightarrow\mathrm{c},TS} +  k_{\mathrm{t}\rightarrow\mathrm{c},TS'} \\
     k_{\mathrm{cis}\rightarrow\mathrm{trans}} &= k_{\mathrm{c}\rightarrow\mathrm{t},TS} +  k_{\mathrm{c}\rightarrow\mathrm{t},TS'} \, .
\end{align}
     \label{eq:sum_rates}
\end{subequations}
In order to calculate Kramers' rate in the moderate-to-high friction limit as in eq.~\ref{main-eq:KramersModerate} or in the high friction limit as in eq.~\ref{main-eq:KramersHigh}, the friction coefficient at the barrier top can be calculated directly from the diffusion profile using:
\begin{equation}
\label{eq:gamma_ddagger}
    \xi_{TS} = \frac{k_BT}{\mu_{TS} D_{TS}}
\end{equation} 
where $D_{TS}=D(\varphi_{TS})$ is the value of the diffusion coefficient at the barrier top taken from the spline interpolation and $\mu_{TS}$ has been approximated by averaging $\mu_\mathrm{cis}$ and $\mu_\mathrm{trans}$. 
The angular frequency at the barrier top $\omega_{TS}$ has been calculated in a similar way as at the reactant states using:
\begin{equation}
\label{eq:omega_ddagger}
    \omega_{TS} = \sqrt{\frac{\kappa_{TS}}{\mu_{TS}}} 
\end{equation} 
where $\kappa_{TS}$ was obtained using a parabolic fit to the free energy surface at the barrier top. 
An identical analysis can be done to obtain the friction coefficient $\xi_{TS'}$ at the other barrier $TS'$.
Again, total rates are obtained by summing rates for both barriers as in eqs.~\ref{eq:sum_rates}.
Calculating isomerization rates over a specific barrier using the Pontryagin equation (eq.~\ref{main-eq:Pontryagin}) was done by nested integration using the calculated free energy profile from MetaD or US as well as the position dependent diffusion from eq.~\ref{eq:diffusion_coefficient}. 
Here, the inner integral was carried out from the barrier peak on the other side of the reactant state. 
Again, rates over individual barriers were combined to describe full thermal isomerization rates using eqs.~\ref{eq:sum_rates}. 
Rates from grid-based models were calculated by discretizing the dihedral CV $\varphi$ in 500 cells of equal size and building the rate matrix according to eq.~\ref{main-eq:Qij_HAA_1}. 
For each cell $i$ with cell middle $\varphi_i$, the population $\pi_i = \pi(\varphi_i)$ was determined by using spline interpolation of the free energy surface as obtained from metadynamics or US, evaluating at $\varphi=\varphi_i$ and applying eq.~\ref{main-eq:eq_distribution}. 
In principle, populations $\pi_i$ need not be normalized since only ratios appear in eq.~\ref{main-eq:Qij_HAA_1}.
The values for the diffusion coefficient $D(\varphi_i)$ were similarly obtained by spline interpolation of the results from application of eq.~\ref{eq:diffusion_coefficient} to the US trajectories and evaluating at the cell middles. 
For the very high barriers we are dealing with, populations $\pi_i$ in cells near the barrier can get very small, and high precision numbers need to be used in the construction of the rate matrix. 
The mpmath\cite{mpmath} python package was used to administer arbitrary precision in building the rate matrix, and the FLINT\cite{FLINT} python package was used to solve for the mean first-passage times in eq.~\ref{main-eq:solve_Q}. 
A precision of 50 digits was used for these calculations.
The initial conditions are enforced by setting $\mathbf{1}[j]=0$ and adapting the rate matrix $\mathbf{Q}[j,:]=0$ and $\mathbf{Q}[j,j]=-1$ for all $j\in B$.
%

%
%
%
\subsubsection{Infrequent Metadynamics}
\begin{figure}[h]
\includegraphics[scale=1]{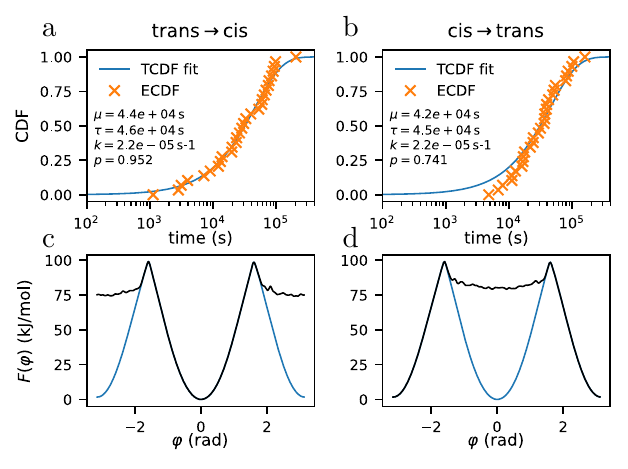}
\caption{
\textbf{a, b:} TCDF fit to results of 30 InMetaD runs for trans-cis (a) and cis-trans (b) transitions. 
$\mu$ is the transition time averaged over 30 runs, $\tau$ is the transition time as obtained from fitting and $k=1/\tau$ the corresponding rate. $p$ is the calculated $p$-value of the KS test. 
\textbf{c, d:} Free energy profiles obtained from metadynamics simulation (as in Fig.~\ref{main-fig:FES_phi}) with the biasing potential at the moment of transitioning for an example run of InMetaD for trans-cis (c) and cis-trans (d) isomerization added on top in black.
}
\label{fig:inmetad}
\centering
\end{figure}

Infrequent metadynamics (InMetaD) were run for both the trans-cis and cis-trans transition in sets of 30 runs and fitted to a Poisson distributions\cite{salvalaglio2014assessing} as described in the Theory section (eq.~\ref{main-eq:TCDF}).
Biasing was done on the C$_{13}$=C$_{14}$ dihedral CV $\varphi$ at a pace of $100\,\mathrm{ps}$ with a Gaussian height of $1.2\,\mathrm{kJ/mol}$, standard deviation of $0.05\,\mathrm{rad}$ and bias factor of 16.
Trajectories for runs from trans to cis were terminated once a value (in radians) of $\varphi\in\left[-\pi/5,\pi/5\right]$ was reached, where the molecule is definitely in the cis state.
The biased transition time $\tau_{\mathrm{t}\rightarrow \mathrm{c}, i}^{\mathrm{InMetaD}}$ was then taken to be the time of the last trajectory point where the configuration can still be considered at the trans side, i.e.~the last trajectory point where $\varphi<-\pi/2$ or $\varphi>\pi/2$.
The unbiased transition times $\tau_{\mathrm{t}\rightarrow \mathrm{c}, i}$ can then be calculated from eq.~\ref{main-eq:InMetaD}.
Trajectories for runs from cis to trans were stopped once a value of $\varphi<-1.9\,\mathrm{rad}$ or $\varphi>1.9\,\mathrm{rad}$ was reached, where the molecule is definitely in the trans state.
The biased transition time $\tau_{\mathrm{c}\rightarrow \mathrm{t}, i}^{\mathrm{InMetaD}}$ was then taken to be the time of the last trajectory point where the configuration can still be considered at the cis side, 
i.e.~the last trajectory point where $\varphi\in\left[-\pi/2,\pi/2\right]$, and unbiased transition times $\tau_{\mathrm{c}\rightarrow \mathrm{t}, i}$ can be calculated from eq.~\ref{main-eq:InMetaD}.
KS tests were done using using a million randomly generated points according to the corresponding TCDF (eq.~\ref{main-eq:TCDF}) for both trans-cis and cis-trans transitions, yielding a $p$-value of 0.95 and 0.74 respectively, which is well above the proposed cutoff of 0.05. 
A graphical representation of the TCDF fit and KS test as well as the biased potential at the moment of the transitioning for example runs can be found in Fig.~\ref{fig:inmetad}.
Average transition times, standard errors, Poisson fitted transition times and corresponding $p$-values can be found in Table \ref{tab:inmetad_times}.
%

%
%
%
\subsubsection{Multidimensional Free Energy and Diffusion surfaces}

\begin{figure}[h]
\includegraphics[scale=1]{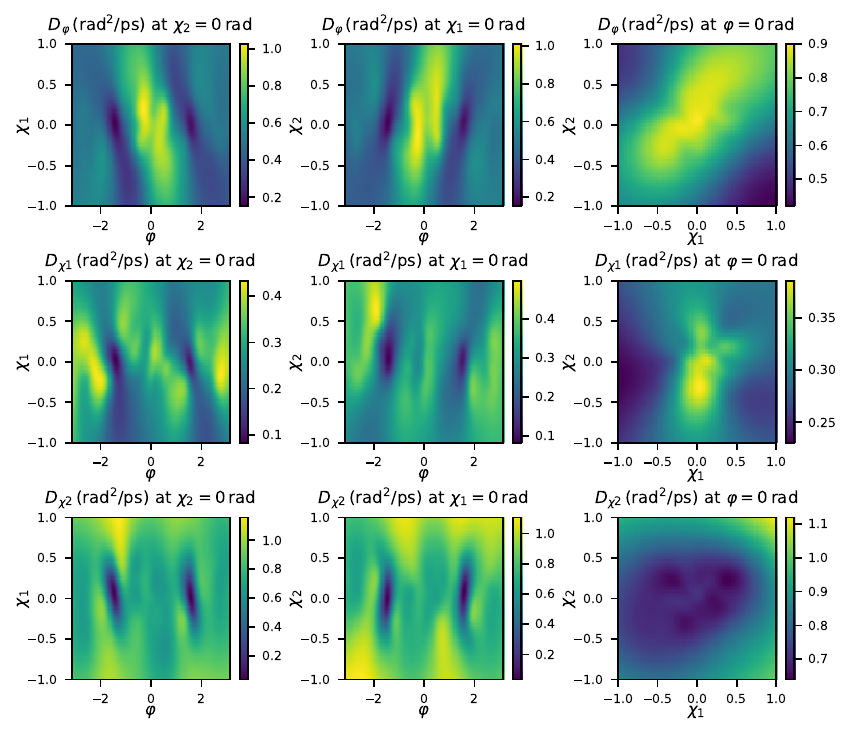}
\caption{Interpolated three-dimensional diffusion surfaces $D_\varphi$, $D_{\chi_1}$ and $D_{\chi_2}$ obtained from \textit{grid1} of three-dimensional harmonic restraints.}
\label{fig:3D_diffusion1}
\centering
\end{figure}

\begin{figure}[h]
\includegraphics[scale=1]{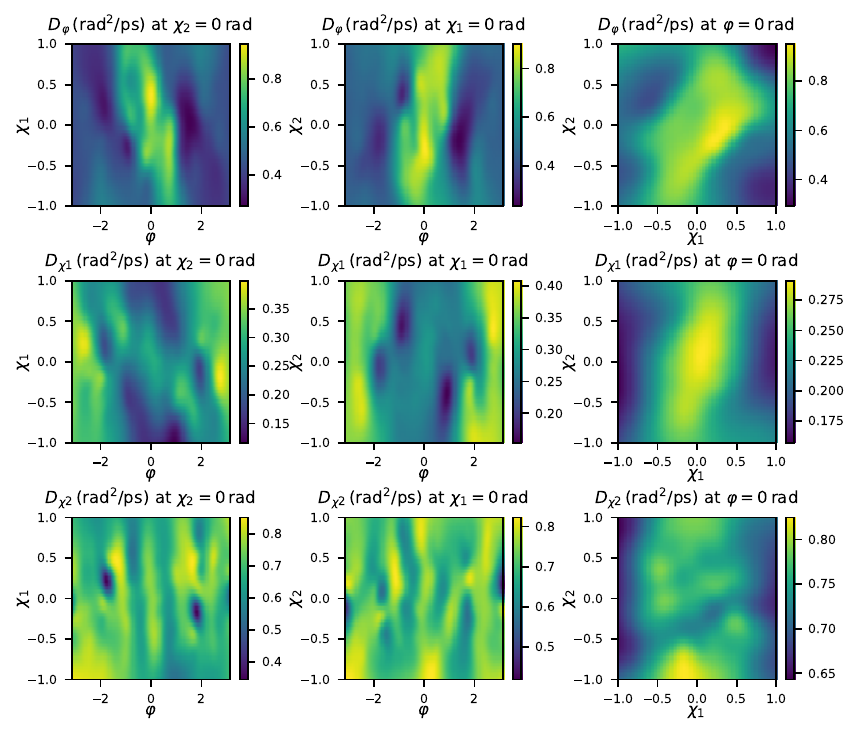}
\caption{Interpolated three-dimensional diffusion surfaces $D_\varphi$, $D_{\chi_1}$ and $D_{\chi_2}$ obtained from \textit{grid2} of three-dimensional harmonic restraints.}
\label{fig:3D_diffusion2}
\centering
\end{figure}

Multidimensional free energy surfaces were calculated from multidimensional metadynamics simulations implemented using a similar setup as for the one-dimensional case. 
Well-tempered metadynamics were run biasing the three-dimensional space spanned by the following collective variables:
\begin{itemize}
    \item $\varphi$: C$_{13}$=C$_{14}$ dihedral angle
    \item $\chi_1$: improper dihedral constituting the out of plane bending of the carbon atom of the methyl group on the C$_{13}$ atom
    \item $\chi_2$: improper dihedral constituting the out of plane bending of the hydrogen on the C$_{14}$ atom.
\end{itemize}
See eqs.~\ref{eq:dihedral_cos_sin}-\ref{eq:dihedral_atan2} for a mathematical definition.
Three-dimensional Gaussians of width $0.07\,\mathrm{rad}$ in each CV were deposited at a pace of $1\,\mathrm{ps}$ and with a bias factor of 12. 
In this case, metadynamics were only carried out for $1\,\mathrm{\mu s}$ because the retinal cofactor was noticed to collapse upon the lysine backbone for larger simulation times.
Since such configurations were not observed during one-dimensional metadynamics or umbrella sampling, and we are not interested in them from a conceptual point of view, the trajectory was cut before they appear, i.e.~after $1\,\mathrm{\mu s}$.
The three-dimensional free energy surface $F(\varphi,\chi_1,\chi_2)$ was calculated by building a three-dimensional histogram from trajectory data and reweighting using the bias obtained at the end of the metadynamics simulation.
Equivalently, two-dimensional free energy surfaces $F(\varphi,\chi_1)$ and $F(\varphi,\chi_2)$ (Fig.~\ref{main-fig:2D_correlations_FES_paths}.b and d) and one-dimensional free energy surface $F(\varphi)$ (Fig.~\ref{main-fig:FES_phi}) can be computed by reweighting two- and one-dimensional histograms respectively, using the same trajectory data and bias.
Convergence of the bias and block error analysis are shown in Fig.~\ref{fig:metad_convergence}.b.
Multidimensional diffusion surfaces $D_\varphi$, $D_{\chi_1}$ and $D_{\chi_2}$ were computed by applying a multidimensional generalization of Hummer's formulation in eq.~\ref{eq:diffusion_coefficient}. 
In our three-dimensional case, a series of trajectories are run with three-dimensional harmonic restraints positioned on a regular grid in collective variable space. 
For each trajectory, one value for each of the diffusion coefficients $D_\varphi$, $D_{\chi_1}$ and $D_{\chi_2}$ can then be calculated by computing correlation functions in each direction (eq.~\ref{eq:diffusion_coefficient}).
Additionally their corresponding average positions $\left<\varphi\right>$, $\left<\chi_1\right>$ and $\left<\chi_2\right>$ are computed, yielding a three-dimensional `grid' (which now might be irregular) in collective variable space, with for each point an associated value for $D_\varphi$, $D_{\chi_1}$ and $D_{\chi_2}$. 
Diffusion surfaces $D_\varphi(\varphi,\chi_1,\chi_2)$, $D_{\chi_1}(\varphi,\chi_1,\chi_2)$ and $D_{\chi_2}(\varphi,\chi_1,\chi_2)$ can then be obtained by three-dimensional interpolation.
In this fashion, two sets of diffusion profiles in each direction were calculated using different grids and different spring constants for the harmonic restraints. 
We will refer to the sets as \textit{grid1} and \textit{grid2}.
For \textit{grid1}, 200 trajectories of $5\,\mathrm{ns}$ were run employing three-dimensional harmonic restraints with  $400\,\mathrm{kJ}.\mathrm{mol}^{-1}.\mathrm{rad}^{-2}$ spring constants in each direction, positioned on a regular $8 \times 5 \times 5$ grid in CV space as follows:
\begin{itemize}
    \item $\varphi$ varies over 8 steps in regular intervals from $-\pi$ to $\pi$
    \item $\chi_1$ varies over 5 steps in regular intervals from -0.5 to 0.5
    \item $\chi_2$ varies over 5 steps in regular intervals from -0.5 to 0.5.
\end{itemize}
Two-dimensional cuts of the resulting three-dimensional diffusion surfaces are shown in Fig.~\ref{fig:3D_diffusion1}.
For \textit{grid2}, 729 trajectories of $2\,\mathrm{ns}$ were run employing three-dimensional harmonic restraints with  $600\,\mathrm{kJ}.\mathrm{mol}^{-1}.\mathrm{rad}^{-2}$ spring constants in each direction, positioned on a regular $9 \times 9 \times 9$ grid in CV space as follows:
\begin{itemize}
    \item $\varphi$ varies over 9 steps in regular intervals from $-\pi$ to $\pi$
    \item $\chi_1$ varies over 9 steps in regular intervals from -1 to 1
    \item $\chi_2$ varies over 9 steps in regular intervals from -1 to 1.
\end{itemize}
Two-dimensional cuts of the resulting three-dimensional diffusion surfaces are shown in Fig.~\ref{fig:3D_diffusion2}.
Multidimensional US simulations were performed by running $5\,\mathrm{ns}$ trajectories on a total of 784 three-dimensional harmonic restraints, positioned on a three-dimensional grid in $(\varphi,\chi_1,\chi_2)$.
%
%
The harmonic restraints had spring constants of $400\,\mathrm{kJ}.\mathrm{mol}^{-1}$ in $\varphi$ directions and $300\,\mathrm{kJ}.\mathrm{mol}^{-1}$ in both $\chi_1$ and $\chi_2$ directions, and were positioned as follows:
\begin{itemize}
    \item $\varphi$ varies over 16 steps in regular intervals from $-\pi$ to $\pi$
    \item $\chi_1$ varies over 7 steps in regular intervals from -1 to 1
    \item $\chi_2$ varies over 7 steps in regular intervals from -1 to 1.
\end{itemize}
We will refer to this grid as \textit{grid3}.
While \textit{grid1} and \textit{grid2} were exclusively used for calculations of position-dependent diffusion profiles, \textit{grid3} was exclusively used for construction of a three-dimensional free energy surface $F(\varphi,\chi_1,\chi_2)$.
This was done employing binless WHAM\cite{tan2012theory,bussi2019analyzing}.
\subsubsection{Adaptive Path Collective Variables}
For a more accurate description of the dynamics, we aim to find a path CV description of the thermal isomerization in the space spanned by the $\varphi$, $\chi_1$ and $\chi_2$ CVs.
Paths have been optimized using the adaptive path collective variable method\cite{leines2012path,perez2019adaptive} implemented in PLUMED under the ADAPTIVE\_PATH module in combination with well-tempered metadynamics. 
In order to correctly handle the periodicity of $\varphi$, the sine and cosine were used rather than including the angle directly. In order to avoid differences in scale of the CVs\cite{perez2019adaptive,ortiz2021simultaneous}, we have also taken the sines of the improper dihedrals. 
Notice that in this case we do not have to include the corresponding cosines as the range of interest of the improper dihedrals doesn't warrant it. 
Thus, in practice, the adaptive path CV algorithm was performed in four dimensions:
\begin{itemize}
    \item sin\_phi: sine of $\varphi$
    \item cos\_phi: cosine of $\varphi$
    \item sin\_improper1: sine of $\chi_1$
    \item sin\_improper2: sine of $\chi_2$.
\end{itemize}
Although in principle cyclic paths can be handled with the adaptive path CV scheme\cite{ortiz2021simultaneous}, we have chosen to study each transition separately, i.e.~we optimized the \textit{cis\_trans1}, \textit{cis\_trans2}, \textit{trans\_cis1} and \textit{trans\_cis2} paths in separate runs.
\begin{itemize}
    \item \textit{trans\_cis1} describes trans-cis isomerization in counterclockwise direction (increasing torsion angle)
    \item \textit{trans\_cis2} describes trans-cis isomerization in clockwise direction (decreasing torsion angle).
    \item \textit{cis\_trans1} describes cis-trans isomerization in counterclockwise direction (increasing torsion angle)
    \item \textit{cis\_trans2} describes cis-trans isomerization in clockwise direction (decreasing torsion angle)
\end{itemize}
The initial and final states for each path, which are kept fixed during the adaptive path CV algorithm, have been chosen as $0$ or $\pm\pi\,\mathrm{rad}$ depending on the transition under consideration. 
As initial guess paths, linear interpolations of $\varphi$ between the initial and final state values were used, while $\chi_1$ and $\chi_2$ were simply set to zero over the full initial guess paths.
The adaptive path CV for each transition was run using 21 path nodes over the course of a $1\,\mu\mathrm{s}$  well-tempered metadynamics trajectory. 
Notice that one of the path CV components, $\chi_2$, encompasses out of plane bending of a hydrogen atom.  
Since LINCS constraints were applied, a smaller time step of $1\,\mathrm{fs}$ was chosen for all dynamics where an (adaptive) path CV is being biased.
Additionally, the actual biasing was done in a more gentle way, decreasing the height of the initial Gaussians to $0.2\,\mathrm{kJ/mol}$ while the width was set at $0.05$ normalized path units and the pace was $0.5\,\mathrm{ps}$. 
We also intended to limit the bias factor. For adaptive path CVs, however, the bias factor is generally preferred to be chosen higher than for general well-tempered metadynamics runs as to optimize the convergence of the path\cite{perez2019adaptive}. 
A factor of 12 turned out to be a good compromise for all transitions except for \textit{trans\_cis2}, where a smaller factor of 10 was used. 
During metadynamics sampling we used a tube restraint of $200\,\mathrm{kJ/mol}$ per normalized units squared to avoid bifurcations\cite{leines2012path,perez2019adaptive}, e.g.~isomerizations in the wrong direction, and a half life of $5\times 10^5$ steps to allow sufficient flexibility in the path adaptive algorithm\cite{perez2019adaptive}.
Furthermore, harmonic walls of $500\,\mathrm{kJ/mol}$ per normalized units squared have been put on the path parameter $\sigma_s$ before the reactant state and behind the product state, that is at $\sigma_s = -0.4$ and $\sigma_s=1.4$.
\subsubsection{Free Energy and Rate Calculations on Path Collective Variables}
\label{section:FES_rates_path_CV}
\begin{figure}[h]
\includegraphics[scale=1]{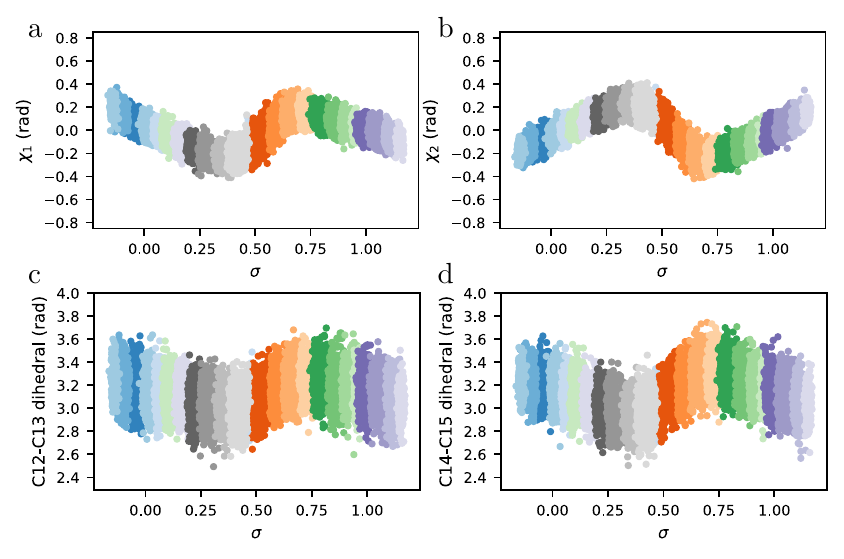}
\caption{\textbf{a, b, c, d:} Scatter plot of improper dihedrals $\chi_1$ (a) and $\chi_2$ (b) as well as proper dihedrals C$_{13}$=C$_{14}$-C$_{15}$=NH (c) and C$_{11}$=C$_{12}$-C$_{13}$=C$_{14}$ (d) versus the path CV for umbrella sampling along the path CV of path \textit{cis\_trans1}.
Clearly the correlation of the improper dihedrals $\chi_1$ and $\chi_2$ is handled more smoothly as the sampling doesn't `jump' at the peaks anymore (compare to Fig.~\ref{main-fig:2D_correlations_FES_paths}).
Interestingly, also the correlation of the proper dihedrals (C$_{14}$-C$_{15}$ and C$_{12}$-C$_{13}$) is handled more smoothly as the sampling doesn't `jump' at the peaks anymore either (compare to Fig.~\ref{fig:phi_proper_correlation} immediately above).}
\label{fig:sss_correlation}
\centering
\end{figure}

Free energy profiles on each of the four paths were calculated using metadynamics and umbrella sampling biasing the path collective variable optimized during the adaptive path sampling described above. 
All dynamics were done using $1\,\mathrm{fs}$ time steps.
Well-tempered metadynamics were run for $1\,\mu\mathrm{s}$ for each path depositing Gaussians of standard deviation of $0.5/21=0.0238$ normalized path units, a height of $1.4\,\mathrm{kJ/mol}$, a pace of $0.5\,\mathrm{ps}$ and a bias factor of 12. 
Tube restraints of $200\,\mathrm{kJ/mol}$ per normalized units squared were used for all profiles.
Furthermore, harmonic walls with a spring constant of $500\,\mathrm{kJ/mol}$ per path units squared were employed before the reactant state at $\sigma_s=-5/21=0.238$ and after the product state at $\sigma_s=26/21=1.238$ to avoid isomerization along different paths from the one being investigated.
Unbiasing weights were calculated using the bias potential at the end of the trajectories, and FES were composed from the accompanying weighted histograms. 
Notice it is also necessary to reweight the tube restraints; we noticed a difference in barrier height of about $2\,\mathrm{kJ/mol}$ if this restraint was not included in the reweighting.
Construction of weighted histograms was done with kernel density estimation (KDE) with Gaussian kernels of bandwidth $0.1/21=0.00476$ for all metadynamics runs.
Error estimation and convergence of the free energy difference are shown for metadynamics simulations for each path in Fig.~\ref{fig:metad_convergence}.c.
Umbrella sampling simulations were carried out using 70 umbrellas of $20\,\mathrm{ns}$, restraint along the path collective variable using harmonic restraints of spring constant $100\,\mathrm{kJ/mol}$ per normalized path units squared, located at regular intervals between $-3.2/21=-0.152$ and $24.4/21=1.162$ normalized path units.
This makes for a total of $1.4\,\mu\mathrm{s}$ simulation time per path. 
Again, tube restraints and harmonic walls before reactant and behind product states were used for all profiles, with the same spring constants and positions as for metadynamics on the path CV described above. 
For some trajectories restrained near the barrier top, LINCS errors occurred. 
This could generally be helped by choosing a more suitable starting configuration or by reducing the time step to $0.5\,\mathrm{fs}$ for these cases.
Notice that sampling along $\sigma_s$ is smoother than sampling along $\varphi$, as there is no `jump' at the barrier, see Fig.~\ref{fig:sss_correlation}.
For all umbrella sampling sets, conventional discrete grids were utilized to construct weighted histograms.
Diffusion profiles were calculated by applying Hummer's method (eq.~\ref{eq:diffusion_coefficient}) to trajectories of each of the umbrellas. 
Rates were calculated similarly as for the C$_{13}$=C$_{14}$ dihedral angle CV described in Section.~\ref{section:rates}. 
Since we have calculated free energies for all paths separately, we are only interested in rates from left to right for each path, i.e.~for increasing value of the path CV $\sigma_s$. 
Reduced masses $\mu_A$ and $\mu_B$ in reactant state $A$ and product state $B$ were calculated from $10\,\mathrm{ns}$ unbiased simulations in the reactant and product states respectively, and subsequent application of eq.~\ref{eq:reduced_mass} monitoring the kinetic energy in the path CV $\sigma_s$ instead of in $\varphi$. 
Similarly $\omega_A$ was obtained using the path CV equivalent of eq.~\ref{eq:omega_A} where the spring constant $\kappa_A$ is obtained by fitting the free energy surface along $\sigma_s$ in the reactant state $A$ to a harmonic potential. 
Angular frequencies $\omega_A$ obtained this way were compared to frequencies obtained from measuring oscillation periods $T_A$ in the reactant states, with both corresponding very well. 
Similarly, $\xi_{TS}$ and $\omega_{TS}$ were calculated in the same way as we did for the dihedral CV, i.e.~trough eqs.~\ref{eq:gamma_ddagger} and \ref{eq:omega_ddagger}, where $\kappa_{TS}$ was obtained by a parabolic fit and $\mu_{TS}$ by averaging $\mu_{A}$ and $\mu_{B}$. 
The obtained values are shown in Tables \ref{tab:path_constants_metad} and \ref{tab:path_constants_us}.
Coefficients for evaluating the friction limit can be found in the same tables.
These constants can be used to calculate the TST and Kramers' rates.  
Pontryagin rate was computed carrying out the nested integration using the free energy and diffusion profile as a function of $\sigma_s$.
Grid-based models were carried out by discretizing the path CV in 500 bins and using high precision libraries\cite{mpmath,FLINT} with 50 digits to build and solve the rate matrix as in eqs.~\ref{main-eq:Qij_HAA_1} and \ref{main-eq:solve_Q}, similarly as for the dihedral CV case.

An overview of all resulting rates from free energy profiles from metadynamics and umbrella sampling along the path CVs can be found in Table \ref{main-tab:rates}.
%

%
\begin{table*}
    \centering
    \begin{tabular}{ c  c  c  c  c  c}
    &  &  \multicolumn{2}{c}{trans$\rightarrow$cis} & \multicolumn{2}{c}{cis$\rightarrow$trans} \\ 
    & \textbf{units}  & $TS$ & $TS'$  & $TS$ & $TS'$ \\ 
    \hline
    \hline
    $\mu_\mathrm{A}$            & [$\mathrm{kg}.\mathrm{m}^2.\mathrm{rad}^{-2}$]
                                & $ 6.81 \times 10^{-47} $
                                & $ 6.81 \times 10^{-47} $
                                & $ 2.41 \times 10^{-47} $
                                & $ 2.41 \times 10^{-47} $
                                \\

    $D_{TS}$              &[$\mathrm{rad}^2.\mathrm{ps}^{-1}$]
                                & $ 0.680 $
                                & $ 0.687 $
                                & $ 0.680 $
                                & $ 0.687 $\\
                                
    $\xi_{TS}$         &[$\mathrm{ps}^{-1}$] 
                                & $ 1.32 \times 10^{2} $
                                & $ 1.31 \times 10^{2} $
                                & $ 1.32 \times 10^{2} $
                                & $ 1.31 \times 10^{2} $\\

    $\omega_\mathrm{A}$         &[$\mathrm{ps}^{-1}$]
                                & $ 4.99 \times 10^{1} $
                                & $ 4.99 \times 10^{1} $
                                & $ 7.72 \times 10^{1} $
                                & $ 7.72 \times 10^{1} $ \\
                                
    $\omega_{TS}$         &[$\mathrm{ps}^{-1}$]
                                & $ 2.44 \times 10^{2} $
                                & $ 2.63 \times 10^{2} $
                                & $ 2.44 \times 10^{2} $
                                & $ 2.63 \times 10^{2} $ \\
                                
    $F^{\ddagger}$              &[$\mathrm{kJ}.\mathrm{mol}^{-1}$]
                                & $ 88.6 $ 
                                & $ 89.3 $
                                & $ 88.8 $
                                & $ 89.5 $\\ 
    $\xi_{TS}/\omega_{TS}$&[-]
                                & $ 0.54 $
                                & $ 0.50 $
                                & $ 0.54 $
                                & $ 0.50 $\\
    $RT/F^{\ddagger}$&[-]
                                & $ 0.028 $
                                & $ 0.028 $
                                & $ 0.028 $
                                & $ 0.028 $\\ 
    \hline
    \hline
    \end{tabular}
    \caption{Retinal: parameters for one-dimensional rate theories calculated for $F(\varphi)$ obtained by umbrella sampling.
    }
    \label{tab:dihedral_constants_us}
\end{table*}

%
\begin{table*}
    \centering
    \begin{tabular}{ c  c  c  c  c  c}
    &  &  \multicolumn{2}{c}{trans$\rightarrow$cis} & \multicolumn{2}{c}{cis$\rightarrow$trans} \\ 
    & \textbf{units}  & $TS$ & $TS'$  & $TS$ & $TS'$ \\ 
    \hline
    \hline
    $\mu_\mathrm{A}$            & [$\mathrm{kg}.\mathrm{m}^2.\mathrm{rad}^{-2}$]
                                & $ 6.81 \times 10^{-47} $
                                & $ 6.81 \times 10^{-47} $
                                & $ 2.41 \times 10^{-47} $
                                & $ 2.41 \times 10^{-47} $
                                \\

    $D_{TS}$              &[$\mathrm{rad}^2.\mathrm{ps}^{-1}$]
                                & $ 0.680 $
                                & $ 0.687 $
                                & $ 0.680 $
                                & $ 0.687 $\\
                                
    $\xi_{TS}$         &[$\mathrm{ps}^{-1}$] 
                                & $ 1.32 \times 10^{2} $
                                & $ 1.31 \times 10^{2} $
                                & $ 1.32 \times 10^{2} $
                                & $ 1.31 \times 10^{2} $\\

    $\omega_\mathrm{A}$         &[$\mathrm{ps}^{-1}$]
                                & $ 5.17 \times 10^{1} $
                                & $ 5.17 \times 10^{1} $
                                & $ 7.99 \times 10^{1} $
                                & $ 7.99 \times 10^{1} $\\
                                
    $\omega_{TS}$         &[$\mathrm{ps}^{-1}$]
                                & $ 2.97 \times 10^{2} $
                                & $ 3.42 \times 10^{2} $
                                & $ 2.97 \times 10^{2} $
                                & $ 3.42 \times 10^{2} $ \\
                                
    $F^{\ddagger}$              &[$\mathrm{kJ}.\mathrm{mol}^{-1}$]
                                & $ 96.9 $
                                & $ 97.3 $
                                & $ 98.6 $
                                & $ 99.0 $ \\ 
    $\xi_{TS}/\omega_{TS}$&[-]
                                & $ 0.45 $
                                & $ 0.38 $
                                & $ 0.45 $
                                & $ 0.38 $\\
    $RT/F^{\ddagger}$&[-]
                                & $ 0.026 $
                                & $ 0.026 $
                                & $ 0.025 $
                                & $ 0.025 $\\ 
    \hline
    \hline
    \end{tabular}
    \caption{Retinal: parameters for one-dimensional rate theories calculated for $F(\varphi)$ obtained by metadynamics.
    }
    \label{tab:dihedral_constants_metad}
\end{table*}

%
\begin{table*}
    \centering
    \begin{tabular}{ c  c  c  c  c  c}
      & \textbf{units}  &  \textit{trans\_cis1} & \textit{trans\_cis2} & \textit{cis\_trans1} & \textit{cis\_trans2}  \\ 
    \hline
    \hline
    $\mu_\mathrm{A}$        &[$\mathrm{kg}.\mathrm{m}^2$]
                                & $ 1.42 \times 10^{-48} $
                                & $ 1.44 \times 10^{-48} $
                                & $ 5.18 \times 10^{-49} $
                                & $ 5.10 \times 10^{-49} $  \\

    $D_{TS}$              &[$\mathrm{ps}^{-1}$]
                                & $ 5.413 $
                                & $ 5.552 $
                                & $ 5.449 $
                                & $ 5.421 $\\
                                
    $\xi_{TS}$         &[$\mathrm{ps}^{-1}$] 
                                & $ 7.92 \times 10^2 $
                                & $ 7.65 \times 10^2 $
                                & $ 7.85 \times 10^2 $
                                & $ 7.90 \times 10^2 $ \\

    $\omega_\mathrm{A}$         &[$\mathrm{ps}^{-1}$]
                                & $ 5.89 \times 10^1 $
                                & $ 5.86 \times 10^1 $
                                & $ 8.86 \times 10^1 $
                                & $ 8.91 \times 10^1 $ \\
                                
    $\omega_{TS}$         &[$\mathrm{ps}^{-1}$]
                                & $ 1.46 \times 10^2 $
                                & $ 1.52 \times 10^2 $
                                & $ 1.54 \times 10^2 $
                                & $ 1.52 \times 10^2 $ \\
                                
    $F^{\ddagger}$              &[$\mathrm{kJ}.\mathrm{mol}^{-1}$]
                                & $ 96.9 $
                                & $ 97.1 $
                                & $ 97.9 $
                                & $ 97.8 $ \\ 
    $\xi_{TS}/\omega_{TS}$&[-]
                                & $ 5.41 $
                                & $ 5.05 $
                                & $ 5.09 $
                                & $ 5.21 $ \\
    $RT / F^{\ddagger}$         &[-]
                                & $ 0.026 $
                                & $ 0.026 $
                                & $ 0.025 $
                                & $ 0.025 $\\ 
    \hline
    \hline
    \end{tabular}
    \caption{Retinal: parameters for one-dimensional rate theories calculated for $F(\sigma_s)$ obtained by umbrella sampling.}
    \label{tab:path_constants_us}
\end{table*}

%
\begin{table*}
    \centering
    \begin{tabular}{ c  c  c  c  c  c}
    & \textbf{units}  &  \textit{trans\_cis1} & \textit{trans\_cis2} & \textit{cis\_trans1} & \textit{cis\_trans2}  \\ 
    \hline
    \hline
    $\mu_\mathrm{A}$            &[$\mathrm{kg}.\mathrm{m}^2$]
                                & $ 1.42 \times 10^{-48} $
                                & $ 1.44 \times 10^{-48} $
                                & $ 5.18 \times 10^{-49} $
                                & $ 5.10 \times 10^{-49} $\\

    $D_{TS}$              &[$\mathrm{ps}^{-1}$]
                                & $ 5.396 $
                                & $ 5.565 $
                                & $ 5.403 $
                                & $ 5.440 $\\
                                
    $\xi_{TS}$         &[$\mathrm{ps}^{-1}$] 
                                & $ 7.94 \times 10^2 $
                                & $ 7.63 \times 10^2 $
                                & $ 7.92 \times 10^2 $
                                & $ 7.87 \times 10^2 $\\

    $\omega_\mathrm{A}$         &[$\mathrm{ps}^{-1}$]
                                & $ 5.91 \times 10^1 $
                                & $ 5.86 \times 10^1 $
                                & $ 8.88 \times 10^1 $
                                & $ 8.94 \times 10^1 $ \\
                                
    $\omega_{TS}$         &[$\mathrm{ps}^{-1}$]
                                & $ 1.54 \times 10^2 $
                                & $ 1.58 \times 10^2 $
                                & $ 1.55 \times 10^2 $
                                & $ 1.55 \times 10^2 $ \\
                                
    $F^{\ddagger}$              &[$\mathrm{kJ}.\mathrm{mol}^{-1}$]
                                & $ 105.2 $
                                & $ 104.9 $
                                & $ 106.7 $
                                & $ 104.8 $\\ 
    $\xi_{TS}/\omega_{TS}$&[-]
                                & $ 5.16 $
                                & $ 4.82 $
                                & $ 5.12 $
                                & $ 5.08 $\\
    $RT/F^{\ddagger}$&[-]
                                &  $ 0.024 $
                                &  $ 0.024 $
                                &  $ 0.023 $
                                &  $ 0.024 $\\ 
    \hline
    \hline
    \end{tabular}
    \caption{Retinal: parameters for one-dimensional rate theories calculated for $F(\sigma_s)$ obtained by metadynamics.}
    \label{tab:path_constants_metad}
\end{table*}
\subsubsection{Multidimensional Discretization of the Fokker-Planck Operator}
\begin{figure*}
\includegraphics[scale=1]{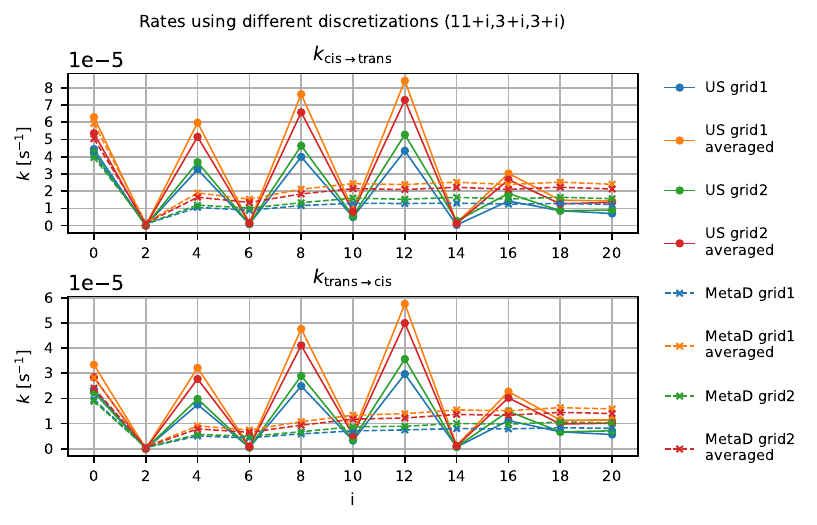}
\caption{Isomerization rates as a function of discretization for 3D grid-based models applied to 3D FES from metadynamics and umbrella sampling (\textit{grid3}) and with averaged and position dependent diffusion for two diffusion schemes \textit{grid1} (Fig.~\ref{fig:3D_diffusion1}) and \textit{grid2} (Fig.~\ref{fig:3D_diffusion2}).}
\label{fig:k_discretization}
\centering
\end{figure*}

Similarly as for the one-dimensional cases, grid-based models can be implemented by discretizing the three-dimensional CV space spanned by $\varphi$, $\chi_1$ and $\chi_2$ and building the rate matrix according to eq.~\ref{main-eq:Qij_HAA_1}. 
This was done for the free energy surface obtained from three-dimensional metadynamics (see above) as well as for the free energy surface obtained from three-dimensional umbrella sampling (see above, \textit{grid3}).
The $\chi_1$ and $\chi_2$ CVs where discretized between $-1$ and $1\,\mathrm{rad}$ for the metadynamics surface and $-0.7$ and $0.7\,\mathrm{rad}$ for the US surface, and were treated as non-periodic. 
For both surfaces, $\varphi$ was discretized between $-\pi$ and $\pi\,\mathrm{rad}$ and treated as periodic just as was done for the one-dimensional case.
Within a choice of discretization, all cells had the same shape and size, i.e.~each CV was discretized in cells of the same length. 
The free energy and diffusion surfaces calculated as detailed above were interpolated using radial basis function (RBF) interpolation as implemented in scipy, and evaluated at the cell middles $q_i$ for each cell $i$ to yield the free energy and diffusion values $\pi_i$ and $D_i$ necessary to build the rate matrix according to eq.~\ref{main-eq:Qij_HAA_1}. 
High precision libraries\cite{mpmath,FLINT} were used to handle the high barriers, just as for the one-dimensional case. 
50 digits were used for all calculations. 
Notice that working with high-precision numbers makes calculations much slower and therefore severely limits the discretization which can be used. 
The discretization used in this work divided the CV space in (31,23,23) blocks in $\varphi$, $\chi_1$ and $\chi_2$ collective variables respectively, yielding a total of 16399 cells. 
Rates from the three-dimensional grid-based models for different diffusion surfaces can be found in Tab.~\ref{main-tab:rates}.
While the discretization is fine enough to yield converged rates for the 3D FES from metadynamics, the rates from 3D US do not converge as quickly (Fig.~\ref{fig:k_discretization}).
Therefore, rates calculated from the 3D US FES are given between brackets in Tab.~\ref{main-tab:rates}, and have to be interpreted with care.
We stress that discretizations could be significantly increased for application to barrier heights that do not necessitate high precision numbers.

\subsubsection{Three-dimensional Infrequent Metadynamics}
Three-dimensional infrequent metadynamics were run for both trans-cis and cis-trans transitions in sets of 30 runs for each.
Biasing was done using three-dimensional Gaussians of height $0.75\,\mathrm{kJ/mol}$ with a standard deviation of $0.07\,\mathrm{rad}$ in all three dimensions, that is, along $\varphi$, $\chi_1$ and $\chi_2$. 
The deposition pace was $20\,\mathrm{ps}$ while a bias factor of 20 was used.
Determining the biased transition time for trans-to-cis and cis-to-trans simulation was done based on the $\varphi$ value alone in the same way as for one-dimensional infrequent metadynamics described above.
The reweighted transition times were fitted to a Poisson distribution and a KS test was done using a million randomly generated points according to the TCDF from the corresponding fits, as described in Ref.~\citenum{salvalaglio2014assessing}.
The corresponding rates can be found in Table \ref{main-tab:rates}.
Average transition times, standard errors, Poisson fitted transition times and corresponding $p$-values can be found in Table \ref{tab:inmetad_times}.

\begin{table*}
    \centering
    \begin{tabular}{ c  c  c  c  c  c  c }
    \multicolumn{7}{c}{\textbf{InMetaD}} \\ 
    \hline
    \hline    
   \textbf{biased CV} &
    \multicolumn{3}{c}{\textbf{trans}$\rightarrow$\textbf{cis}} 
    &\multicolumn{3}{c}{\textbf{cis}$\rightarrow$\textbf{trans}}\\
    &\multicolumn{1}{c}{$\mu \pm \mathrm{S.E.}$ [s]}
    &\multicolumn{1}{c}{$\tau$ [s]}
    &\multicolumn{1}{c}{$p$-value}
    &\multicolumn{1}{c}{$\mu \pm \mathrm{S.E.}$ [s]}
    &\multicolumn{1}{c}{$\tau$ [s]}
    &\multicolumn{1}{c}{$p$-value}\\
    \hline
    \hline
    $   \varphi $& 
    $   4.45  \times 10^{4}\pm7.70\times 10^{3}$&   
    $   4.58  \times 10^{4}$&
    $   0.95 $&
    $   4.16  \times 10^{4}\pm6.39\times 10^{3}$&
    $   4.48  \times 10^{4}$&
    $   0.74 $\\ 
    $\varphi$, $\chi_1$, $\chi_2$& 
    $   4.06  \times 10^{4}\pm6.07\times 10^{3}$&   
    $   4.51  \times 10^{4}$&
    $   0.52 $&
    $   4.27  \times 10^{4}\pm8.98\times 10^{3}$&
    $   3.85  \times 10^{4}$&
    $   0.98 $\\ 
    \hline
    \hline

    \end{tabular}
    \caption{Average transition times ($\mu$), corresponding standard errors (S.E.) and transition times from Poisson fit ($\tau$) from infrequent metadynamics for classical model system with 1D biasing using $\varphi$ as collective variable and with 3D biasing in the CV space spanned by $\varphi$, $\chi_1$ and $\chi_2$.
    Rates mentioned in Table \ref{main-tab:rates} correspond to $1/\tau$.
    }
    \label{tab:inmetad_times}
\end{table*}

\subsubsection{Transition State Search}
\begin{figure*}
\includegraphics[scale=1]{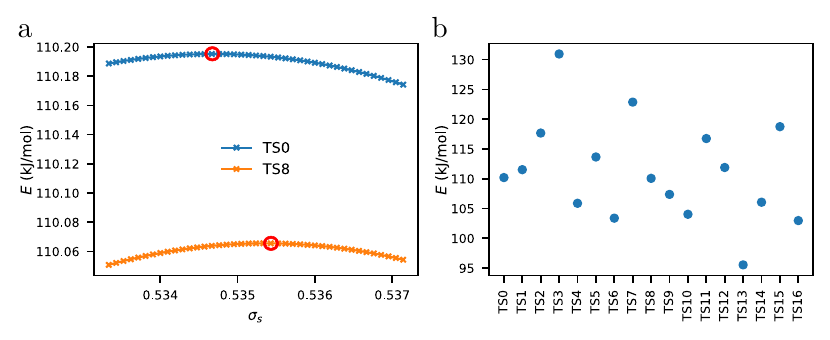}
\caption{\textbf{a:} Potential energies of relaxed scan along the \textit{cis\_trans1} path collective variable for two example candidates (TS0 and TS8). The configurations at the potential energy maxima are circled in red and are chosen to represent the transition state in further analysis (Eyring and high-temperature TST). \\
\textbf{b:} Potential energies of the transition state configurations obtained from relaxed scan for seventeen candidates (TS0-TS16) in reference to the potential energy of the optimized structure in the trans state. Transition states differ up to $35\,\mathrm{kJ}\,\mathrm{mol}^{-1}$ in potential energy.}
\label{fig:TS_energies}
\centering
\end{figure*}

Since GROMACS does not have a method for transition state (i.e.~first-order saddle point) search implemented, transition state configurations were estimated using relaxed scans along one of the path CVs.
From the umbrella sampling simulations along the \textit{cis\_trans1} path (SI section \ref{section:FES_rates_path_CV}), seventeen candidate configurations were selected for which $\varphi\in\left[\pi/2-0.001,\pi/2+0.001\right]$, $\chi_1\in\left[-0.01,0.01\right]$ and $\chi_2\in\left[-0.01,0.01\right]$.
For each of the candidates, a relaxed scan along the \textit{cis\_trans1} path CV was performed.
More specifically, a series of constrained optimizations was carried out, restraining the path parameter $\sigma_s$ at specific values near the transition state in addition to the backbone restraints that were already used in all dynamics simulations.
For each candidate, 41 constrained optimizations were performed, with restraints on $\sigma_s$ carried out between 11.20/21 and 11.28/21 in steps of 0.002/21 normalized path units, each time with a spring constant of $1500\,\mathrm{kJ/mol}$ per normalized path units squared.
The optimizations were performed using GROMACS' limited-memory Broyden-Fletcher-Goldfarb-Shanno quasi-Newtonian\cite{byrd1995limited} minimizer (l-bfgs).
For each of the seventeen candidates, the optimized configuration in the scan at the $\sigma_s$-value with the highest potential energy was chosen to represent the transition state configuration (see Fig.~\ref{fig:TS_energies}.a for candidates TS0 and TS8).
Hessian matrices and the corresponding eigenvalues were calculated on the obtained transition state structures using GROMACS' normal-mode analysis functionalities.
The path parameter restraints used in the constrained optimization were not included for the Hessian calculations.
Additionally, and contrary to the molecular dynamics simulations performed before, the constrained optimization as well as the normal-mode analysis were performed without inclusion of LINCS constraints, as these cannot be handled in GROMACS' normal-mode analysis functionalities.
All structures obtained from the relaxed scans (labeled TS0 to TS16) roughly corresponded to first-order saddle points, where one eigenvalue was large and negative and the subsequent six eigenvalues corresponding to translational and rotational degrees of freedom were small (corresponding to wavenumbers under $40\,\mathrm{cm}^{-1}$).
In parallel with the transition states, structures were energy minimized using the same minimizer and without LINCS constraints in the trans and the cis state as representations of the reactant states.
Equivalent as for the transition states, Hessian matrices and corresponding eigenvalues were calculated, with all eigenvalues found to be positive and the six smallest eigenvalues found to be small.
Potential energies of the transition states (i.e.~at the potential energy maxima of the relaxed scan) in reference to the potential energy of the optimized structure in the trans state are shown in Fig.~\ref{fig:TS_energies}.b for all seventeen candidates (TS0-TS16).
Notice the potential energies of the obtained maxima of the candidates still vary quite a bit (up to $35\,\mathrm{kJ/mol}$), indicating the relaxed PES scan is not an optimal tool for finding the lowest-lying transition state.

\subsubsection{Eyring Transition State Theory and the High-Temperature Limit}
\label{section:SI_Eyring}
\begin{figure*}
\includegraphics[scale=1]{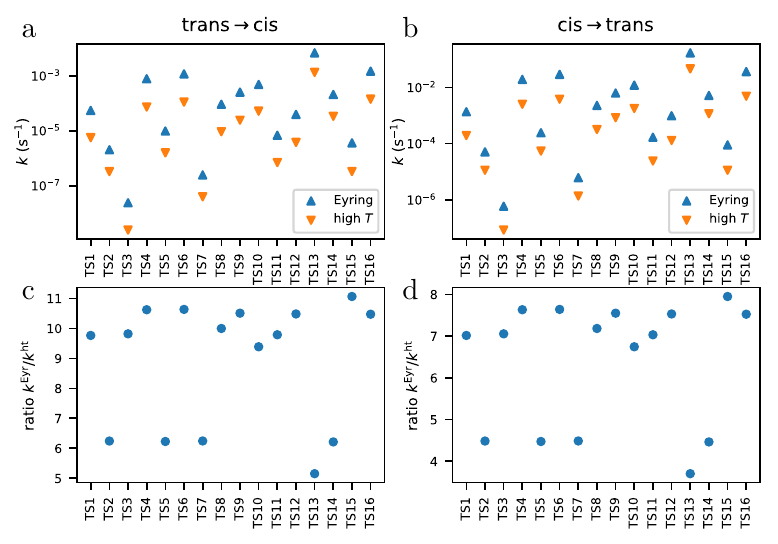}
\caption{\textbf{a,b}: Rates from Eyring TST  and the high temperature (high $T$) approximation for trans $\rightarrow$ cis (\textbf{a}) and cis $\rightarrow$ trans (\textbf{b}) isomerization. Rates for different transition states vary over multiple orders of magnitude.
\textbf{c,d}: ratio of rate from Eyring ($k^\mathrm{Eyr}$) over rate from the high temperature approximation ($k^\mathrm{ht}$) for trans $\rightarrow$ cis (\textbf{c}) and cis $\rightarrow$ trans (\textbf{d}) isomerization. Eyring and high $T$ rates approximately differ by a factor between 4 and 11.}
\label{fig:Eyring_high_T}
\centering
\end{figure*}

When configurations for the reactant (minimum on the PES) and transition state (first-order saddle point on the PES) are available, rates can be estimated by Eyring's transition state theory:
\begin{align}
    k_{AB}^\mathrm{Eyr} 
    = \frac{RT}{h} \frac{\widetilde{q}_{\mathrm{AB}^\ddagger}}{q_{A}} \exp\left(-\frac{E_b}{RT}\right) 
    \label{eq:EyringTST}
\end{align}
where $E_b$ is the potential energy difference between reactant state $A$ and transition state $AB^\ddagger$ and $q_{A}$ and $\widetilde{q}_{AB}^\ddagger$ are the partition functions of the reactant and transition states respectively.
The tilde over the transition state partition function indicates that the degree of freedom associate with the negative eigenvalue of the first-order saddle point should be excluded.
Partition functions are commonly factorized in their translational, rotational, vibrational and electronic contributions.
For unimolecular reactions such as cis-trans isomerization, the translational contribution to the partition function remains unchanged, and thus cancels in eq.~\ref{eq:EyringTST}. 
Rotational contributions to the partition function are also expected to not change much between reactant and transition states, as strong positional restraints keep the backbone in place and the molecule remains relatively linear over the course of isomerization.
Therefore, rotational contributions to $\widetilde{q}_{\mathrm{AB}^\ddagger}/q_{A}$ and thus $k_{AB}^\mathrm{Eyr}$ are neglected in our analysis.
Furthermore, we assume only the electronic ground state is involved during thermal cis-trans isomerization, and thus also electronic contributions are ignored.
Consequently, only vibrational contributions to the partition functions in eq.~\ref{eq:EyringTST} are considered here.
The quantum mechanical partition functions for vibrational degrees of freedom are given by
\begin{align}
	q_{A; \mathrm{vib}} 
	= \prod_{k=1}^{3N-6} \frac{\exp\left(-\frac{h\nu_{A,k}}{2RT}\right)}{1-\exp\left(-\frac{h\nu_{A,k}}{RT}\right)} \quad\mathrm{and}\quad 
 \widetilde{q}_{AB^{\ddagger}; \mathrm{vib}} 
	 \prod_{k=1, k\ne r}^{3N-6} \frac{\exp\left(-\frac{h\nu_{AB^{\ddagger},k}}{2RT}\right)}{1-\exp\left(-\frac{h\nu_{AB^{\ddagger},k}}{RT}\right)}
\label{eq:q_vib} 
\end{align}
where $N$ is the amount of atoms in the system. 
Frequencies $\nu_{A,k}$ and $\nu_{AB^{\ddagger},k}$ are obtained from the square-rooted eigenvalues of the mass-weighted Hessian of the reactant and transition states respectively\cite{ghysbrecht2023thermal,ochterski1999vibrational}. 
Rates calculated by combining eqs.~\ref{eq:EyringTST} and \ref{eq:q_vib} are given in Fig.~\ref{fig:Eyring_high_T}.a and b in blue.

Instead of using the quantum mechanical partition function, one can also use classical partition functions for the vibrational degrees of freedom in a so-called high-temperature approximation.
The high-temperature equivalents of eq.~\ref{eq:q_vib} are given by\cite{ghysbrecht2023thermal}
\begin{align}
    q_{A;\mathrm{ht},\mathrm{vib}} = \prod_{k=1}^{3N-6}\frac{RT}{h\nu_{A,k}} \quad \mathrm{and} \quad \widetilde{q}_{AB^\ddagger;\mathrm{ht},\mathrm{vib}} = \prod_{k=1,k\ne r}^{3N-6}\frac{RT}{h\nu_{AB^{\ddagger},k}}
\end{align}
and the corresponding high-temperature approximation of Eyring's TST
\begin{align}
    k_{AB}^\mathrm{ht} &= \frac{RT}{h} \frac{\widetilde{q}_{\mathrm{AB}^\ddagger;\mathrm{ht},\mathrm{vib}}}{q_{A;\mathrm{ht},\mathrm{vib}}} \exp\left(-\frac{E_b}{RT}\right)  \\
    &= \frac{\prod_{k=1}^{3N-6}  \nu_{A,k}} {\prod_{k=1, k\ne r}^{3N-6} \nu_{AB^{\ddagger},k}}  \exp\left(-\frac{E_b}{RT}\right)\, .
    \label{eq:highT_TST}
\end{align}
Rates from eq.~\ref{eq:highT_TST} are given in Fig.~\ref{fig:Eyring_high_T}.a and b in orange.
The high-temperature limit estimates the rate corresponding to sampling from a fully classical dynamics on the potential energy surface given by the force field.
This is the same rate as estimated by the rate methods used above, as all of these are based in classical MD simulations, i.e.~calculated from simulations integrating Newton's laws of motion.
When using the quantum partition functions in eq.~\ref{eq:q_vib}, however, quantization of the vibrational degrees of freedom is taken into account.
Comparing rates from Eyring's TST to rates using the high-temperature limit thus gives us an idea of the impact of this quantization.
From Fig.~\ref{fig:Eyring_high_T}.c and d, quantization is expected to increase rates by a factor between 4 to 11.

\clearpage
\newpage

\section{Additional tables and figures}

\begin{table}[h!]
    \centering
    \begin{tabular}{|l | c || c | c || c | c |}

     \textbf{method}      & \textbf{equation}   &\multicolumn{2}{c||}{\textbf{Moderate friction}} & \multicolumn{2}{c|}{\textbf{High friction}} \\ 
    \hline
        \hline
         \multicolumn{2}{|c||}{} &\multicolumn{4}{c|}{\textbf{Small barrier}} \\       
    \hline    
    \multicolumn{2}{|c||}{}
                & $k_{AB}$ [ps$^{-1}$]     
                & $k_{BA}$ [ps$^{-1}$]   
                & $k_{AB}$ [ps$^{-1}$]
                & $k_{BA}$ [ps$^{-1}$] \\
    \hline
    \hline
    Simple TST                &(\ref{main-eq:simpleTST})
                                & $  1.86  \times 10^{-  2 }$ 
                                & $  1.94  \times 10^{-  2 }$ 
                                & $  1.86  \times 10^{-  2 }$ 
                                & $  1.94  \times 10^{-  2 }$ \\
    \hline
    Kramers (weak friction)      &(\ref{main-eq:KramersWeakLimit})
                                & $  6.55  \times 10^{-  2 }$
                                & $  3.53  \times 10^{-  2 }$
                                & $  2.09  \times 10^{-  1 }$
                                & $  1.13  \times 10^{-  1 }$ \\
    \hline
    Kramers (moderate friction)      &(\ref{main-eq:KramersModerate})
                                & $  1.44  \times 10^{-  2 }$
                                & $  6.77  \times 10^{-  3 }$
                                & $  8.84  \times 10^{-  3 }$
                                & $  4.13  \times 10^{-  3 }$ \\
                                
    \hline
    Kramers (high friction)      &(\ref{main-eq:KramersHigh})
                                & $  3.11  \times 10^{-  2 }$
                                & $  1.48  \times 10^{-  2 }$
                                & $  9.73  \times 10^{-  3 }$
                                & $  4.63  \times 10^{-  3 }$ \\
                                
    \hline
    Grid-based                        
    &(\ref{main-eq:Qij_SqRA_1})
                                & $  3.03  \times 10^{-  2 }$
                                & $  1.43  \times 10^{-  2 }$
                                & $  9.47  \times 10^{-  3 }$
                                & $  4.49  \times 10^{-  3 }$ \\

    \hline
     Direct simulation                 &                                 
                                & {\begin{tabular}[c]{@{}l@{}}
                                $  1.48  \times 10^{- 2 }$ \\ 
                                $  \pm$\\
                                $  1.53  \times 10^{- 2 }$ 
                                \end{tabular}}
                                
                                & {\begin{tabular}[c]{@{}l@{}}
                                $  6.52  \times 10^{- 3 }$ \\ 
                                $  \pm$\\
                                $  6.59  \times 10^{- 3 }$ 
                                \end{tabular}}
                                
                                & {\begin{tabular}[c]{@{}l@{}}
                                $  1.05  \times 10^{- 2 }$ \\ 
                                $  \pm$\\
                                $  9.72  \times 10^{- 3 }$ 
                                \end{tabular}}
                                
                                & {\begin{tabular}[c]{@{}l@{}}
                                $  5.04  \times 10^{- 3 }$ \\ 
                                $  \pm$\\
                                $  4.95  \times 10^{- 3 }$ 
                                \end{tabular}} \\
    \hline
    \hline
         \multicolumn{2}{|c||}{} &\multicolumn{4}{c|}{\textbf{High barrier}} \\       
    \hline
    \hline
    Simple TST                &(\ref{main-eq:simpleTST})
                                & $  4.58  \times 10^{-  9 }$ 
                                & $  4.63  \times 10^{-  9 }$ 
                                & $  4.58  \times 10^{-  9 }$  
                                & $  4.63  \times 10^{-  9 }$  \\
    \hline
    Kramers (weak friction)      &(\ref{main-eq:KramersWeakLimit})
                                & $  7.10  \times 10^{-  9 }$ 
                                & $  3.32  \times 10^{-  9 }$ 
                                & $  4.26  \times 10^{-  7 }$ 
                                & $  1.99  \times 10^{-  7  }$  \\
    \hline
    Kramers (moderate friction) &(\ref{main-eq:KramersModerate})
                                & $  4.47  \times 10^{-  9 }$ 
                                & $  2.02  \times 10^{-  9 }$ 
                                & $  1.40  \times 10^{-  9 }$ 
                                & $  6.38  \times 10^{- 10  }$  \\
    \hline
    Kramers (high friction)     &(\ref{main-eq:KramersHigh})
                                & $  9.33  \times 10^{-  8 }$ 
                                & $  4.22  \times 10^{-  8 }$ 
                                & $  1.55  \times 10^{-  9 }$ 
                                & $  7.04  \times 10^{- 10  }$  \\
                                
    \hline
    Grid-based                        
    &(\ref{main-eq:Qij_SqRA_1})
                                & $  8.40  \times 10^{-  8 }$ 
                                & $  3.80  \times 10^{-  8 }$ 
                                & $  1.40  \times 10^{-  9 }$ 
                                & $  6.34  \times 10^{- 10  }$  \\

    \hline
     Infrequent metadynamics    
     &(\ref{main-eq:InMetaD})
                                & {\begin{tabular}[c]{@{}l@{}}
                                $  4.60  \times 10^{- 9 }$ \\ 
                                $  \pm$\\
                                $  4.26  \times 10^{-  9  }$ \end{tabular}}
                                
                                & {\begin{tabular}[c]{@{}l@{}}
                                $  1.84  \times 10^{- 9 }$ \\ 
                                $  \pm$\\
                                $  1.37  \times 10^{-  9 }$ \end{tabular}}
                                
                                & {\begin{tabular}[c]{@{}l@{}}
                                $  1.24  \times 10^{- 9 }$ \\ 
                                $  \pm$\\
                                $  4.05  \times 10^{-  10  }$ \end{tabular}}
                                
                                & {\begin{tabular}[c]{@{}l@{}}
                                $  6.29  \times 10^{- 10 }$ \\ 
                                $  \pm$\\
                                $  6.72  \times 10^{-  10  }$ \end{tabular}} \\
    \hline
    \hline
         \multicolumn{2}{|c||}{} &\multicolumn{4}{c|}{\textbf{Interpolated potential}} \\       
    \hline
    \hline
    Simple TST                &(\ref{main-eq:simpleTST})
                                & $  6.08  \times 10^{- 13 }$ 
                                & $  1.58  \times 10^{- 13 }$ 
                                & $  6.08  \times 10^{- 13 }$  
                                & $  1.58  \times 10^{- 13 }$  \\
    \hline
    Kramers (weak friction)      &(\ref{main-eq:KramersWeakLimit})
                                & $  1.47  \times 10^{- 12 }$ 
                                & $  4.11  \times 10^{- 13 }$ 
                                & $  5.91  \times 10^{- 11 }$ 
                                & $  1.64  \times 10^{- 11  }$  \\                                
    \hline
    Kramers (moderate friction)      &(\ref{main-eq:KramersModerate})
                                & $  5.92  \times 10^{- 13 }$ 
                                & $  1.55  \times 10^{- 13 }$ 
                                & $  2.46  \times 10^{- 13 }$ 
                                & $  6.59  \times 10^{- 14  }$  \\
                                
    \hline 
    Kramers (high friction)      &(\ref{main-eq:KramersHigh})
                                & $  9.31  \times 10^{- 12 }$ 
                                & $  2.20  \times 10^{- 12 }$ 
                                & $  2.32  \times 10^{- 13 }$ 
                                & $  5.50  \times 10^{- 14 }$  \\
                                
    \hline
    Grid-based                        
    &(\ref{main-eq:Qij_SqRA_1})
                                & $  1.17  \times 10^{- 11  }$ 
                                & $  3.18  \times 10^{- 12  }$ 
                                & $  2.94  \times 10^{- 13  }$ 
                                & $  7.95  \times 10^{- 14  }$  \\

    \hline
     Infrequent metadynamics    
     &(\ref{main-eq:InMetaD})
                                & {\begin{tabular}[c]{@{}l@{}}
                                $  5.21  \times 10^{- 13 }$ \\ 
                                $  \pm$\\
                                $  6.39  \times 10^{- 13 }$ \end{tabular}}
                                
                                & {\begin{tabular}[c]{@{}l@{}}
                                $  1.76  \times 10^{-  13  }$ \\ 
                                $  \pm$\\
                                $  2.18  \times 10^{-  13  }$ \end{tabular}}
                                
                                & {\begin{tabular}[c]{@{}l@{}}
                                $  2.07  \times 10^{-  13  }$ \\ 
                                $  \pm$\\
                                $  1.98  \times 10^{-  13  }$ \end{tabular}}
                                
                                & {\begin{tabular}[c]{@{}l@{}}
                                $  5.28  \times 10^{-  14  }$ \\ 
                                $  \pm$\\
                                $  4.95  \times 10^{-  14  }$ \end{tabular}} \\
    \hline
    \hline
    \end{tabular}
    \caption{One-dimensional potential. 
    Kinetic rates estimated at specific friction values: 
    $\xi = 2.5\, \mathrm{ps}^{-1}$ (moderate friction - small barrier); 
    $\xi = 8\, \mathrm{ps}^{-1}$ (high friction - small barrier); $\xi = 0.5\, \mathrm{ps}^{-1}$ (moderate friction - high barrier);
    $\xi = 30\, \mathrm{ps}^{-1}$ (high friction - high barrier);
    $\xi = 2.5\, \mathrm{ps}^{-1}$ (moderate friction - interpolated potential);
    $\xi = 100\, \mathrm{ps}^{-1}$ (high friction - interpolated potential).
    }
    \label{tab:rates1D}
\end{table}

%

\clearpage
\subsection{Optimized reaction coordinate}

\begin{figure}[h]
\includegraphics[scale=1]{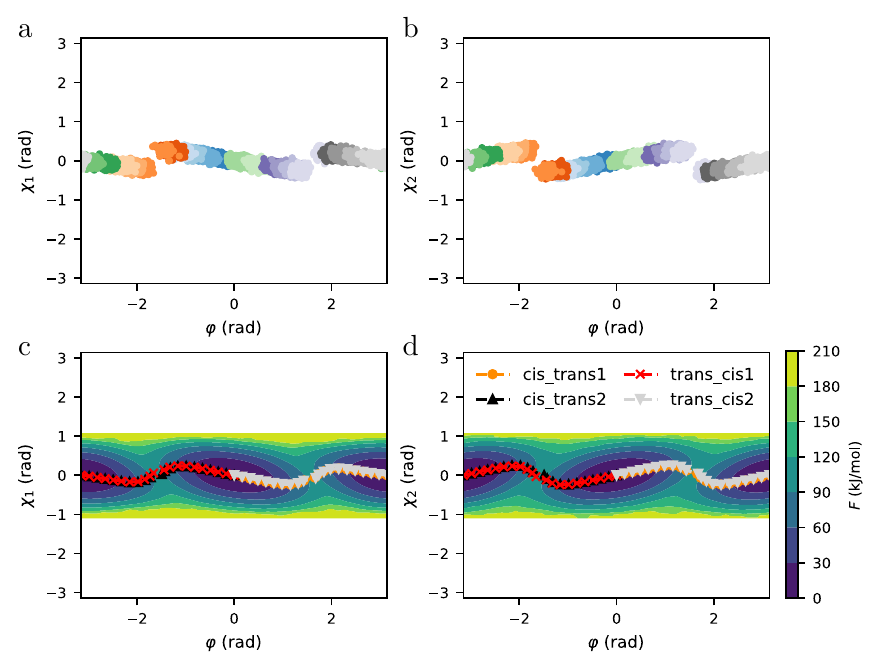}
\caption{Unscaled version of Fig.~\ref{main-fig:2D_correlations_FES_paths} to indicate how small the out of plane bending of the improper dihedral substituents really are.}
\label{fig:unscaled_2D}
\centering
\end{figure}

\begin{figure}[h]
\includegraphics[scale=1]{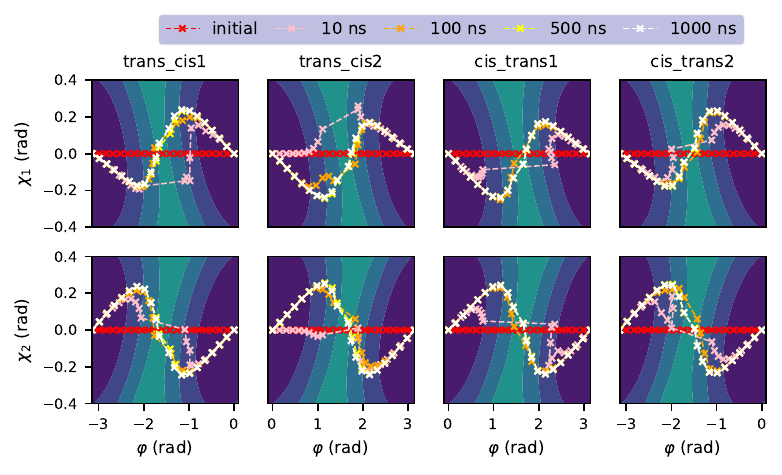}
\caption{Time evolution of 21 nodes of path collective variables for four trajectories of $1\,\mu\mathrm{s}$ of metadynamics, each trajectory representing one transition: \textit{cis\_trans1}, \textit{cis\_trans2}, \textit{trans\_cis1} and \textit{trans\_cis2}. 
\textbf{Top row:} path evolution represented in 2D space spanned by $\varphi$ and $\chi_1$. Underlying contour plot taken from 2D-reweighted free energy surface from 3D metadynamics simulation (see also Fig.~\ref{main-fig:2D_correlations_FES_paths} left). 
\textbf{Bottom row:} path evolution represented in 2D space spanned by $\varphi$ and $\chi_2$. Underlying contour plot taken from 2D-reweighted free energy surface from 3D metadynamics simulation (see also Fig.~\ref{main-fig:2D_correlations_FES_paths} right). 
}
\label{fig:2D_path_convergence}
\centering
\end{figure}

\begin{figure}
\includegraphics[scale=1]{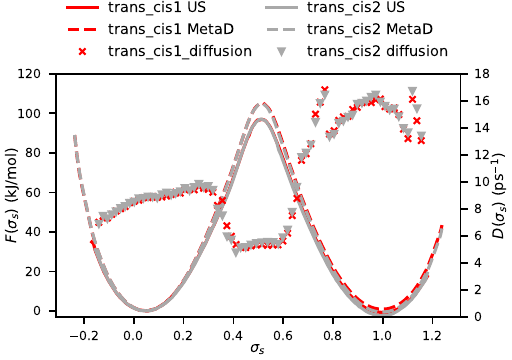}
\caption{Free energy profiles from metadynamics and umbrella sampling as well as diffusion profiles for optimized trans to cis path collective variable.
}
\label{fig:path_profiles_trans_cis}
\centering
\end{figure}

\clearpage
\subsection{Umbrella sampling vs. metadynamics}
\label{sec:US_vs_metad}
\begin{figure*}
\includegraphics[scale=1.]{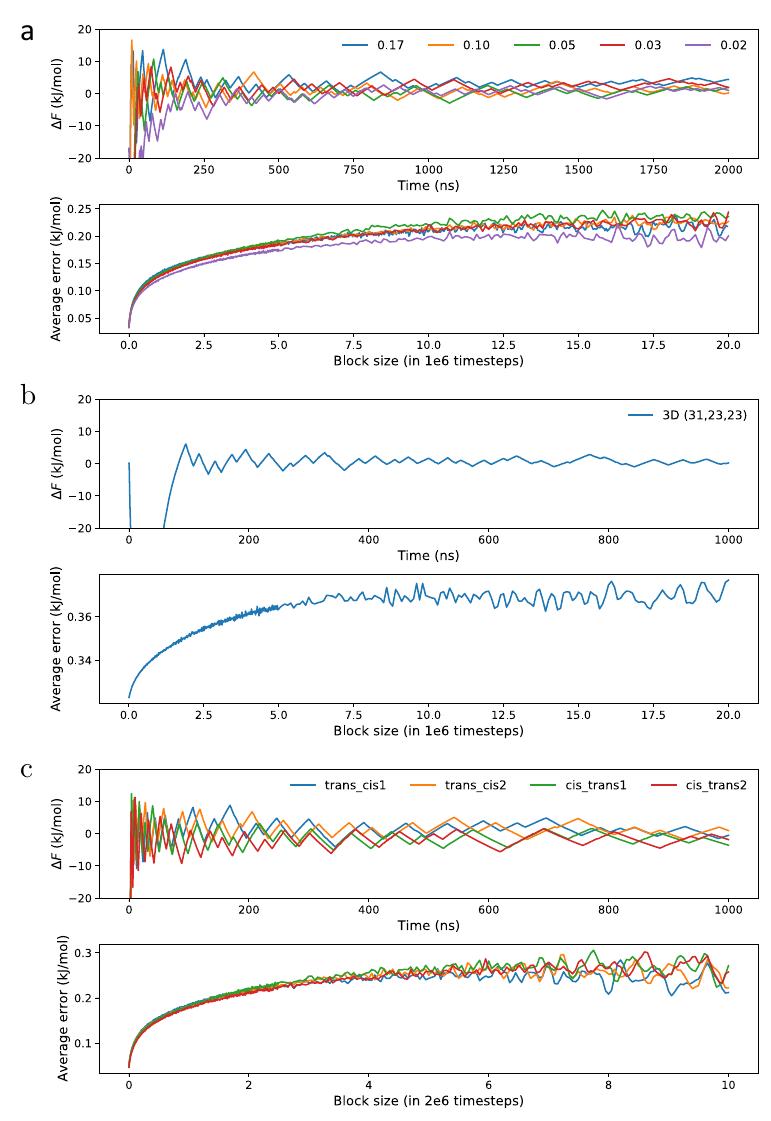}
\caption{
\textbf{Convergence of the metadynamics bias}
Convergence of free energy difference between trans and cis for metadynamics bias and error convergence from block analysis for:
\textbf{a:} MetaD simulations biasing the C$_{13}$=C$_{14}$ dihedral angle $\varphi$ for different Gaussian standard deviations (in radians). 
All these simulations were run for $2\,\mathrm{\mu s}$ with a deposition pace of $1\,\mathrm{ps}$ and a biasing factor of 10.
\textbf{b:} 3D MetaD simulation. Error convergence from block analysis for 3D free energy surface was done using discretization (31,23,23).
Simulation was run for $1\,\mathrm{\mu s}$ with a deposition pace of $1\,\mathrm{ps}$ and a biasing factor of 12.
\textbf{c:} MetaD simulations along path CVs. 
Simulation was run for $1\,\mathrm{\mu s}$ with a deposition pace of $0.5\,\mathrm{ps}$ and a biasing factor of 12.
}
\label{fig:metad_convergence}
\centering
\end{figure*}
Convergence of the free energy difference between the cis and trans state $\Delta F=F_\mathrm{cis}-F_\mathrm{trans}$ in the metadynamics biases are given in Fig.~\ref{fig:metad_convergence} for different metadynamics runs.
For the corresponding details about parameter sets, see SI section \ref{sec:methods_MD}.
For the final free energy surfaces, see Figs.~\ref{main-fig:FES_phi} and \ref{main-fig:2D_correlations_FES_paths}.
The free energy difference at a certain simulation time is calculated by determining the FES corresponding to the bias at that time (i.e.~from the scaled upside-down bias, see Refs.~\citenum{branduardi2012metadynamics,bussi2019analyzing,bussi2020using}). 
This FES is used to calculate the relative probabilities of being in cis versus being in trans.
Using Eq.~\ref{eq:config_Boltzmann_density}:
\begin{equation}
    \pi_\mathrm{cis} = \int_{-\pi/2}^{\pi/2}\mathrm{d}\varphi \, \pi(\varphi) = \int_{-\pi/2}^{\pi/2}\mathrm{d}\varphi \, \exp\left(-\frac{F(\varphi)}{RT}\right)
\end{equation}
and equivalent for trans in $\varphi<-\pi/2$ and $\varphi>\pi/2$.
For the 3D FES, the integration is additionally carried out over $\chi_1$ and $\chi_2$ over their full range.
The free energy of a state can then be calculated using $F_\mathrm{cis}=-RT\ln \pi_\mathrm{cis}$ and equivalent for trans, and the free energy difference
\begin{align}
    \Delta F = F_\mathrm{cis} - F_\mathrm{trans} = -RT\ln\frac{\pi_\mathrm{cis}}{\pi_\mathrm{trans}}  \, .
\end{align}

\begin{figure}[h]
\includegraphics[scale=1]{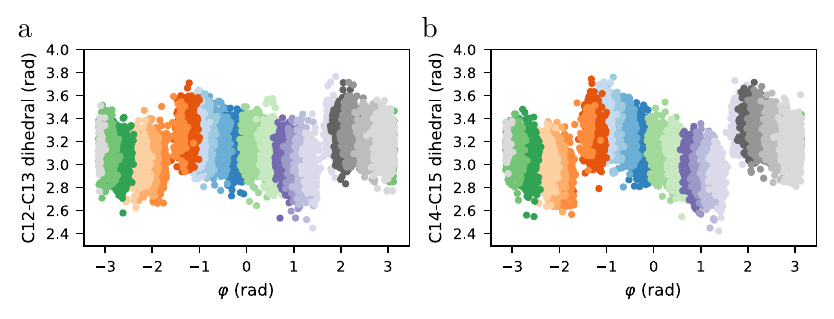}
\caption{\textbf{a, b:} Scatter plot of proper dihedrals C$_{13}$=C$_{14}$-C$_{15}$=NH (a) and C$_{11}$=C$_{12}$-C$_{13}$=C$_{14}$ (b) versus $\varphi$ for umbrella sampling along $\varphi$ (US \textit{set2}).
These proper dihedrals are also correlated and also cause hysteresis when using $\varphi$ as a reaction coordinate (compare to Fig.~\ref{main-fig:2D_correlations_FES_paths}).}
\label{fig:phi_proper_correlation}
\centering
\end{figure}

\clearpage
\subsection{Multidimensional models}

\clearpage 
\section{References}
\bibliographystyle{unsrt}
\bibliography{literature}

\makeatletter\@input{yy.tex}\makeatother